\documentclass[graybox, usenatbib]{svmult}

\usepackage{mathptmx}       % selects Times Roman as basic font
\usepackage{helvet}         % selects Helvetica as sans-serif font
\usepackage{courier}        % selects Courier as typewriter font
\usepackage{type1cm}        % activate if the above 3 fonts are
\usepackage{makeidx}         % allows index generation
\usepackage{graphicx}        % standard LaTeX graphics tool
\usepackage{multicol}        % used for the two-column index
\usepackage[bottom]{footmisc}% places footnotes at page bottom
\usepackage{natbib}

\makeindex             % used for the subject index
\def\gtsima{$\; \buildrel > \over \sim \;$}
\def\ltsima{$\; \buildrel < \over \sim \;$}
\def\gtrsim{\lower.5ex\hbox{\gtsima}}
\def\lesssim{\lower.5ex\hbox{\ltsima}}
\newcommand\msun{\, \rm M_\odot} 
\newcommand\kms{{\, \rm km\,s^{-1}}}
\newcommand\mpc{{\, \rm mpc}}
\newcommand\pcc{{\, \rm pc^{-3}}}

\begin{document}

\title*{Star Formation and Dynamics in the Galactic Centre}
% Use \titlerunning{Short Title} for an abbreviated version of
% your contribution title if the original one is too long
\author{Michela Mapelli and Alessia Gualandris}
% Use \authorrunning{Short Title} for an abbreviated version of
% your contribution title if the original one is too long
\institute{Michela Mapelli\at INAF, Osservatorio Astronomico di Padova, Vicolo dell'Osservatorio 5, I--35122, Padova, Italy \email{michela.mapelli@oapd.inaf.it}
\and Alessia Gualandris \at Department of Physics, University of Surrey, Guildford GU2 7XH, United Kingdom \email{a.gualandris@surrey.ac.uk}}
%
% Use the package "url.sty" to avoid
% problems with special characters
% used in your e-mail or web address
%
\maketitle

\abstract*{The centre of our Galaxy is one of the most studied and yet enigmatic places in the Universe. At a distance of about 8 kpc from our Sun, the Galactic centre (GC) is the ideal environment to study the extreme processes that take place in the vicinity of a supermassive black hole (SMBH). Despite the hostile environment, several tens of early-type stars populate the central parsec of our Galaxy. A fraction of them lie in a thin ring with mild eccentricity and inner radius $\sim{}0.04$ pc, while the S-stars, i.e. the $\sim{}30$ stars closest to the SMBH ($\lesssim{}0.04$ pc), have randomly oriented and highly eccentric orbits. The formation of such early-type stars has been a puzzle for a long time: molecular clouds should be tidally disrupted by the SMBH before they can fragment into stars. We review the main scenarios  proposed to explain the formation and the dynamical evolution of the early-type stars in the GC. In particular, we discuss the most popular {\it in situ} scenarios (accretion disc fragmentation and molecular cloud disruption) and {\it migration} scenarios (star cluster inspiral and Hills mechanism). We focus on the most pressing challenges that must be faced to shed light on the process of star formation in the vicinity of a SMBH.}
\abstract{The centre of our Galaxy is one of the most studied and yet enigmatic places in the Universe. At a distance of about 8 kpc from our Sun, the Galactic centre (GC) is the ideal environment to study the extreme processes that take place in the vicinity of a supermassive black hole (SMBH). Despite the hostile environment, several tens of early-type stars populate the central parsec of our Galaxy. A fraction of them lie in a thin ring with mild eccentricity and inner radius $\sim{}0.04$ pc, while the S-stars, i.e. the $\sim{}30$ stars closest to the SMBH ($\lesssim{}0.04$ pc), have randomly oriented and highly eccentric orbits. The formation of such early-type stars has been a puzzle for a long time: molecular clouds should be tidally disrupted by the SMBH before they can fragment into stars. We review the main scenarios  proposed to explain the formation and the dynamical evolution of the early-type stars in the GC. In particular, we discuss the most popular {\it in situ} scenarios (accretion disc fragmentation and molecular cloud disruption) and {\it migration} scenarios (star cluster inspiral and Hills mechanism). We focus on the most pressing challenges that must be faced to shed light on the process of star formation in the vicinity of a SMBH.}

%Understanding the formation of such early-type stars has been a challenge for a long time
%we discuss the role of precession and relaxation processes on the dynamical evolution of stellar orbit in the GC, giving particular attention to the intriguing open questions that make 
%about 8 kpc far from
%The centre of our Galaxy is one of the most studied and yet enigmatic places of the Universe. At a distance of about 8 kpc from our Sun, the Galactic centre (GC) is the ideal environment to study the extreme processes that take place in the vicinity of a supermassive black hole (SMBH). Despite the hostile environment, several tens of early-type stars populate the central parsec of our Galaxy. A fraction of them lie in a thin ring with mild eccentricity and inner radius $\sim{}0.04$ pc, while the S-stars, i.e. the $\sim{}30$ stars closest to the SMBH ($\lesssim{}0.04$ pc), have randomly oriented and highly eccentric orbits. The formation of such early-type stars has been a puzzle for a long time: molecular clouds should be tidally disrupted by the SMBH before they can fragment into stars. We review the main scenarios  proposed to explain the formation and the dynamical evolution of the early-type stars in the GC. In particular, we discuss the most popular {\it in situ} scenarios (accretion disc fragmentation and molecular cloud disruption) and {\it migration} scenarios (star cluster inspiral and Hills mechanism). We focus on the most pressing challenges that must be faced to shed light on the process of star formation in the vicinity of a SMBH.}

\section{Introduction: the Galactic centre as a laboratory for both dynamics and star formation under extreme conditions}
\label{sec:intro}

The Galactic centre (GC) is a unique laboratory to study physical
processes in the vicinity of a supermassive black hole (SMBH). In
fact, the GC hosts the only concentration of mass
($\approx{}4\times{}10^6\msun$) that can be identified with a
SMBH beyond reasonable doubt (\citealt{Schodel02};
\citealt{Ghez03}). Furthermore, its distance from our Sun
($\approx{}8$ kpc) is several orders of magnitude smaller than the
distance from the other SMBH candidates.  Despite the hostile
environment due to the presence of a SMBH, the GC is an overwhelmingly
crowded environment: the observations have revealed the presence of
molecular, atomic and ionized gas, of a cusp of late-type stars, and
of $\sim{}100-200$ early-type stars\index{Early-type stars}. About 20-50 \% of the early-type
stars\index{Early-type stars} lie in a relatively thin ring (with inner radius $\sim{}0.04$
pc) and follow a top-heavy mass function (MF, e.g. \citealt{Paumard06,Bartko09,Lu13,Yelda2014})\index{Mass function}. The $\sim{}30$ stars closest
($\lesssim{}0.04\,{}{\rm pc}\sim{}1$ arcsec) to SgrA$^\ast{}$
(i.e. the radio source that is associated with the central SMBH) are B
stars, with an age $<100$ Myr. These, named the S-stars, have very
eccentric and randomly oriented orbits. The presence of the early-type
stars\index{Early-type stars} in the central parsec is particularly puzzling, because the
gravitational shear exerted by the SMBH disrupts molecular clouds
before they can fragment into stars.

Because of its unique characteristics, the GC has been the subject of
a plethora of studies and of a few dedicated reviews
(e.g. \citealt{Morris96, Genzel10}) over the last $\sim{}20$
years. Our review does not pretend to be either more complete or
detailed than previous ones. Rather, it looks at the GC from a slightly
different perspective: it focuses on the {\it young stars} that
populate the GC, and on the {\it theoretical scenarios} that have been
proposed to explain {\it their formation and their dynamical
  evolution}.

The review is structured as follows. In Sect.~\ref{sec:2}, we
briefly summarize the state-of-the-art knowledge about the GC from an
observational point of view, focusing on those aspects that are more
relevant for the formation of the early-type stars\index{Early-type stars}.  In
Sect.~\ref{sec:3}, we discuss the main scenarios that have been
proposed for the formation of the early-type stars\index{Early-type stars} (including disc
fragmentation\index{Fragmentation}, molecular cloud disruption\index{Tidal disruption}, inspiral of a star cluster
and tidal break-up of binaries). Sect.~\ref{sec:4} is devoted to the
dynamical evolution of the early-type stars\index{Early-type stars}, considering both
different relaxation mechanisms and secular processes.  Finally,
Sect.~\ref{sec:5} deals with the main theoretical scenarios which
have been proposed to explain the nature of one of the most peculiar
objects that have been observed in the GC: the dusty object G2.

\section{A crowded environment}
\label{sec:2}
In this Section, we briefly review the most updated observations of
the main components of the GC: the SMBH (\ref{subsec:2.1}), the young
and old stars (\ref{subsec:2.2}), the gas component (\ref{subsec:2.3})
and the recently discovered, very puzzling G2 cloud
(\ref{subsec:2.4}). We also discuss the possibility that the GC hosts
one or more intermediate-mass black holes
(IMBHs\index{Intermediate-mass black hole}), i.e. black holes with
mass in the $10^2-10^5\msun$ range (\ref{subsec:2.5}).

 In the next Sections (\ref{sec:3}--\ref{sec:5}),
we will focus on the theoretical interpretation of such observations,
and in particular on the processes that drive the formation and
evolution of stars in the GC.

\subsection{The supermassive black hole}
\label{subsec:2.1}
The very first hints for the presence of a SMBH\index{Supermassive black hole|textbf} candidate in the centre of the Milky Way (MW) came from the detection of a very compact radio source  (\citealt{balick74}) in the innermost parsec.

The first attempts to estimate the mass enclosed in the central parsec are radial velocity
measurements of ionized gas located in the structure which
 is known as minispiral
(\citealt{Lacy80}). On the other hand, the radial velocity of ionized
gas may be affected by a plethora of processes besides gravity. Thus,
the first strong claim for a dark mass in the centre of the MW came
from radial velocity measurements of stars, obtained by means of
near-infrared (NIR) spectra of the stellar population in the central
parsecs (e.g. \citealt{McGinn89}; \citealt{Sellgren90};
\citealt{Haller96}). These early measurements indicated the presence
of $\sim{}3\times{}10^6\msun$ confined in $\sim{}0.1$ pc,
corresponding to a minimum density of $\sim{}3\times{}10^9$ M$_\odot$
pc$^{-3}$. Such density is still consistent with a cluster of compact
stellar remnants (e.g. \citealt{Maoz98}). The first measurements of
stellar proper motions with diffraction-limited NIR observations
(\citealt{Genzel97}; \citealt{Eckart97}; \citealt{Ghez98})
strengthened the constraints significantly, indicating a
$2.6\pm{}0.6\times{}10^6\msun$ dark mass confined within $\sim{}0.01$
pc, corresponding to a minimum density $\sim{}10^{12}$ M$_\odot$
pc$^{-3}$. This density excludes the star cluster of compact remnants
(\citealt{Genzel97}) and leaves only two possible candidates: either a
SMBH or a fermion ball (e.g. \citealt{Tsiklauri98}). Tracing the orbit
of the so called S2 (or S0-2) star (with an orbital period $T_{\rm
  orb}=15.9$ yr) led to the measurement of
$3.7\pm{}1.5\times{}10^6\msun$ (\citealt{Schodel02}) and
$4.0\pm{}0.6\times{}10^6\msun$ (\citealt{Ghez03}, see also
\citealt{Ghez05} and \citealt{Ghez08}) in the inner 0.0006
pc. Finally, the most recent estimate of the S2 orbit leads to $m_{\rm
  BH} = 4.30\pm{} 0.20_{\rm stat} \pm{}0.30_{\rm sys} \times{}
10^6\msun$ (where $m_{\rm BH}$ is the mass of the SMBH,
\citealt{Gillessen09b}).  This value comes from a joint fit of New
Technology Telescope (NTT), Very Large Telescope (VLT) and Keck
astrometric data ranging from 1992 to 2003 (see Table~1 of
\citealt{Gillessen09b}). The largest source of uncertainty in this
measurement is our distance from the GC ($=8.28\pm{}0.15_{\rm
  stat}+0.29_{\rm sys}\,{}$km, \citealt{Gillessen09a};
\citealt{Gillessen09b}; see also \citealt{Morris12}).

One of the open questions about the SMBH is its possible past
activity. The strongest hint for a past activity is represented by
fluorescent X-ray line emission (e.g. \citealt{Sunyaev93}), especially
the 6.4 keV Fe K$\alpha{}$ line. This line is emitted by various
molecular clouds\index{Molecular cloud} in the GC (e.g. \citealt{Ponti10}). The lines emitted
from different clouds might be triggered by different sources
(e.g. different X-ray binaries), but this possibility is not supported
by observations of currently active X-ray sources. Thus, if the
fluorescent X-ray line emission comes from a single source, such
source must have been powerful enough: it might be the `echo' of an
energetic flaring event of Sgr~A$^\ast{}$ that occurred several
hundreds years ago, such as the tidal disruption\index{Tidal disruption} of a star or of a
smaller body (e.g. \citealt{Koyama96}; \citealt{Yu11}; see
\citealt{Morris12} for a recent review on this and related topics).

Recently, \cite{Rea2013} reported the discovery of a young magnetar
(SGR J1745$-$2900) at $2.4\pm{}0.3$ arcsec projected distance from
SgrA$^\ast{}$. The probability that the magnetar is a foreground or
background object is very low ($\sim{}10^{-6}$), while the probability
that it is on a bound orbit around the SMBH is non-negligible. If SGR
J1745$-$2900 is on a bound orbit, the fluorescent X-ray line emission
in the GC might be easily explained by a past ($\sim{}100$ year ago)
giant flare by the magnetar. This scenario is a non-unlikely
alternative to a past flare by Sgr~A$^\ast{}$.

%{\bf MM: aggiungere magnetar PEsposito} Fe K\alpha{}
%continuum-subtracted mosaic image of the different EPIC-pn
%observations of the bridge region. A brightening of the bridges 1, 2,
%3, and 4 is clear. Such variation occurs in a time-scale of about 2-4
%years, but in a spatial scale of about 15 light years. This apparent
%superluminal motion can be explained if the bridge MC is illuminated
%by a bright (L > 1.3 × 1038 erg s–1) and distant (>15 pc) X-ray
%source active for several years. Either a flare from Sgr A* or a
%bright and long outburst of a X-ray binary can be the source of such
%a phenomenon. The observation of superluminal echo cannot be
%explained by either a single internal source or by low energy cosmic
%ray irradiation. It is also highly unlikely that the variations is
%produced by several uncorrelated sources.  MASSA FROM S-stars

%QUIESCENZA e attivita nel passato

%PROVA PER LA RELATIVITa?

\subsection{The stars: old stars, early-type stellar disc(s), and S-stars}
\label{subsec:2.2}
\subsubsection{The nuclear star cluster}
The ensemble of the (both young and old) stars in the central few
parsecs is often referred to as the nuclear star cluster\index{Nuclear star cluster} 
(NSC) of the MW. NSCs are located at the photometric and dynamical
centre of almost all spiral galaxies (e.g. \citealt{Cote06} and
references therein), but the NSC of the MW is the only one where
single stars can be resolved and their proper motions measured
(\citealt{Genzel03}; \citealt{Schoedel07}; \citealt{Trippe08};
\citealt{Schoedel09}; \citealt{Schoedel10}). \cite{Eckart93} and
\cite{Genzel96} derived number density counts from high-resolution NIR
speckle imaging observations
%at the diffraction limit of a 4-m class telescope 
between 1 and 20 arcsec and found that the stellar density scales as
$\rho{}\propto{}r^{-2}$ (isothermal profile). Some indication for a
cusp (rather than a cored) central density was reported by
\cite{Eckart95} and by \cite{Alexander99}.
 %%%%%%%%%%%%%%%%%%%%%%%%%%%%%%%%FIGURE 1 %%%%%%%%%%%%%%%%%%%%%%%%%%%%%%%%%%%%%%%%%
\begin{figure}[t]
\sidecaption[t]
\includegraphics[height=6.5cm]{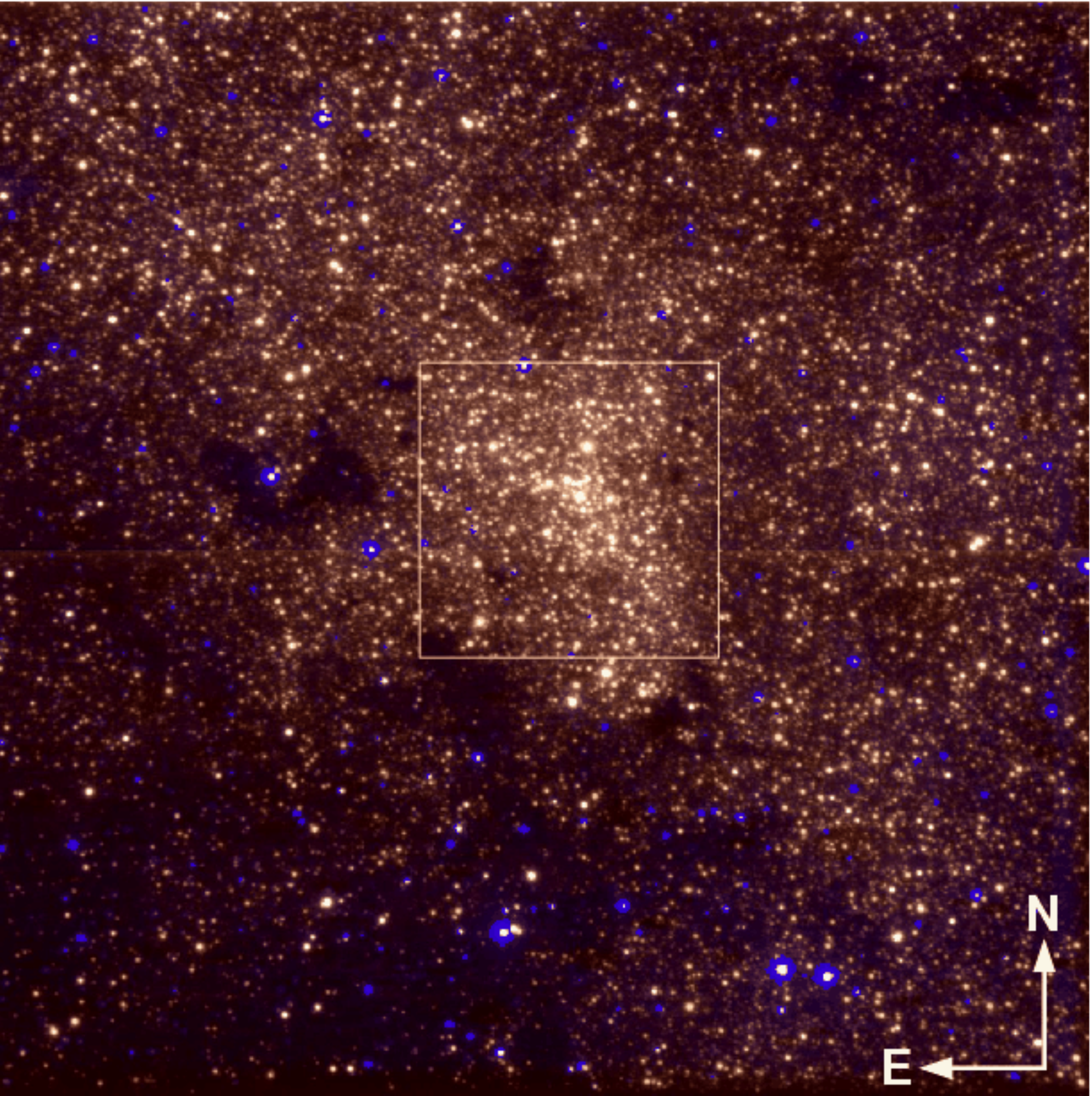}
\caption{Colour image composed of ISAAC imaging observations at 2.09
  $\mu{}$m and in the J-band. The field-of-view is $150''\times
  150''$. The field of about $40''\times 40''$ that was observed with
  adaptive-optics observations is marked by a square. The galactic
  plane runs approximately southwest-northeast across the image. From
  Fig.~1 of \cite{Schoedel07}. }
\label{fig:schoedel1}
\end{figure}
%%%%%%%%%%%%%%%%%%%%%%%%%%%%%%%%FIGURE %%%%%%%%%%%%%%%%%%%%%%%%%%%%%%%%%%%%%%%%%

\cite{Genzel03} combined high-resolution stellar number counts from
NACO\footnote{The adaptive optics module NAOS and the NIR camera
  CONICA (abbreviated as NACO) are mounted at the ESO 8 m-class VLT
  unit telescope 4 on Cerro Paranal, Chile.} $H-$ and $K-$ band
imaging data of the very central region ($0.1-10$ arcsec), with lower
resolution number counts from speckle imaging observations at
$10\le{}R/\textrm{arcsec}\le{}100$ (where $R$ is the projected
distance from Sgr~A$^\ast$). These data are best-fitted by a broken
power-law
\begin{equation}
\rho_{\ast{}}=1.2\times{}10^6\,{}\textrm{M}_\odot{}\,{}\textrm{pc}^{-3}\,{}\left(\frac{R}{10 \textrm{ arcsec}}\right)^{-\alpha{}},
\end{equation}
with $\alpha{}=2.0\pm{}0.1$ ($\alpha{}=1.4\pm{}0.1$) at $R\ge{}10$ arcsec ($R<10$ arcsec). 

\begin{figure}[t]
%\sidecaption[t]
\includegraphics[width=12.5cm]{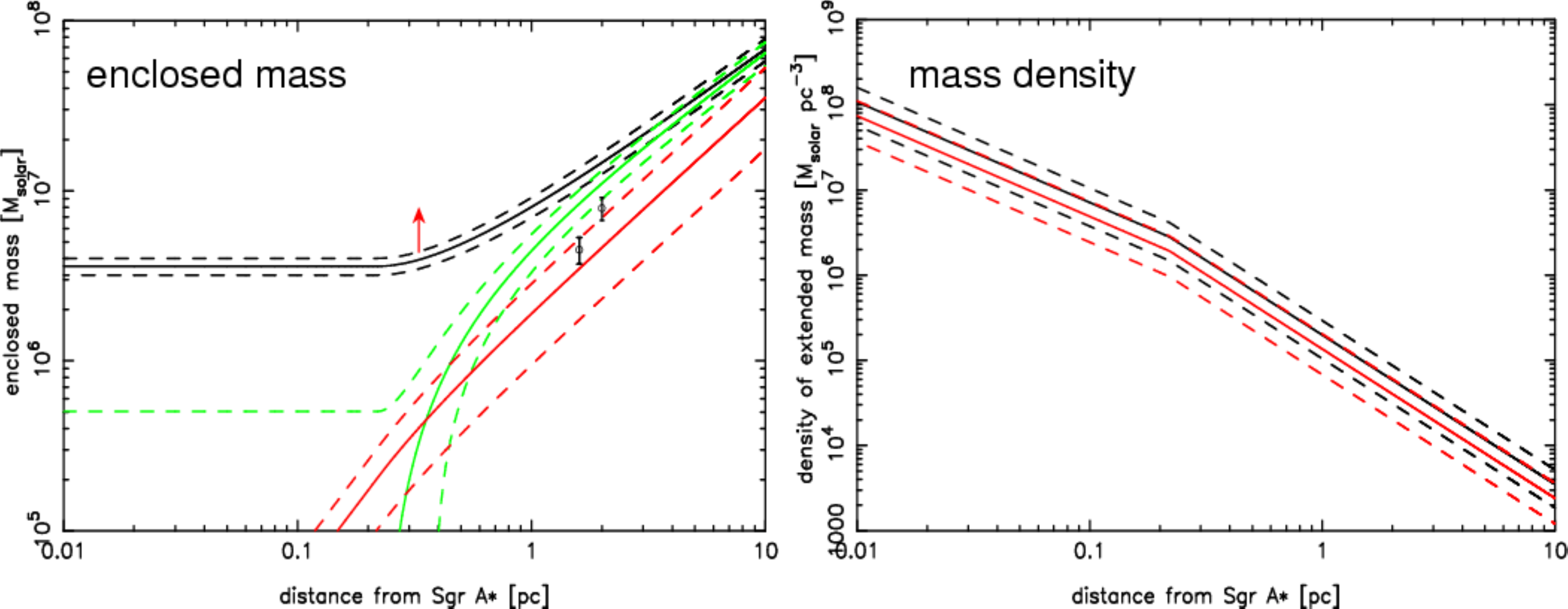}
\caption{Left-hand panel: estimate of the enclosed mass versus
  projected distance (black line), derived with the Bahcall-Tremaine
  (BT, \citealt{Bahcall81}) mass estimator, assuming a broken
  power-law structure of the stellar cluster and a constant
  line-of-sight velocity dispersion outside of the break radius (see
  text for details). The up-pointing red arrow is the enclosed mass
  estimate based on the bright star IRS 9 (\citealt{Reid2007}). The
  circle at 1.6 pc is the mass estimate based on the assumption that
  the circumnuclear ring (CNR)\index{Circumnuclear ring} is a rotating ring with a rotation
  velocity of 110 km s$^{-1}$ and a radius of 1.6 pc
  (\citealt{Christopher05}). The circle at 2.0 pc is the mass estimate
  based on the assumption that the CNR\index{Circumnuclear ring} is a rotating ring with a
  rotation velocity of 130 km s$^{-1}$ and a radius of 2.0 pc
  (\citealt{Rieke88}; \citealt{Gusten87}). Green line: enclosed mass after
  subtraction of the SMBH mass, derived from the BT mass estimator
  (black). Red line: estimated mass of the visible stellar
  cluster. The dashed lines indicate the $1\sigma $
  uncertainties. Right-hand panel: density of the enclosed mass, after
  subtraction of the SMBH mass (black). The red line indicates the
  mass density of the stellar cluster. The dashed lines indicate the
  $1\sigma $ uncertainties. From Fig.~19 of \cite{Schoedel07}.}
\label{fig:schoedel2}
\end{figure}

\cite{Schoedel07} confirm and refine this result, by means of an
homogeneous sample of high-resolution data (using the NIR camera and
spectrometer ISAAC at the ESO VLT UNIT telescope 4 on Paranal, see
Fig.~\ref{fig:schoedel1}). They find a best-fitting power-law
\begin{equation}\label{eq:schoedelc}
\rho_{\ast{}}=2.8\pm{}1.3\times{}10^6\,{}\textrm{M}_\odot{}\,{}\textrm{pc}^{-3}\,{}\left(\frac{R}{6 \textrm{ arcsec}}\right)^{-\alpha{}},
\end{equation}
with $\alpha{}=1.75$ ($\alpha{}=1.2$) at $R\ge{}6$ arcsec ($R<6$
arcsec). Thus, the updated break of the power law is $R_{\rm
  break}=6\pm{}1$ arcsec$=0.22\pm{}0.04$ pc. This implies that the NSC
contains about twice the SMBH mass in $<2$ pc (see
Fig.~\ref{fig:schoedel2}). The main assumptions that have been done to
obtain this result are (i) that the velocity dispersion is constant
outside 0.22 pc; (ii) that the NSC is spherically symmetric, does not
rotate and is isotropic; (iii) that the \cite{Bahcall81} mass
estimator can be used in the case of NSC; (iv) that the mass-to-light
ratio is 2 M$_\odot/$L$_\odot$ at 2$\mu{}$m (\citealt{Haller96}), to
estimate the unresolved stellar component.

%%%%%%%%%%%%%%%%%%%%%%%%%%%%%%%%FIGURE %%%%%%%%%%%%%%%%%%%%%%%%%%%%%%%%%%%%%%%%%
%%%%%%%%%%%%%%%%%%%%%%%%%%%%%%%%FIGURE %%%%%%%%%%%%%%%%%%%%%%%%%%%%%%%%%%%%%%%%%

\begin{figure}[t]
%\sidecaption[t]
\includegraphics[width=12.5cm]{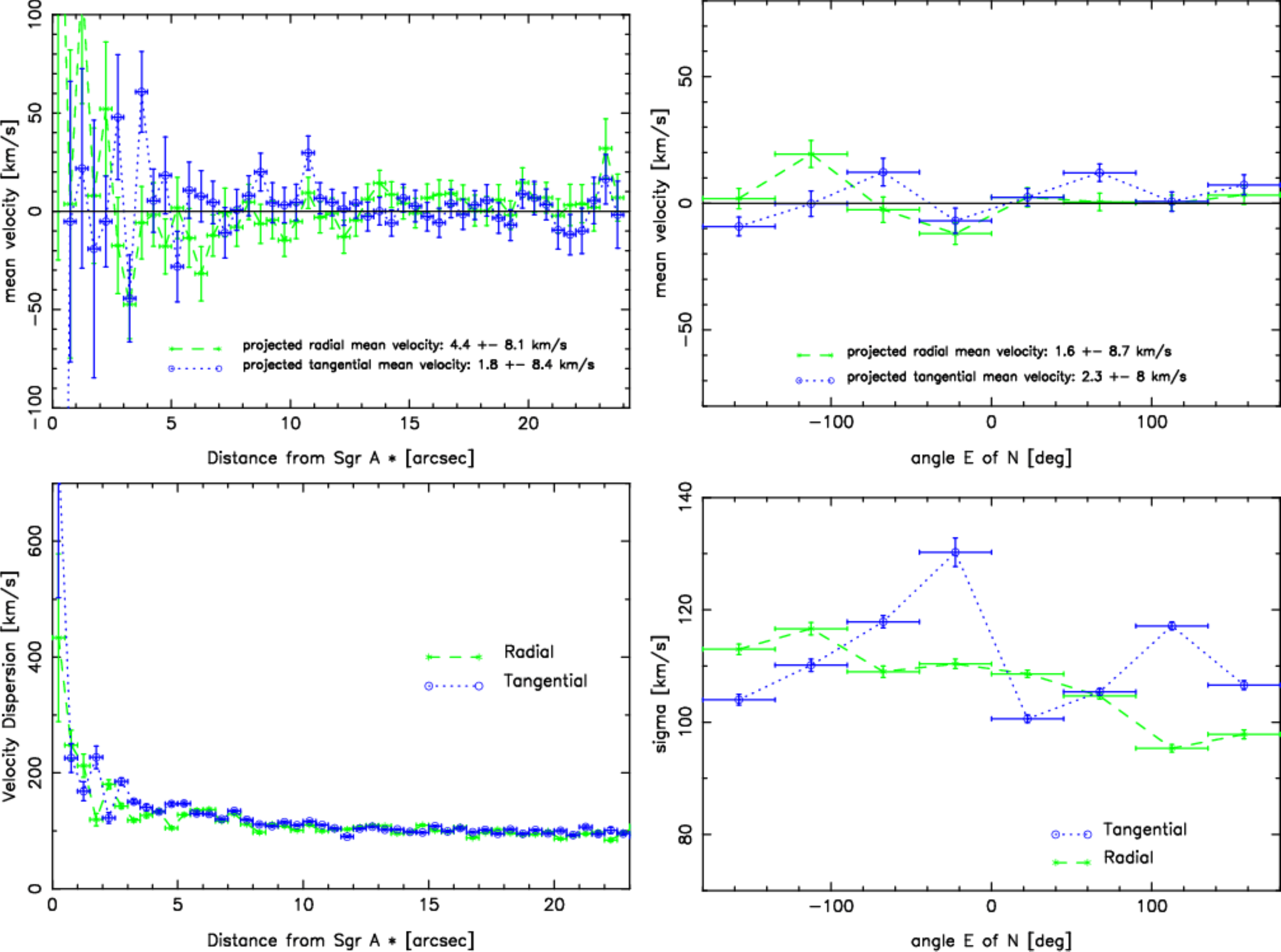}
\caption{Top: mean projected radial and tangential velocities versus
  projected distance from Sgr~A$^\ast$ (left) and versus angle east of
  north (right).  Bottom: projected radial (green) and tangential
  (blue) velocity dispersions in the GC NSC versus projected distance
  from Sgr A$^\ast$ (left) and versus angle east of north
  (right). From Fig.~6 of \cite{Schoedel09}.}
\label{fig:schoedel3}
\end{figure}

\cite{Schoedel09} use multi-epoch adaptive-optics assisted NIR
observations, obtained with NACO at VLT, to study the proper motions
of $>6000$ stars in the central parsec of the MW (with uncertainties
$<25$ km s$^{-1}$). They find that stellar velocities are purely
Keplerian only in the inner $\lesssim{}0.3$ pc, while the velocity
dispersion is nearly constant at $r>0.5$ pc (see
Fig.~\ref{fig:schoedel3}). Furthermore, \cite{Schoedel09} suggest that the velocity dispersion is isotropic. This result has been recently revised by \cite{Chatzopoulos14}, who claim that there are significant differences between proper motion dispersions along different axes, due to a flattening of the NSC. 
%, according to the analysis by \cite{Schoedel09} (but \citealt{cha.} These findings confirm the validity of some of the
%crucial assumptions done to derive the enclosed NSC mass (\citealt{Schoedel07})
In addition, the NSC is found to rotate
parallel to Galactic rotation (\citealt{Trippe08}; \citealt{Schoedel09}; \citealt{Chatzopoulos14}).

%%%%%%%%%%%%%%%%%%%%%%%%%%%%%%%%FIGURE %%%%%%%%%%%%%%%%%%%%%%%%%%%%%%%%%%%%%%%%%
%%%%%%%%%%%%%%%%%%%%%%%%%%%%%%%%FIGURE %%%%%%%%%%%%%%%%%%%%%%%%%%%%%%%%%%%%%%%%%

%%%%%%%%%%%%%%%%%%%%%%%%%%%%%%%%FIGURE %%%%%%%%%%%%%%%%%%%%%%%%%%%%%%%%%%%%%%%%%
\begin{figure}[t]
\sidecaption[t]
\includegraphics[width=6.5cm]{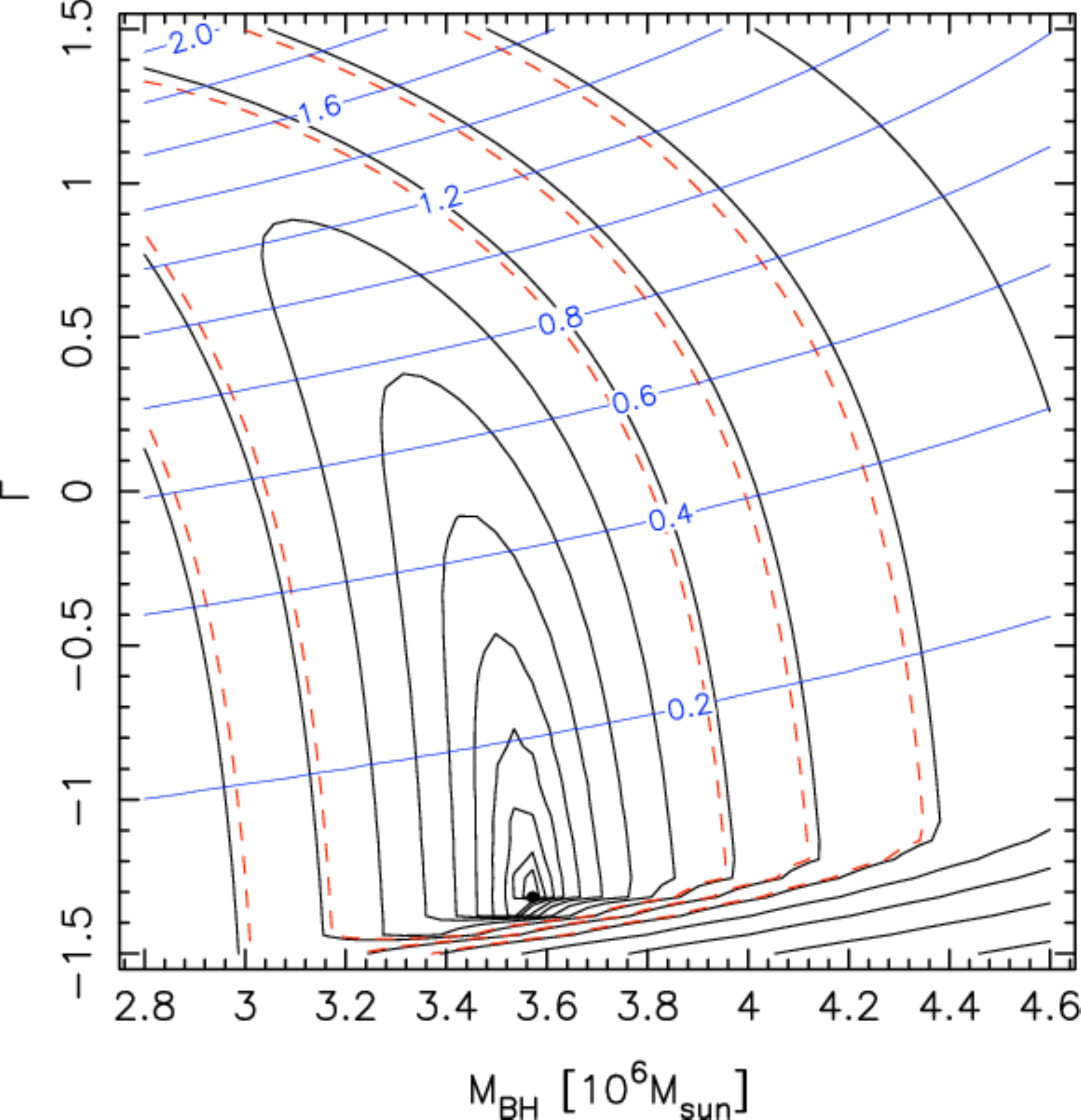}
\caption{Results of isotropic modelling of the NSC. The three free
  parameters ( $m_{\rm BH}, M_\star (r<1\textrm{ pc}), \Gamma $) were
  varied in comparing the fit of the model to the velocity dispersion
  data. Black thick curves are contours of constant $\chi ^{2}$,
  separated by a constant factor of 100.3; dashed red curves indicate
  (68\%, 90\% and 99\%) confidence intervals. Blue thin curves are
  contours of the best-fit value of $M_\star(r<1~{\rm pc})$ at each
  value of ( $m_{\rm BH}, \Gamma $); these curves are labelled by
  $M_\star/10^6\msun$. The overall best-fit model is indicated
  by the filled circle. From Fig.~14 of \cite{Schoedel09}.}
\label{fig:schoedel4}
\end{figure}
%%%%%%%%%%%%%%%%%%%%%%%%%%%%%%%%FIGURE %%%%%%%%%%%%%%%%%%%%%%%%%%%%%%%%%%%%%%%%%

The mass of the central SMBH\index{Supermassive black hole} is not sufficient to explain the observed
proper motions. In particular, \cite{Schoedel09} model the mass
distribution as
\begin{equation}\label{eq:schoedel}
M(r)=m_{\rm BH}\,{}+\,{}4\,{}\pi{}\,{}\int_0^r{\rm d}\tilde{r}\,{}\tilde{r}^2\,{}\rho{}(\tilde{r}),
\end{equation}
where
\begin{equation}
\rho{}(r)=\rho_0\,{}\left(\frac{r}{5\textrm{ pc}}\right)^{-\Gamma{}}\,{}\left(1+\frac{r}{5\textrm{ pc}}\right)^{\Gamma{}-4}
\end{equation}
and minimize the $\chi{}^2$ of the proper-motion measurements with
three free parameters: $m_{\rm BH}$, $\Gamma{}$ and
$M_\ast{}(<1\textrm{ pc})$ (where $M_\ast{}(<1\textrm{ pc})$ is the
mass of stars inside 1 pc). For $\Gamma{}\ge{}0$ and
$3.5\lesssim{}m_{\rm BH}/(10^6\,{}\textrm{M}_{\odot{}})\lesssim{}4.5$,
$M_\ast{}>0.4\times{}10^6\,{}\textrm{M}_\odot$ (see
Fig.~\ref{fig:schoedel4}). Similar results can be found assuming an
anisotropic distribution of the velocity dispersion. This result
strengthens the evidence for a massive NSC. We notice that the
best-fitting value for the mass of the SMBH ($m_{\rm
  BH}=3.6^{+0.2}_{-0.4}\times{}10^6\msun$, at 68\% confidence
level) is smaller than the one derived from the orbits of the S-stars
(\citealt{Gillessen09a}), although the former is marginally consistent
with the latter. Furthermore, even values of $\Gamma{}<0$
(i.e. `centrally evacuated' mass models) are allowed by the fit shown
in Fig.~\ref{fig:schoedel4}. 

 Recently, \cite{Chatzopoulos14} did a similar analysis using 2500 line-of-sight velocities and 10000 proper motions obtained with VLT instruments, and 200 maser velocities (see \citealt{fritz14} for a description of the data sample).
Using axisymmetric Jeans modeling to fit the proper motion and line-of-sight velocity dispersions, \cite{Chatzopoulos14} obtain new best estimates for the NSC mass, black hole mass, and distance $M_\ast{}(r<100'')=(9.26\pm{}0.31|_{\rm stat}\pm{}0.9|_{\rm syst})\times{}10^6{\rm M}_{\odot{}}$, $m_{\rm BH}=(3.88\pm{}0.14|_{\rm stat}\pm{}0.4|_{\rm syst})\times{}10^6{\rm M}_{\odot{}}$, and $R_0=8.30\pm{}0.09|_{\rm stat}\pm{}0.1|_{\rm syst}$ kpc, respectively.

\subsubsection{The disc(s) of early-type stars\index{Early-type stars|textbf}}
%%%%%%%%%%%%%%%%%%%%%%%%%%%%%%%%FIGURE %%%%%%%%%%%%%%%%%%%%%%%%%%%%%%%%%%%%%%%%%
\begin{figure}[t]
\sidecaption[t]
\includegraphics[width=6.5cm]{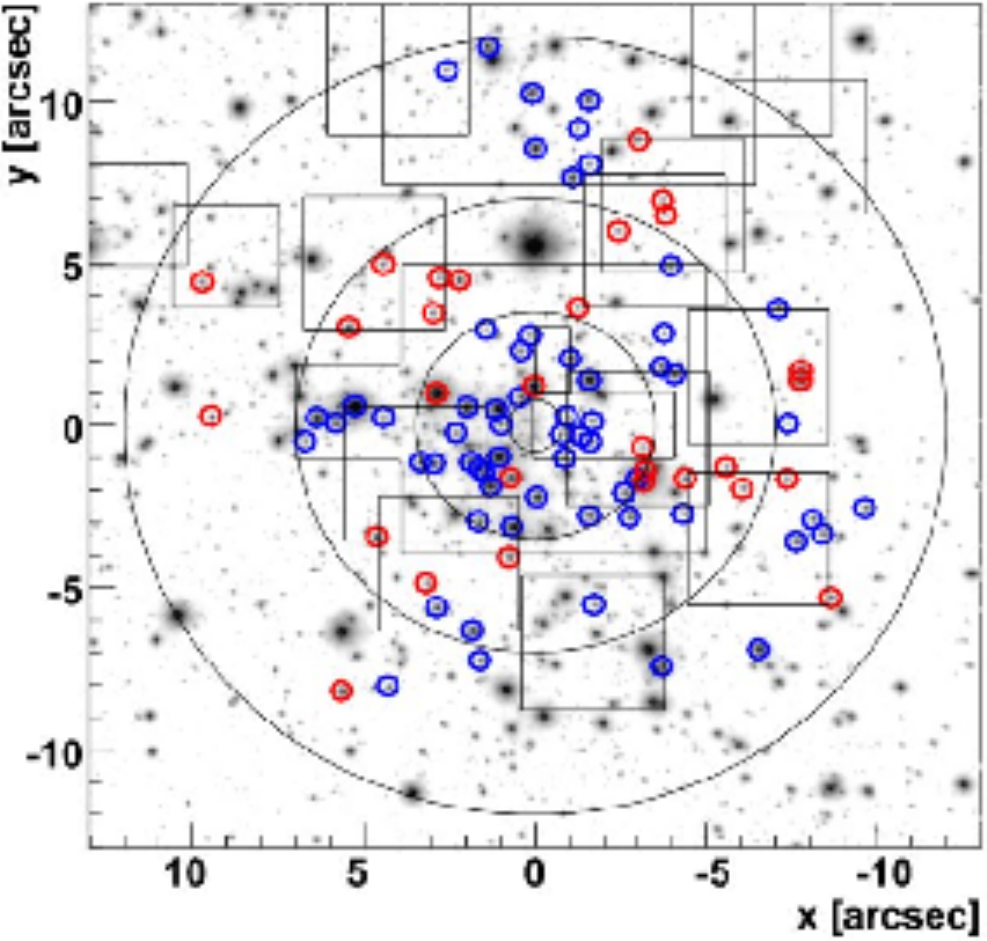}
\caption{Sample of 90 WR/O stars ($m_{\rm K} < 14$ and
  $\Delta{}(v_{\rm z}) \le{} 100$ km s$^{-1}$) in the central 0.5 pc
  of our Galaxy: blue circles indicate CW orbits (61 WR/O stars) and
  red circles indicate counterclockwise orbits (29 WR/O stars). The
  black circles show projected distances of 0.''8, 3.''5, 7'', and
  12'' from Sgr~A$^\ast$. Squares indicate the exposed fields with
  SINFONI in the 25 mas pixel$^{-1}$ and 100 mas pixel$^{-1}$
  scale. The whole inner 0.5 pc region is contained in lower
  resolution (250 mas pixel$^{-1}$ scale) SINFONI observations
  (\citealt{Paumard06}). From Fig.~1 of \citealt{Bartko09}.}
\label{fig:bartko1}
\end{figure}
%%%%%%%%%%%%%%%%%%%%%%%%%%%%%%%%FIGURE %%%%%%%%%%%%%%%%%%%%%%%%%%%%%%%%%%%%%%%%%

The presence of young massive stars in the central parsec of the MW
has been discussed for a long time (\citealt{Lacy82};
\citealt{Allen87}; \citealt{Rieke89}; \citealt{Allen90}; see
\citealt{Morris96} for a review).  So far, more than a hundred young
massive stars have been observed in the vicinity of Sgr~A$^\ast{}$
(\citealt{Krabbe91}; \citealt{Morris93}; \citealt{Genzel94};
\citealt{Blum95a}; \citealt{Blum95b};
\citealt{Eckart95};\citealt{Krabbe95};\citealt{Libonate95};
\citealt{Tamblyn96}; \citealt{Genzel03}; \citealt{Paumard06};
\citealt{Bartko09}, see Fig.~\ref{fig:bartko1}).  Many of them are
O-type and Wolf-Rayet (WR) stars.  Radial velocity and spectral type
of these stars have been thoroughly investigated thanks to spectroscopy,
while proper motions and brightness have been provided by
photometry. The most recent spectroscopic data include observations
with the integral field spectrograph SINFONI (\citealt{Bartko09} and
references therein) at the ESO/VLT, and with the OH-Suppressing
Infrared Imaging Spectrograph (OSIRIS) at the Keck II telescope
(\citealt{Do13}). The most recent photometric data include
observations with NACO at the ESO/VLT (\citealt{Trippe08};
\citealt{Bartko09}) and with NIRC2 at the Keck II telescope
(\citealt{Do13}).

The analysis of orbital angular momentum\index{Angular momentum} directions shows that some of
the early-type stars\index{Early-type stars} lie in a disc (\citealt{Paumard06};
\citealt{Bartko09}; \citealt{Lu09}; \citealt{Yelda12}; \citealt{Do13};
\citealt{Lu13}). This disc is called clockwise (CW) disc\index{Clockwise disc|textbf}, because it
shows CW motion when projected on the plane of the sky
(\citealt{Genzel03}; \citealt{Paumard06}). The fraction of early-type
stars\index{Early-type stars} that actually belong to the CW disc\index{Clockwise disc} is still debated: the recent
study by \cite{Yelda2014} indicates that only $\sim{}20$ per cent of
early-type stars\index{Early-type stars} lie in the CW disc\index{Clockwise disc}, while previous studies
(e.g. \citealt{Do13}; \citealt{Lu13}) suggest a higher fraction
($\sim{}50$ per cent).

\cite{Bartko09} compute significance maps from the sky maps of the
density of reconstructed angular momentum\index{Angular momentum} directions of the observed
stars ($\rho_{\rm obs}$), by defining the significance for each bin
of the sky map as
\begin{equation}\label{eq:bartko}
\textrm{significance}=\frac{\rho{}_{\rm obs}-\langle{}\rho{}_{\rm iso}\rangle{}}{\rho{}_{\rm iso,\,{}rms}},
\end{equation}
where $\langle{}\rho{}_{\rm iso}\rangle{}$ and $\rho{}_{\rm
  iso,\,{}rms}$ are the mean density and the root mean square density
(of angular momentum\index{Angular momentum} directions) for a set of simulated stars
following an isotropic distribution, respectively.

%%%%%%%%%%%%%%%%%%%%%%%%%%%%%%%%FIGURE %%%%%%%%%%%%%%%%%%%%%%%%%%%%%%%%%%%%%%%%%
\begin{figure}[t]
\sidecaption[t]
\includegraphics[width=12.5cm]{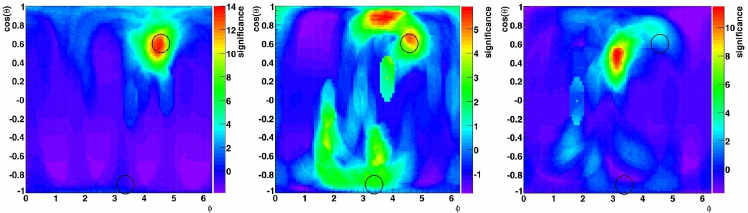}
\caption{Cylindrical equal area projections of the distributions of
  significance in the sky for three radial bins: 32 WR/O stars with
  projected distances in the bin 0.''8$-$3.''5 (left-hand panel), 30
  WR/O stars in the bin 3.''$5-7$'' (central panel), and 28 WR/O stars
  in the bin 7''$-12$'' (right-hand panel). The position of the CW
  disc\index{Clockwise disc} and of the (possible) counterclockwise disc as derived by
  \cite{Paumard06} are marked with black circles.  In the inner bin
  there is a maximum excess significance of 13.9 $\sigma{}$ at
  $(\phi{},\,{}\theta{}) = (256^{\circ{}}, 54^{\circ})$, compatible
  with the CW system of \cite{Paumard06}. The significance map
  in the middle interval shows two extended excesses, one for
  CW and one for counterclockwise orbits. The CW excess
  has a local maximum significance of 5.4 $\sigma{}$ at
  $(\phi{},\,{}\theta{}) = (262^{\circ{}}, 48^{\circ})$, compatible
  with the orientation of the CW system of \cite{Paumard06},  
  but a global maximum significance of 5.9 $\sigma{}$ at a clearly
  offset position: $(\phi{},\,{}\theta{}) = (215^{\circ{}},
  28^{\circ})$. The significance map in the outer bin shows a maximum
  excess significance of 11.5 $\sigma{}$ at yet another position
  $(\phi{},\,{}\theta{}) = (179^{\circ{}}, 62^{\circ})$. The
  morphology of the excesses in the CW system may indicate a
  smooth transition of the excess centre with projected radius.  From
  Fig.~11 of \cite{Bartko09}.}
\label{fig:bartko11}
\end{figure}
%%%%%%%%%%%%%%%%%%%%%%%%%%%%%%%%FIGURE %%%%%%%%%%%%%%%%%%%%%%%%%%%%%%%%%%%%%%%%%

From left to right, the three panels of Fig.~\ref{fig:bartko11} show
the significance maps (derived as described above) for stars with
distance 0.''8$-$3.''5, 3.''$5-7$'' and 7''$-12$'' (i.e. 0.032$-$0.14
pc, 0.14$-$0.28 pc and 0.28$-$0.48 pc) from Sgr~A$^{\ast}$. We recall
that a razor-thin disc is expected to define an infinitely small
circle in these maps. It is apparent that only the stars in the bin
closest to Sgr~A$^\ast$ define a unique disc, consistent with the CW
disc\index{Clockwise disc}. In the intermediate bin, various features are present. One of
these features is still consistent with the CW disc\index{Clockwise disc}, while the other
features may be interpreted as a second dismembered disc or as
outliers of the CW disc\index{Clockwise disc}. A relevant portion of these `outliers' shows
counterclockwise motion, which has been claimed to indicate the
presence of a second dissolving disc (\citealt{Lu06}, \citealt{Lu09};
\citealt{Bartko09}). Finally, the stars in the outer bin mostly belong
to a single disc, but offset with respect to the inner bin.

The results shown in Fig.~\ref{fig:bartko11} have the following
crucial implications.
\begin{itemize}
\item[](i) Only a fraction of the early-type stars\index{Early-type stars} in the central
  parsec are members of the CW disc\index{Clockwise disc}.
\item[](ii) The probability for an early-type star\index{Early-type stars} to be member of the CW disc\index{Clockwise disc} decreases with increasing distance from the centre.
\item[] (iii) The CW disc\index{Clockwise disc} is likely warped and/or tilted, as the orientation of its normal vector changes by several degrees ($\sim{}60^\circ{}$, \citealt{Bartko09}) from its inner to its outer edge.
\end{itemize}

\begin{figure}
\sidecaption[t]
\includegraphics[width=10.0cm]{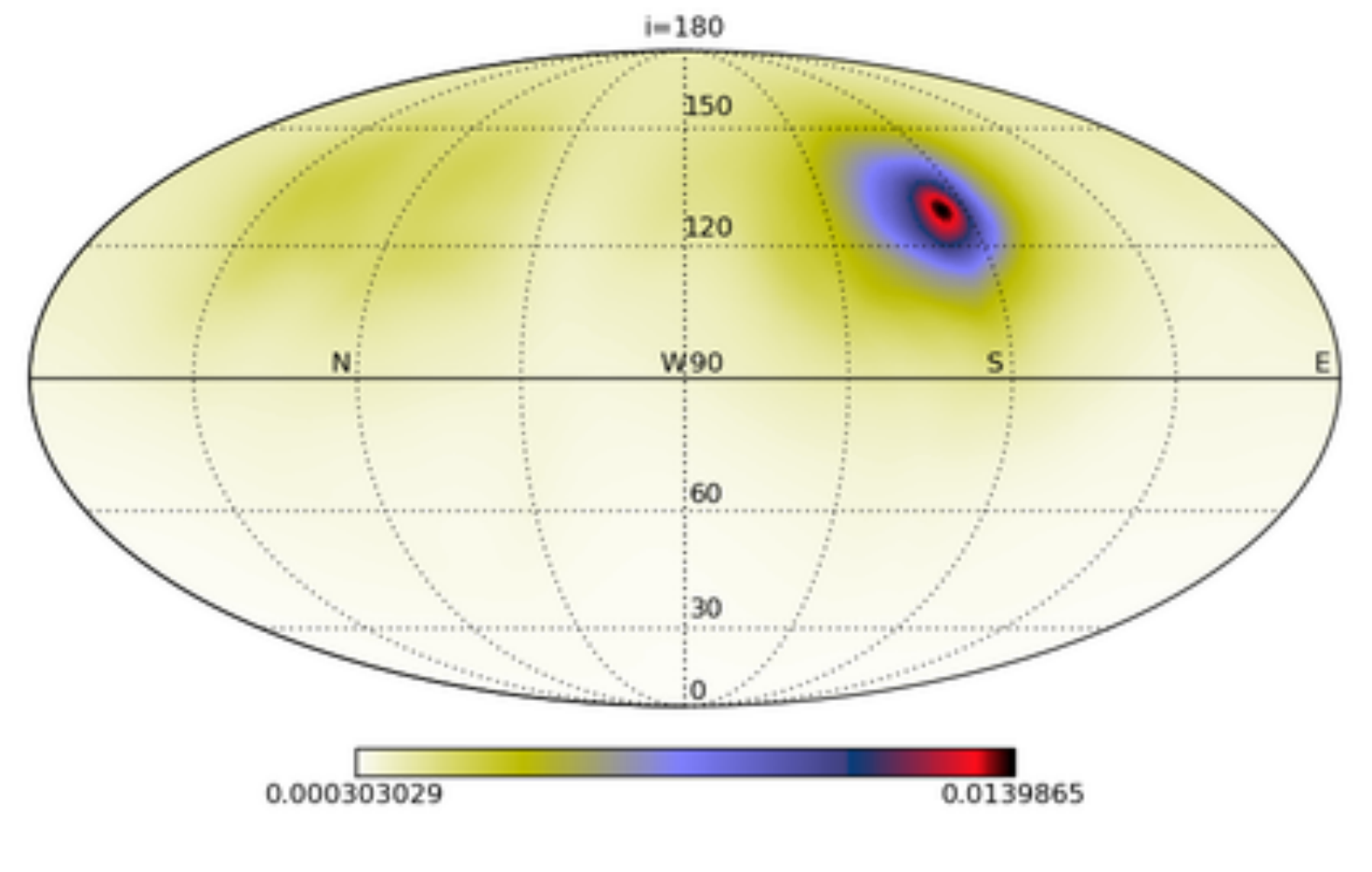}
\includegraphics[width=10.0cm]{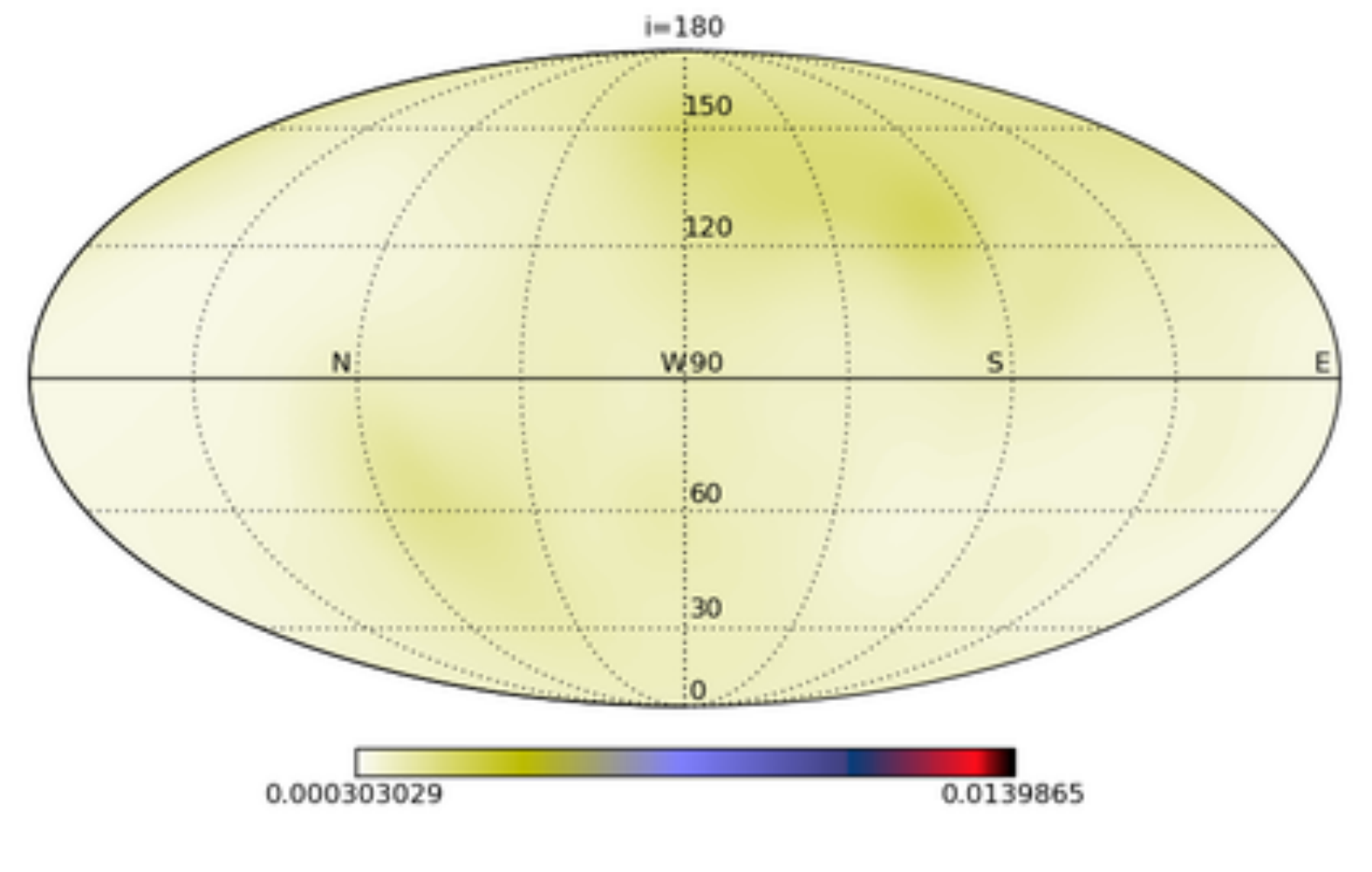}
\includegraphics[width=10.0cm]{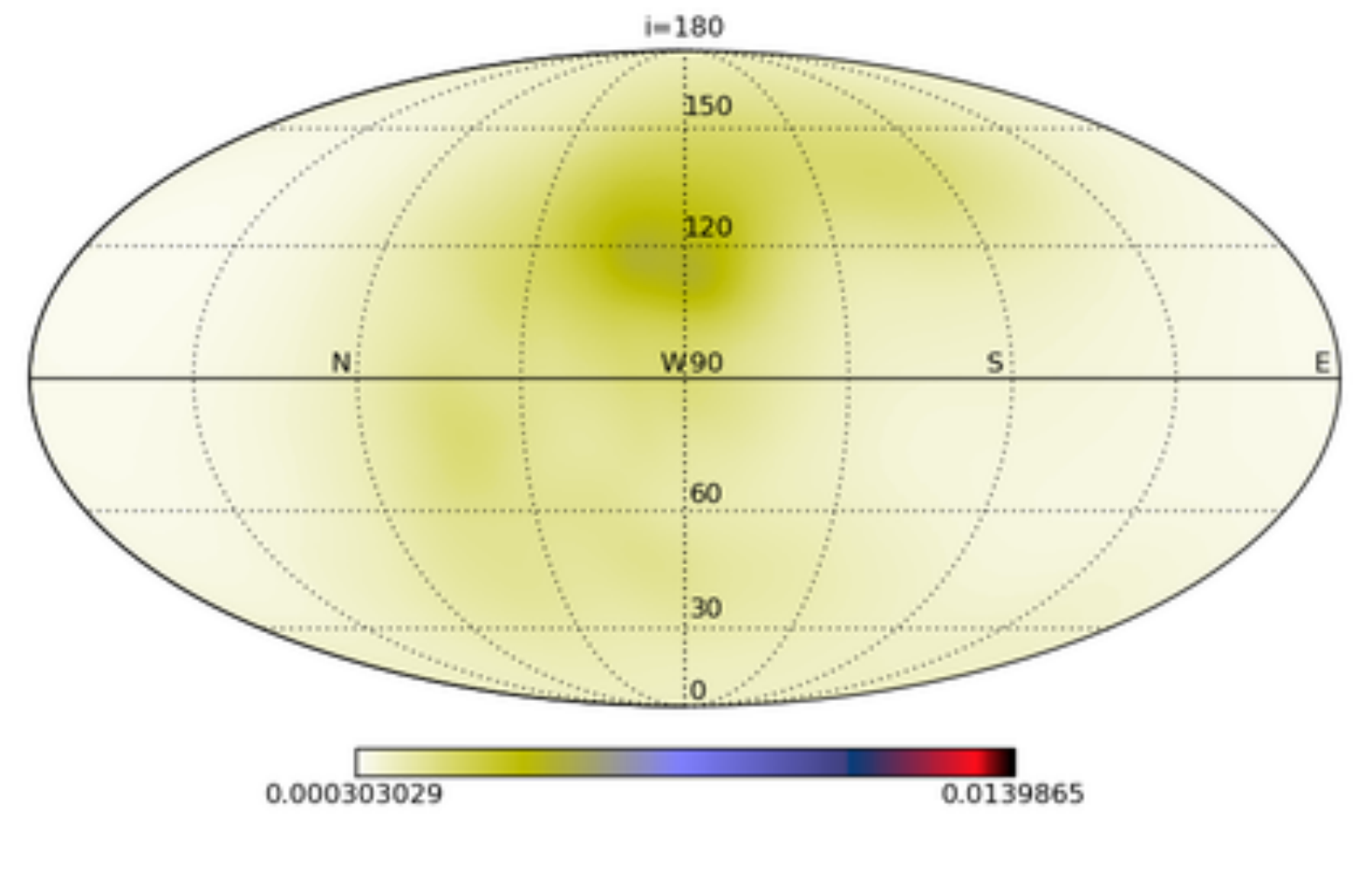}
\caption{Density of normal vectors for stars in the three separate radial
bins: 0.''8-3.''2 ({\em top}), 3.''2-6."5 ({\em middle}),
and 6."5-13."3 ({\em bottom}). The CW disc\index{Clockwise disc} feature at
($i$, $\Omega$) = (130$^{\circ}$, 96$^{\circ}$) is prominent in the inner 
radial bin and shows a decrease in density with radius.  The degenerate orbital 
solutions associated with the CW disc\index{Clockwise disc} stars are seen as the slight density 
enhancement near ($i$, $\Omega$) $\sim$ (130$^{\circ}$, 300$^{\circ}$) in the top panel.
The middle radial interval shows hints of the CW disc\index{Clockwise disc} and extended structure 
around this location.  In the outermost radial bin, a density
enhancement is seen at ($i$, $\Omega$) = (117$^{\circ}$, 192$^{\circ}$).
The same scaling is used in each plot to show the relative strength of the features. The horizontal lines represent $i$ and are spaced 30$^{\circ}$ apart
and the longitudinal lines represent $\Omega$ and are spaced 45$^{\circ}$ apart, with the line marked E representing $\Omega$ = 0$^{\circ}$. Fig.~14 of \cite{Yelda2014}.}
\label{fig:yelda14}
\end{figure}

Recently, \cite{Yelda2014} consider a sample of 116 stars, for which they measure both
proper motions and, in a few cases, accelerations. 
\cite{Yelda2014} compute significance maps from the sky
maps of the density of reconstructed angular momentum\index{Angular momentum} directions,
using a formula very similar to Eq.~\ref{eq:bartko}
(\citealt{Yelda2014} normalize the significance to the standard deviation rather than
to the root mean square density). Similarly to \cite{Bartko09}, they
group the stars into three radial bins: 0''.8--3''.2, 3''.2--6''.5 and
6''.5--13''.3 (i.e. 0.032--0.128 pc, 0.128--0.26 pc and 0.26--0.532
pc). Fig.~\ref{fig:yelda14} shows the resulting density of normal vectors for the three bins. The main results are:
\begin{itemize}
\item[](i) there are no statistically significant signatures of a
  counterclockwise disc. It seems that the two discs scenario is
  definitely dead.
\item[](ii) The existence of a CW disc\index{Clockwise disc} is confirmed with high
  significance in the inner bin, but there is no clear evidence that
  the CW disc\index{Clockwise disc} extends to the two outermost radial bins. Thus, the
  outer radius of the CW disc\index{Clockwise disc} might be as small as $\sim{}0.13$ pc
  (rather than $\sim{}0.5$ pc, as discussed by \citealt{Bartko09}).
\item[] (iii) Since the CW disc\index{Clockwise disc} extends only to $\sim{}0.13$ pc, it is
  neither significantly warped nor tilted.
%\item[] (iv) 
\end{itemize}
The results by \cite{Yelda2014}, if confirmed, will 
%quite revolutionize 
significantly change our previous picture of the early-type stars\index{Early-type stars} in the GC.

%%%%%%%%%%%%%%%%%%%%%%%%%%%%%%%%FIGURE %%%%%%%%%%%%%%%%%%%%%%%%%%%%%%%%%%%%%%%%%
%%%%%%%%%%%%%%%%%%%%%%%%%%%%%%%%FIGURE %%%%%%%%%%%%%%%%%%%%%%%%%%%%%%%%%%%%%%%%%

\begin{figure}[t]
\sidecaption[t]
\includegraphics[width=12.5cm]{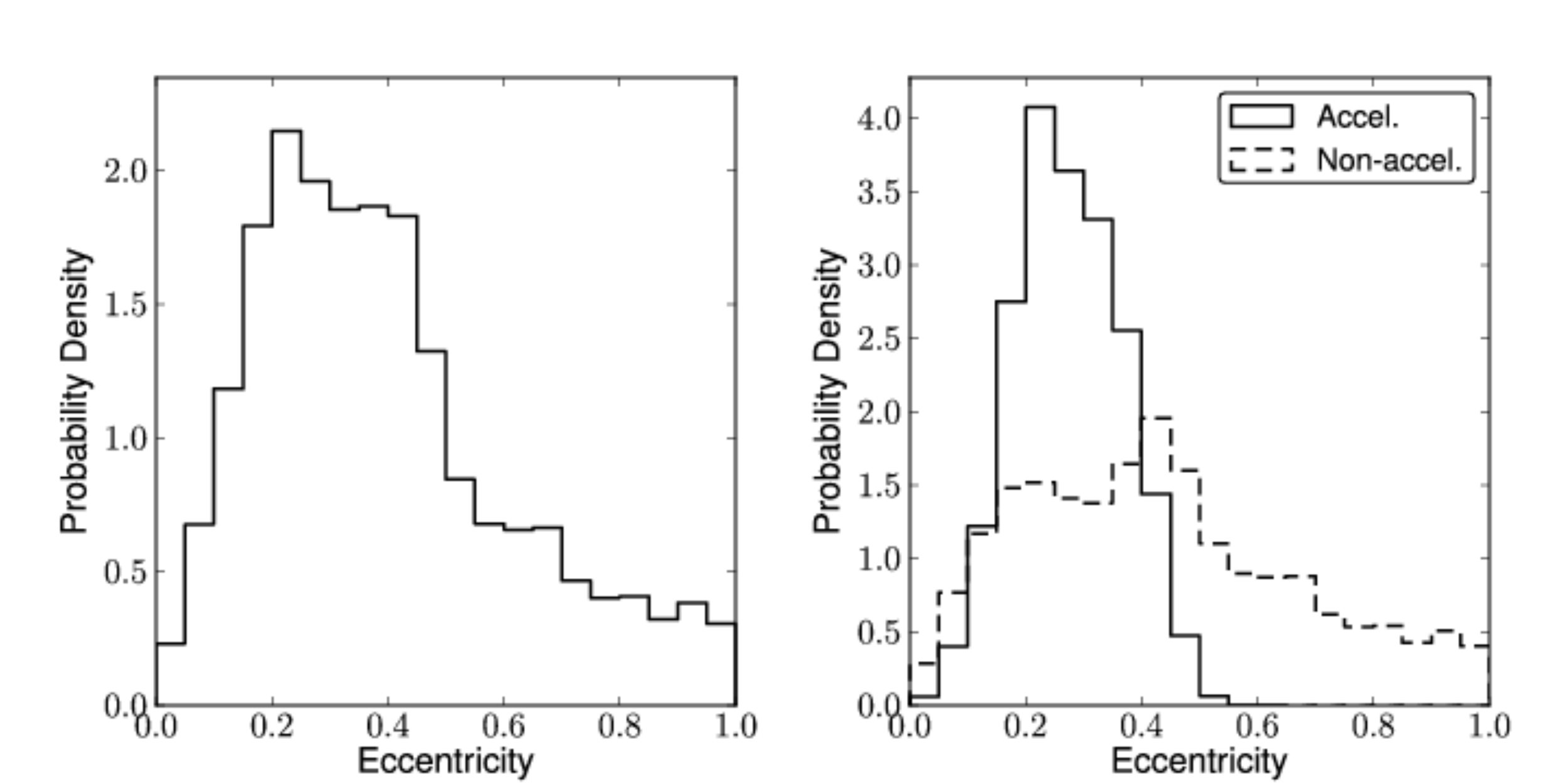}
\caption{{\em Left:} Eccentricity distribution of the CW disc\index{Clockwise disc}.
All orbital solutions falling within 15.2$^{\circ}$ of the disc are included, thereby  weighting the distributions by disc membership probability. 
{\em Right:} Eccentricity distributions shown separately for likely disc
members with acceleration detections ({\em solid}) and without ({\em dashed}).  From Fig.~12 of \cite{Yelda2014}.}
\label{fig:yelda12}
\end{figure}

Furthermore, \cite{Yelda2014} measure the orbital eccentricity of
stars in their sample (Fig.~\ref{fig:yelda12}). They confirm that the
peak of the eccentricity distribution is at $e\sim{}0.2-0.4$ and the
distribution of eccentricities is quite broad, as found in previous
studies (\citealt{Bartko09}; \citealt{Lu09}; \citealt{Yelda12};
\citealt{Do13}; \citealt{Lu13}). On the other hand, \cite{Yelda2014}
show that the distribution of eccentricities is much narrower if only
stars with detected acceleration are considered
(Fig.~\ref{fig:yelda12}). In particular, the resulting average
eccentricity is $\langle{}e\rangle{}=0.27\pm{}0.07$ and the
high-eccentricity tail disappears.
%%%%%%%%%%%%%%%%%%%%%%%%%%%%%%%%FIGURE %%%%%%%%%%%%%%%%%%%%%%%%%%%%%%%%%%%%%%%%%
%%%%%%%%%%%%%%%%%%%%%%%%%%%%%%%%FIGURE %%%%%%%%%%%%%%%%%%%%%%%%%%%%%%%%%%%%%%%%%

\begin{figure}[th]
\sidecaption[t]
\includegraphics[width=12.0cm]{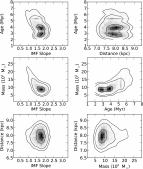}
\caption{Two-dimensional posterior probability distribution functions
  (PDFs) for the observed properties of the early-type stars\index{Early-type stars} (from \citealt{Lu13}). The over-plotted
  contours give 68\%, 95\%, and 99\% confidence intervals. Weak
  correlations exist between age, mass, and initial MF (IMF)\index{Mass function}
  slope. The correlation between the total mass and the age or IMF\index{Mass function}
  slope is a consequence of the age$-$IMF\index{Mass function} slope relationship since, at
  older ages, the most massive stars have disappeared and the total 
  mass must increase to match the observed numbers of stars brighter
  than $K_{\rm P} = 15.5$. From Fig.~10 of \cite{Lu13}.}
\label{fig:lu2013_10}
\end{figure}

The most recent estimate of the age of the early-type stars\index{Early-type stars} is $t_{\rm
  age}\approx{}2.5-6$ Myr (\citealt{Lu13}). This result comes from
integral-field spectroscopy (using the OSIRIS spectrometer on Keck
II), with a completeness of 50\% down to magnitude {\it K'=15.5}
(i.e. stellar mass $\sim{}10\msun$), combined with photometry
using the NIRC2 instrument on Keck II (\citealt{Do13}). The analysis
of the data is based on Bayesian inference methods (\citealt{Lu13},
see Fig.~\ref{fig:lu2013_10}). A previous estimate indicated $t_{\rm
  age}=6\pm{}2$ Myr (\citealt{Paumard06}). Furthermore,
\cite{Yusef-zadeh13} have found possible indications of gas outflows,
suggesting recent star formation (10$^{4-5}$ yr) within 0.6 pc of
SgrA$^\ast$.

The MF\index{Mass function} of the early-type stars\index{Early-type stars} has been claimed to be
very top-heavy for a long time. \cite{Paumard06} suggest an MF\index{Mass function} similar
to ${\rm d}N/{\rm d}m\sim{}m^{-\alpha{}}$, with $\alpha=0.85$ (we
recall that the Salpeter MF\index{Mass function} has $\alpha=2.35$, \citealt{Salpeter55})
and a total mass $\sim{}10^4\msun$. The result of
\cite{Paumard06} was obtained from the luminosity function of the most
massive WR and O-type stars and suffered from lack of sensitivity for
magnitude $K>13$ (i.e. stellar mass $<20\msun$). \cite{Bartko10}
find an even flatter mass-function, with best-fitting slope
$\alpha=0.45\pm{}0.3$.

Recently, \cite{Lu13} use the same data and the same Bayesian approach
as in \cite{Do13}. They derive a new best-fitting slope
$\alpha{}=1.7\pm{}0.2$ (see Fig.~\ref{fig:lu2013_10}), still flatter
than a Salpeter MF\index{Mass function}, but considerably steeper than previous
estimates. Consequently, the total mass of the early-type stars\index{Early-type stars} is
also revised, suggesting a value in the $1.4-3.7\times{}10^{4}\msun$
range (extrapolated down to stars with mass $1\msun$).

 Finally, by means of stellar evolution models, \cite{lockmann2010kroupa} showed that the total observed  luminosity in the central parsec of the NSC is better matched by a continuous star formation over the Galaxy's lifetime, following a \cite{kroupa2001} MF, than by a long-standing top-heavy MF. This suggests that, if the early-type stars follow a top-heavier MF than the rest of the NSC, the circumstances that led to the formation of the early-type stars must be very peculiar, since these have not affected most of the NSC. 

%%%%%%%%%%%%%%%%%%%%%%%%%%%%%%%%FIGURE %%%%%%%%%%%%%%%%%%%%%%%%%%%%%%%%%%%%%%%%%
%%%%%%%%%%%%%%%%%%%%%%%%%%%%%%%%FIGURE %%%%%%%%%%%%%%%%%%%%%%%%%%%%%%%%%%%%%%%%%
\vspace{1cm}

The existence of very young (a few Myr old) stars in the inner parsec has
been an enigma for a long time. The observed MF\index{Mass function} (flatter than the
Salpeter MF\index{Mass function}) and the orbits of these early-type stars\index{Early-type stars} (belonging to
one or two discs plus a number of possible outliers) open several
additional questions. The new results by \cite{Yelda2014}, which 
indicate that only 20 per cent of the early-type stars\index{Early-type stars} are members of
the CW disc\index{Clockwise disc} (see also \citealt{SanchezBermudez2014}) and that the CW disc\index{Clockwise disc} may be much smaller than previously
thought, open further issues.

The early-type stars\index{Early-type stars} that cannot be considered members of the CW disc\index{Clockwise disc},
because of the different angular momentum\index{Angular momentum} direction (in some cases,
they are even counterclockwise), might be either genuine outliers
(i.e. stars that were born outside the CW disc\index{Clockwise disc}) or former members of
the CW disc or even members of other (partially dismembered) discs.
The existence of other stellar discs (in addition to the CW disc\index{Clockwise disc}) is
still an open question. The mechanisms that can either dismember a
disc or perturb the orbits of some of its members are even more
debated.  In the next sections (Sect.~\ref{sec:3} and \ref{sec:4}), 
we will review which theoretical scenarios have been proposed to explain these open questions.

\subsubsection{The S-stars}
The few stars whose orbits are (totally or partially) inside the
innermost arcsecond ($\sim{}0.04$ pc) are referred to as the S-star
cluster\index{S-cluster|see {S-stars}} (\citealt{Schodel03}; \citealt{Ghez03}, \citealt{Ghez05};
\citealt{Eisenhauer05}; \citealt{Gillessen09a}). The orbits of
$\sim{}25-30$ S-stars are known with accuracy, by means of NIR imaging
and spectroscopy. In particular, the motion of the S-stars\index{S-stars} has been tracked
since 1992 at NTT and VLT, and since 1995 at Keck.
Most of the S-stars have been classified as B$0-9$ V stars, with ages
between 6 and 400 Myr (\citealt{Eisenhauer05}). \cite{Gillessen09a}
recently derived the orbital solutions of 28 S-stars: 22 early-type
stars\index{Early-type stars} and 6 late-type S-stars (S17, S21, S24, S27, S38 and
S111). These are the first late-type S-stars with measured
orbits. Thus, most of the S-stars (but not all of them) are early-type
stars\index{Early-type stars}.

%%%%%%%%%%%%%%%%%%%%%%%%%%%%%%%%FIGURE %%%%%%%%%%%%%%%%%%%%%%%%%%%%%%%%%%%%%%%%%
\begin{figure}[t]
\sidecaption[t]
\includegraphics[width=6.5cm]{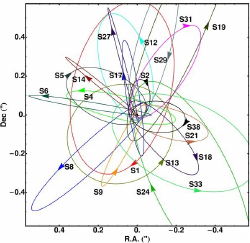}
\caption{Stellar orbits of the stars in the central arcsecond for
  which \cite{Gillessen09a} determined the orbits. The coordinate
  system was chosen such that Sgr A$^\ast{}$ is at rest. From Fig.~16
  of \cite{Gillessen09a}.}
\label{fig:Gillessen2009_16}
\end{figure}
%%%%%%%%%%%%%%%%%%%%%%%%%%%%%%%%FIGURE %%%%%%%%%%%%%%%%%%%%%%%%%%%%%%%%%%%%%%%%%

 Six of the 28 S-stars studied by \cite{Gillessen09a} appear to be
 members of the CW disc\index{Clockwise disc}: they have semi-major axis $\approx{}1''$,
 eccentricity $e\approx{}0.2-0.4$ and angular distance to the CW disc\index{Clockwise disc}
 between 9$^\circ$ and 21$^\circ$. The orbits of the 22 remaining S-stars 
 do not lie in a disc: they are consistent with a random
 distribution in space (see Fig.~\ref{fig:Gillessen2009_16}).

\begin{figure}[t]
\sidecaption[t]
\includegraphics[width=6.5cm]{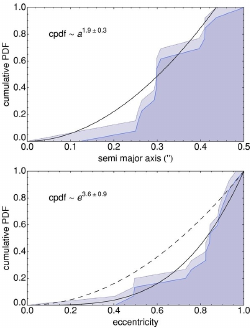}
\caption{Top: Cumulative PDF for the semi-major axis of the early-type
  stars\index{Early-type stars} with $a < 0.''5$. The two curves correspond to the two ways to
  plot a cumulative PDF, with values ranging either from 0 to
  $(N-1)/N$ or from $1/N$ to 1. Solid line: best fit ($n(a) \sim
  a^{0.9\pm{}0.3}$). Bottom: Cumulative PDF for the eccentricities of
  the early-type stars\index{Early-type stars} that are not identified as disc members. As in
  the top panel, the two curves correspond to the two ways to plot a
  cumulative PDF. Dashed line: $n(e) \sim{} e$; Solid line: best fit
  ($n(e) \sim{} e^{2.6\pm{}0.9}$). From Figs.~20 and 21 of
  \cite{Gillessen09a}.}
\label{fig:Gillessen09_2021}
\end{figure}

The distribution of semi-major axes of the 22 `true' S-stars (see
Fig. ~\ref{fig:Gillessen09_2021}) is best-fit (using a log-likelihood
fit) by $n(a)\sim{}a^{0.9\pm{}0.3}$ (\citealt{Gillessen09a}),
corresponding to a number density $n(r)\sim{}r^{-1.1\pm{}0.3}$,
consistent with the mass profile (\citealt{Genzel03};
\citealt{Schoedel07}).

The distribution of eccentricities of the 22 `true' S-stars (see
Fig. ~\ref{fig:Gillessen09_2021}) is best-fit (using a log-likelihood
fit) by $n(e)\sim{}e^{2.6\pm{}0.9}$ (\citealt{Gillessen09a}). This
means that the eccentricities of S-stars are much larger than those of
the CW disc\index{Clockwise disc}. The best-fit distribution is somewhat 
skewed toward larger eccentricity with respect to the thermal distribution
($n(e)\sim{}e$), typical of two-body relaxed systems.

%%%%%%%%%%%%%%%%%%%%%%%%%%%%%%%%FIGURE %%%%%%%%%%%%%%%%%%%%%%%%%%%%%%%%%%%%%%%%%
%%%%%%%%%%%%%%%%%%%%%%%%%%%%%%%%FIGURE %%%%%%%%%%%%%%%%%%%%%%%%%%%%%%%%%%%%%%%%%

Among the S-stars, the S2 star is particularly important because so far it has 
provided the strongest constraints on the SMBH\index{Supermassive black hole} mass 
(e.g. \citealt{Ghez08}; \citealt{Gillessen09a};
\citealt{Gillessen09b}). S2 has been classified as a B0-2.5 V
main sequence star with an estimated zero-age main sequence mass of
$19.5\msun$ (\citealt{Martins08}). It is bright ($K\approx{}14$)
and has a very short orbital period (15.9 years). Astrometric
data taken from 1992 to 2003 (see Sect.~\ref{subsec:2.1}) allowed to
track one entire orbit. Unfortunately, during pericentre passage
(2002) S2 showed a puzzling photometry, which might be due to
confusion with a fainter star (see \citealt{Gillessen09a} for this
issue). For sake of curiosity, S2 is not the shortest-known-period
star orbiting the SMBH: S102 has a period of only 11.5 years
(\citealt{Meyer12}). Astrometric data (NIRC on Keck) covered one
entire orbit of S102. On the other hand, S102 is a factor of 16
fainter than S2. %, and this explains why the strongest constraints on
%SMBH mass come from S2.

 The S-star cluster\index{S-cluster} is one of the most enigmatic
components of the GC: most of the S-stars are
early-type stars\index{Early-type stars} and cannot have formed {\it in situ}, with a pericentre so close to the SMBH (this is the so called
`paradox of youth'\index{Paradox of youth}, \citealt{Ghez03}). Furthermore, their eccentricities are
very high, but these stars are too young to have undergone two-body
relaxation. They have different orbital properties with respect to the
early-type stars\index{Early-type stars} in the CW disc\index{Clockwise disc}, because of the larger eccentricities
and because of the random orientation of their orbital planes. Does this
necessarily mean that the S-stars are a different population with
respect to the CW disc\index{Clockwise disc}? If they are a different population, where do
they come from?  Alternatively, is there any perturbation which can
affect the stars in the CW disc\index{Clockwise disc} and change their orbital properties
till they match those of the S-stars? These questions and the main
scenarios for the formation of S-stars will be discussed in
Sect.~\ref{sec:3} and \ref{sec:4}.

\subsection{The molecular gas and the ionized gas}
\label{subsec:2.3}
The GC is a very crowded environment not only from the point of view of
the stellar population, but also for the gas. The central $\sim{}20$
parsecs of the MW are rich in molecular, atomic and ionized gas, which
form very peculiar structures.
%%%%%%%%%%%%%%%%%%%%%%%%%%%%%%%%FIGURE %%%%%%%%%%%%%%%%%%%%%%%%%%%%%%%%%%%%%%%%%
\begin{figure}[t]
\sidecaption[t]
\includegraphics[width=7cm]{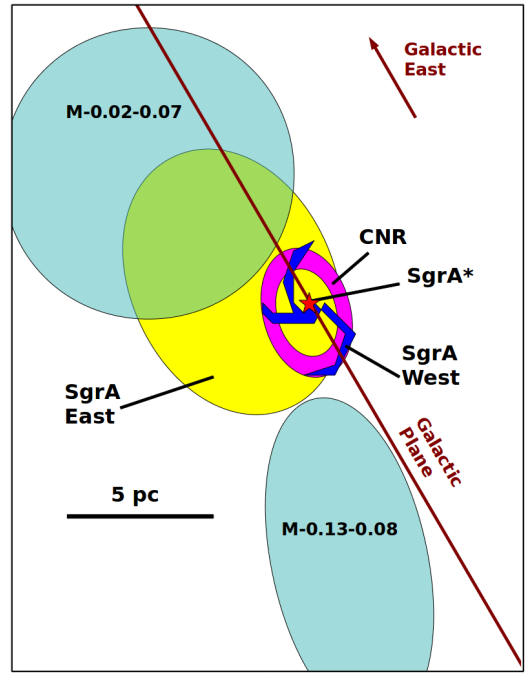}
\caption{Schematic diagram showing the sky locations and rough sizes
  and shapes of GC sources discussed in Sect.~\ref{subsec:2.3}. Red
  star: Sgr~A$^\ast$; magenta ring: CNR\index{Circumnuclear ring}; blue spiral: the minispiral
  (i.e. Sgr~A West); yellow ellipse: Sgr~A East; the two turquoise
  ellipses: the M--0.02--0.07 and the M--0.13--0.08 cloud. A solid red
  line indicating the orientation of the Galactic plane has been drawn
  through the position of Sgr~A$^\ast$. The Galactic eastern 
  direction is indicated. One arcminute corresponds to
  about 2.3 pc at the distance of 8 kpc. This diagram has been
  inspired by Fig.~1 of \cite{Novak00}.}
\label{fig:novak00}
\end{figure}
%%%%%%%%%%%%%%%%%%%%%%%%%%%%%%%%FIGURE %%%%%%%%%%%%%%%%%%%%%%%%%%%%%%%%%%%%%%%%%

The main reservoirs of ionized gas are Sgr~A East and Sgr~A West
(\citealt{Novak00}; \citealt{Zhao09}), both observed in radio and both
overlapped with Sgr~A$^\ast$ (see the schematic illustration in
Fig.~\ref{fig:novak00}). Sgr~A East is a non-thermal elliptical shell
source elongated along the Galactic plane with a major axis of length
$\sim{}10$ pc (\citealt{Downes71}; \citealt{Yusef-zadeh87a};
\citealt{Novak00}). Its centre is displaced by 2.5 pc (in projection)
with respect to Sgr~A$^\ast$. Sgr~A East is generally thought to be a
supernova remnant.

Sgr~A West is a spiral-shaped thermal
radio source (\citealt{Ekers83}; \citealt{Scoville03}), which
surrounds Sgr~A$^\ast$. For its shape, Sgr~A West is often called the
`minispiral'. The three main arms of the minispiral are called the
`Northern Arm' (pointing towards North), the `Eastern Arm' (pointing
towards East) and the `Western Arc' (pointing towards West).
 The nature of the minispiral is very debated. According to a popular scenario, the minispiral arms might be streams associated with molecular gas falling in towards the centre (\citealt{Lo83}; \citealt{Zhao09}; \citealt{Zhao10}).

%%%%%%%%%%%%%%%%%%%%%%%%%%%%%%%%FIGURE %%%%%%%%%%%%%%%%%%%%%%%%%%%%%%%%%%%%%%%%%
\begin{figure}[t]
%\sidecaption[t]
\vspace{-0.2cm}
\hspace{1.0cm}
\begin{tabular}{ p{4.0cm} p{4.0cm} }
\includegraphics[width=5.4cm]{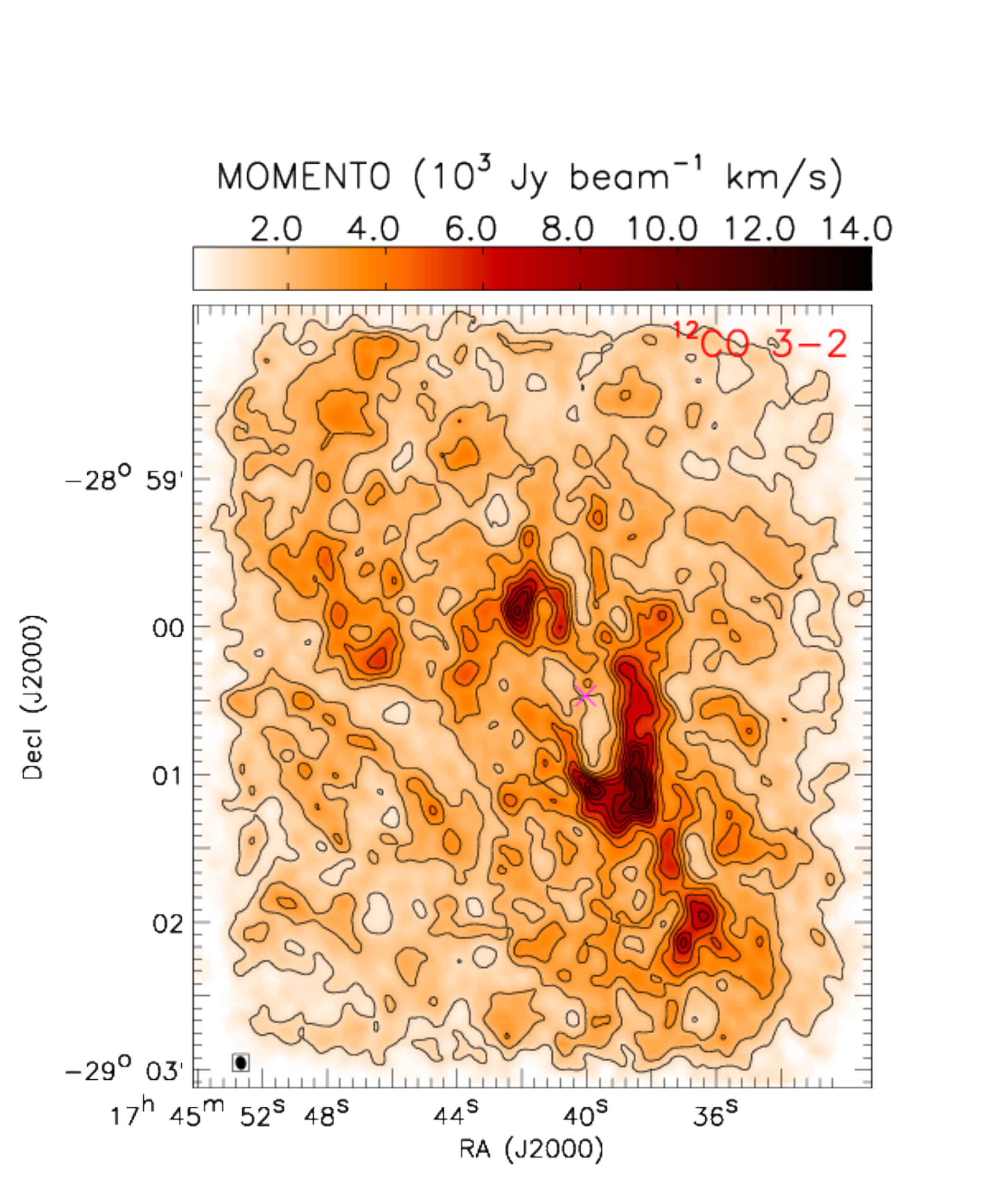} & \includegraphics[width=5.4cm]{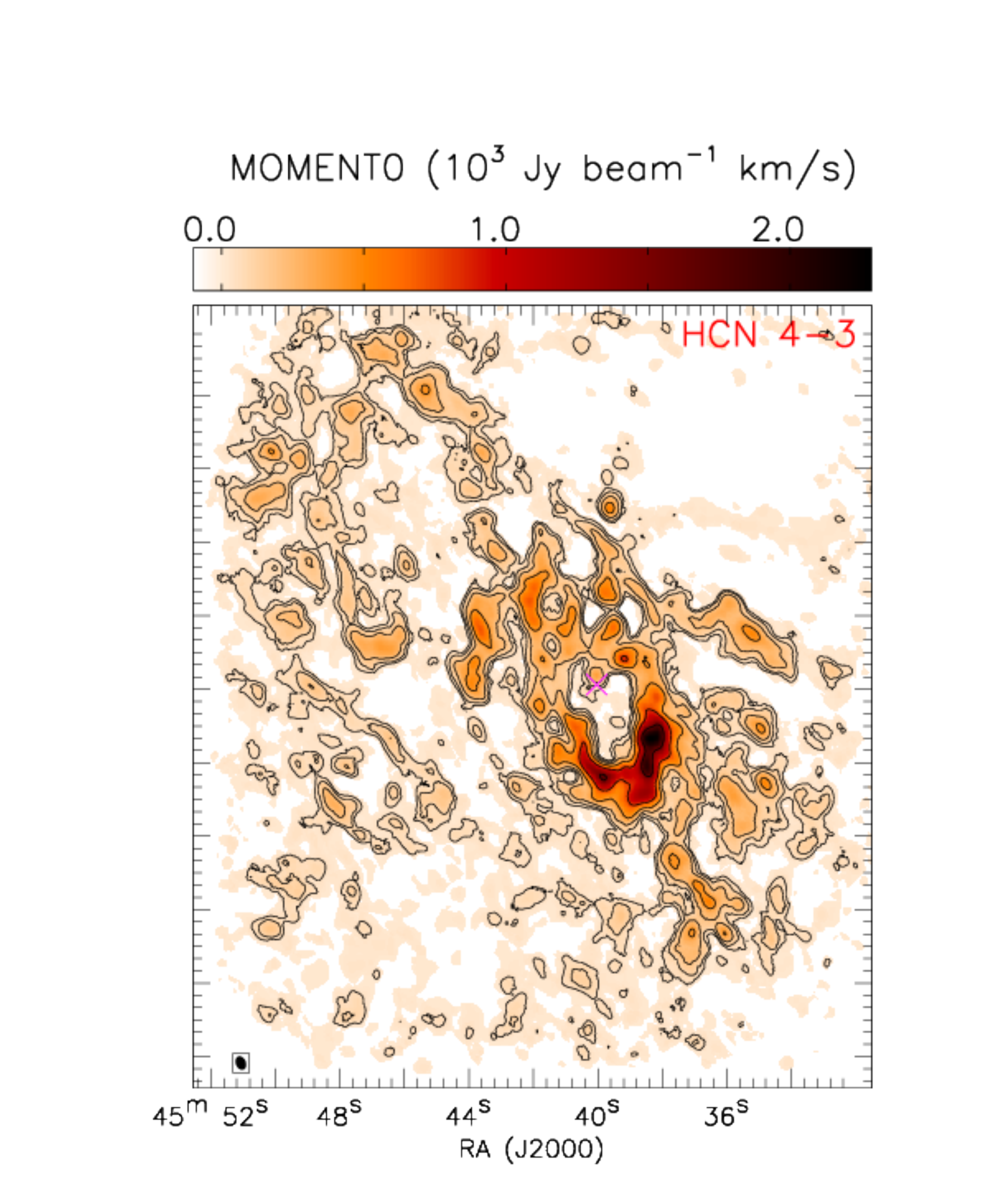}  \\
\end{tabular}

\vspace{-0.8cm}

\hspace{1.0cm}
\begin{tabular}{ p{4.0cm} p{4.0cm} }
\includegraphics[width=5.4cm]{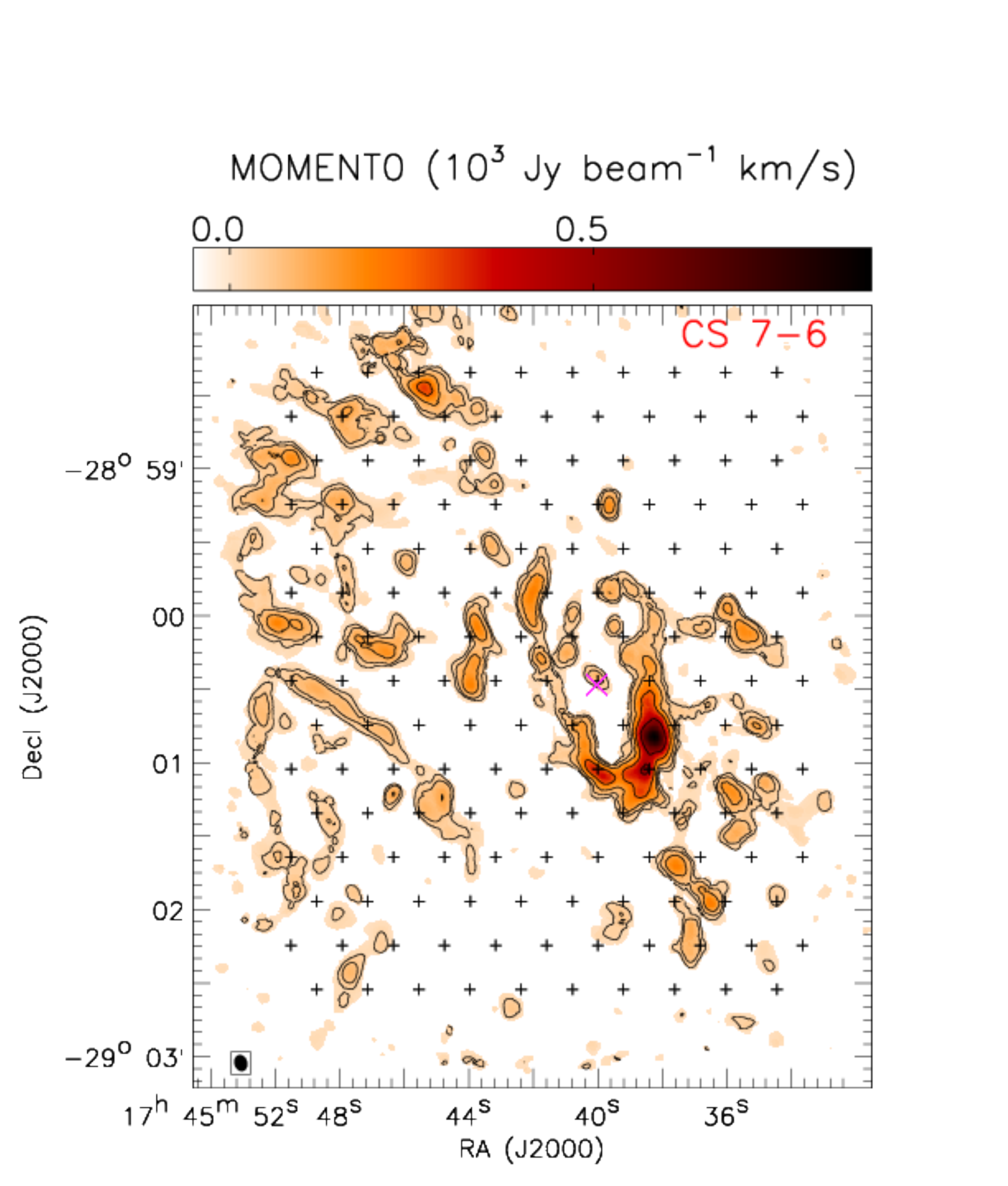} & \includegraphics[width=5.4cm]{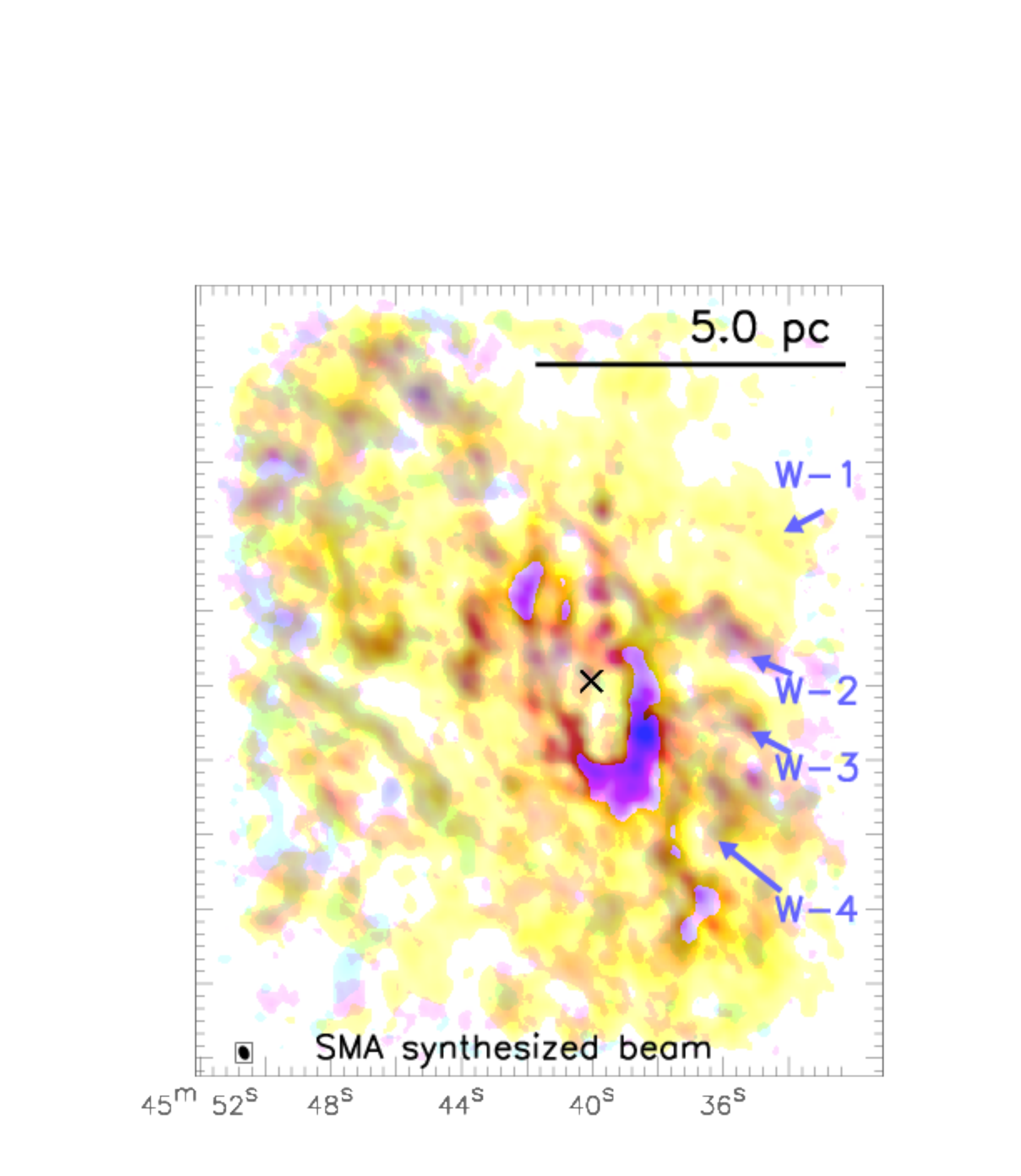}  \\
\end{tabular}
\caption{Velocity integrated (i.e., moment 0) images of the $^{12}$CO
  3--2, HCN 4--3, and CS 7--6 transitions in the region of the CNR\index{Circumnuclear ring}.
  The synthesized beam of the Submillimeter Array (SMA) observations
  is shown in the bottom right.  The contours of the $^{12}$CO 3--2
  image start at the value 1000\,Jy\,beam$^{-1}$km\,s$^{-1}$, and are
  drawn at intervals of 1000\,Jy\,beam$^{-1}$km\,s$^{-1}$.  The
  contours of the HCN 4--3 and CS 7--6 images are
  50\,Jy\,beam$^{-1}$km\,s$^{-1}$$\times$[1, 2, 4, 8, 16, 32] and
  30\,Jy\,beam$^{-1}$km\,s$^{-1}$$\times$[1, 2, 4, 8, 16],
  respectively.  Integration of the signal over a 20 km\,s$^{-1}$
  velocity range has an rms noise level of 5.8
  Jy\,beam$^{-1}$km\,s$^{-1}$ (2.2 K\,km\,s$^{-1}$).  The bottom right
  panel shows an overlay of these lines, in yellow ($^{12}$CO 3--2),
  magenta (HCN 4--3) and cyan (CS 7--6) colors. W-1, W-2, W-3 and W-4
  in the bottom right panel indicate the four western streamers of the
  CNR\index{Circumnuclear ring}. Crosses in the CS image mark the pointing centers of the SMA
  mosaic observations. From Fig.~6 of \cite{baobab12}.}
\label{fig:baobab}
\end{figure}
%%%%%%%%%%%%%%%%%%%%%%%%%%%%%%%%FIGURE %%%%%%%%%%%%%%%%%%%%%%%%%%%%%%%%%%%%%%%%%

A clumpy, inhomogeneous and kinematically disturbed ring of molecular
gas, known as the circumnuclear ring (CNR)\index{Circumnuclear ring|textbf} or the circumnuclear disc (CND), surrounds the minispiral (Fig.~\ref{fig:baobab}). The CNR\index{Circumnuclear ring} was
discovered about 30 years ago (\citealt{Becklin82}) via detection of
double-lobed emission at 50 and 100 $\mu{}$m, from dust. After the
discovery, the CNR\index{Circumnuclear ring} has been observed extensively at radio to infrared
wavelengths (e.g. \citealt{Gatley86}; \citealt{Serabyn86}; 
\citealt{Gusten87}; \citealt{Zylka88}; \citealt{DePoy89}; \citealt{Sutton90};
\citealt{Jackson93}; \citealt{Marr93}; \citealt{Telesco96}; \citealt{Chan97}; 
\citealt{Coil99}; \citealt{Coil00}; \citealt{Wright01};
\citealt{Vollmer01}; \citealt{Yusef-zadeh04}; \citealt{Christopher05};
\citealt{Donovan06}; \citealt{Montero09}; \citealt{Oka11}; \citealt{Martin12}; \citealt{Mills13}).

The observations indicate that the CNR\index{Circumnuclear ring} is a ring of molecular gas and
dust with an inclination of $\sim{}50-70^\circ$ with respect to the
observer. The ring is nearly complete in HCN (Fig.~\ref{fig:baobab}),
but with a large gap in the north (corresponding to the position of
the Northern Arm of the minispiral) and other smaller gaps. The inner
radius of the ring is $\sim{}1.5$ pc (de-projected) and it is quite
sharp, while the outer radius is less defined: HCN, CO and HCO$^+$
were observed out to $\sim{}7$ pc, but recent studies
(e.g. \citealt{Wright01}) suggest an outer edge at $3-4$ pc. The CNR\index{Circumnuclear ring}
has a thickness of $\sim{}0.4$ pc at the inner edge
(\citealt{Jackson93}) and expands to $\sim{}2$ pc in the outer parts
(\citealt{Vollmer01}). 
%The total estimated mass of the CNR\index{Circumnuclear ring} is of the
%order of $10^6\msun$ (\citealt{Christopher05}). 
 The total mass of the CNR\index{Circumnuclear ring} is highly uncertain.
Measurements based on the dust thermal emission indicate a total mass of $\sim{}2\times{}10^4$ M$_\odot$ (\citealt{Mezger89}; \citealt{baobab13}, but see \citealt{Christopher05} for a different estimate).

The CNR\index{Circumnuclear ring} rotates
with a velocity $\sim{}110$ km s$^{-1}$ (\citealt{Marr93};
\citealt{Christopher05}), but the velocity field shows local
perturbations, which may indicate a warp or the presence of different
streamers. This is the reason why previous studies proposed that the
CNR\index{Circumnuclear ring} formed through the collision of two molecular clouds\index{Molecular cloud}
(e.g. \citealt{Gusten87}) or through the assembly of multiple
dynamically different streamers (e.g. \citealt{Jackson93}).

Recently, \cite{baobab12} made wide-field images ($\sim{}5''$
resolution) of three high-excitation molecular gas tracers ($^{12}$
CO3--2, HCN 4--3, CS7--6) in the region of the CNR\index{Circumnuclear ring}
($\sim{}5'\times{}5'$ field-of-view), using the Submillimeter Array
(SMA). They also made a 20'' resolution CS~1--0 image using the
National Radio Astronomy Observatory (NRAO) Green Bank Telescope. The
high-excitation lines observed with the SMA trace the dense and warm
gas ($>10^5$ cm$^{-3}$, $>30$ K), while the CS 1--0 traces the less
dense and cooler gas ($\sim{}5\times{}10^4$ cm$^{-3}$, $<10$
K). \cite{baobab12} find that several $\sim{}5-20$ pc-scale gas
streamers either directly connect to the CNR\index{Circumnuclear ring} or penetrate inside it
(see\footnote{Liu and collaborators have found that they misplaced
  CS7--6 as CS$^{34}$7--6 in their paper (B. Liu private
  communication). This has been fixed in Fig.~\ref{fig:baobab} with
  respect to the original figure published on ApJ. See Liu et al. (in
  preparation) for details.} Fig.~\ref{fig:baobab}). Thus, the CNR\index{Circumnuclear ring}
appears to be the centre of an inflow, quite reminiscent of the molecular gas streaming in the nucleus of NGC~1068 (\citealt{Muller2009}).  \cite{baobab12} speculate that
the CNR\index{Circumnuclear ring} may be dynamically evolving, continuously fed via gas
streamers and in turn feeding gas toward the centre.
%{\bf RICORDARSI CITARE BAOBAB LIU}

The observations also indicate an ongoing interaction between the CNR\index{Circumnuclear ring}
and the minispiral (\citealt{Christopher05}). The strongest
interactions likely occur along the Western Arc and the Northern Arm
of the minispiral. The ionized gas in the Western Arc is oriented
along the CNR\index{Circumnuclear ring} and it is immediately interior to the CNR. For this
reason, the Western Arc has been proposed to be the inner edge of the
CNR\index{Circumnuclear ring}, ionized by the central stellar cluster. This idea is confirmed by
the velocity field (\citealt{Christopher05}). Furthermore, the
minispiral Northern Arm may connect with the northeastern extension of
the CNR\index{Circumnuclear ring} to form a single collimated structure
(\citealt{Christopher05}).

\cite{Christopher05} identify 26 resolved molecular gas cores within
the CNR\index{Circumnuclear ring}. These have a characteristic diameter of $\sim{}0.25$ pc, a
typical density of a few $\times{}10^7$ cm$^{-3}$ and a typical mass
of a few $\times{}10^4\msun$. The density of the molecular cores
is sufficient to prevent tidal disruption\index{Tidal disruption} at $\sim{}1-2$ pc distance
from Sgr~A$^{\ast{}}$, indicating that the CNR\index{Circumnuclear ring} may be a long-lived
structure and may be able to form stars.  In fact, recent observations
with the Green Bank Telescope detect maser lines and both narrow (0.35
km s$^{-1}$) and broad ($30-50$ km s$^{-1}$) methanol emission from
the CNR\index{Circumnuclear ring} (\citealt{Yusef-zadeh08}). This has been interpreted as a
signature of massive star formation in its early phases. In the
following sections (Sect.~\ref{sec:3} and \ref{sec:4}), we will see
that the CNR\index{Circumnuclear ring} may have a crucial role for the formation and for the
secular evolution of the young stars in the GC. 

 Furthermore, several hundreds solar masses of atomic gas ($>300$ M$_\odot$) might exist inside the CNR (\citealt{Jackson93}; \citealt{Goicoechea13}).

\vspace{1cm}

Finally, a number of giant molecular clouds\index{Molecular cloud} are close to the GC (\citealt{Whiteoak74}; \citealt{Guesten80}; \citealt{Dent93}; \citealt{Coil99}; \citealt{Coil00}; \citealt{Pierce-Price00}; \citealt{Herrnstein02}; \citealt{McGary02}; \citealt{Karlsson03}; \citealt{Herrnstein05}; \citealt{Tsuboi09}; \citealt{Amo-Baladron11}; \citealt{Tsuboi11}; \citealt{Tsuboi12}; \citealt{Ao13}; \citealt{Minh13}). Two
molecular clouds\index{Molecular cloud} (the M--0.02--0.07 and the M--0.13--0.08 cloud,
\citealt{Solomon72}, \citealt{Novak00}) lie within 20 pc of the GC
(see Fig.~\ref{fig:novak00}).  M--0.02--0.07 and M--0.13--0.08 have
comparable masses ($\sim{}5\times{}10^5\msun$, \citealt{Lis94}),
and linear dimensions ($10-15$ pc). The centre of M--0.02--0.07 lies
$\sim{}7$ pc away from Sgr~A$^\ast$ (projected
distance). Morphological and kinematic evidence shows that Sgr~A East
has expanded into M--0.02--0.07, compressing portions of this cloud
into a `curved ridge' (\citealt{Ho85}; \citealt{Genzel90};
\citealt{Serabyn92}, see Fig.~\ref{fig:novak00}).  The centre of
M--0.13--0.08 lies $\sim{}13$ pc away from Sgr~A$^\ast$ (projected
distance). The cloud is highly elongated. A finger-like structure
extends from this cloud toward the Galactic eastern direction, and
apparently feeds the CNR\index{Circumnuclear ring} (\citealt{Okumura91}; \citealt{Ho91}).

\subsection{The G2 cloud}
\label{subsec:2.4}
  In the last two years, there has been much excitement about G2: a faint dusty object orbiting the SMBH with a very eccentric orbit ($\sim{}0.98$) and an extremely small pericentre ($\sim{}200$ AU$\sim{}2000$ Schwarzschild radii). The detection of G2 was reported in 2012 (\citealt{Gillessen12}), but the first VLT NIR images where G2 can be seen date back to $\sim{}$2003. G2 immediately raised the expectations of the astrophysical community: was this object going to be tidally disrupted by the SMBH? What is its nature? 

\begin{figure}[t]
%\sidecaption[t]
\includegraphics[width=12.5cm]{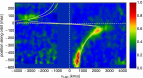}
\caption{Position--velocity diagram of G2\index{G2 cloud}, extracted from 2013 April
  SINFONI data set along the orbit projected into the cube. This
  diagram is a co-add around the lines Br$-\gamma{}$, Helium I, and
  Paschen$-\alpha{}$. The yellow line delineates the $L'-$band based
  orbit, the white line the Br$-\gamma{}$ based one. Fig.~2 of
  \cite{Gillessen13b}.}
\label{fig:fig2gillessen13b}
\end{figure}

The observation of a blue-shifted ($-3000$ km s$^{-1}$) component in April 2013 indicated that a part of G2\index{G2 cloud} had already passed pericentre  (see Fig.~\ref{fig:fig2gillessen13b}). The bulk of G2 transited at pericentre in Spring 2014, and was not completely disrupted during its close-up with the SMBH: the object is still point-like (consistent with the point-spread function), even if with a tail of disrupted material. Several authors predicted an enhancement of the X-ray and near-infrared activity of the SMBH in correspondence of G2 pericentre passage, but no significant event has yet been observed (\citealt{Haggard14}). 
%was observed so far (\citealt{Haggard14}).

G2 has been observed in $L'$ continuum ($3.8$ $\mu{}$m, $m_{L'}\sim{}14$), in the Br-$\gamma{}$ line of hydrogen recombination (Br-$\gamma{}$ luminosity $\sim{}$ a few $\times{}10^{30}$ erg s$^{-1}$, emission measure $\sim{}$ 10$^{57}$ cm$^{-3}$), in Paschen-$\alpha{}$ (1.875 $\mu{}$m) and Helium I (2.058 $\mu{}$m). Its luminosity has remained nearly constant (within a factor of two) since the first observations (\citealt{Pfuhl14, Witzel14}).
The $L'$ continuum emission (corresponding to a luminosity of $\sim{}2\times{}10^{33}$ erg s$^{-1}$, \citealt{Pfuhl14, Witzel14}) is consistent with the thermal emission of dust at $\approx{}$560 K (\citealt{Gillessen12}; \citealt{Eckart13}; \citealt{Gillessen13a}; \citealt{Gillessen13b}). The combination of line emission and NIR continuum indicates that the cloud is composed mainly of ionized gas ($\sim{}10^4$ K) plus some amount of relatively cool dust.
%G2 has been observed in the 2.058 $\mu{}$m HeI line and in the hydrogen recombination lines Bracket-$\gamma{}$, Bracket-$\delta{}$ (hereafter Br-$\gamma{}$ and Br-$\delta{}$, respectively) and Paschen-$\alpha{}$. These emission lines are likely excited by the ultraviolet radiation from the NSC. G2 is also visible in NIR, in the $L'$ band (centred at $\lambda{}=3.8\,{}\mu{}$m),
%corresponding to the thermal emission of dust at $\approx{}$600 K
%(\citealt{Gillessen12}; \citealt{Eckart13}; \citealt{Gillessen13a};
%\citealt{Gillessen13b}). 

%The combination of line emission and NIR
%continuum indicates that the cloud is composed mainly of ionized gas
%($\sim{}10^4$ K) plus some amount of relatively cool
%dust. %Interestingly, the luminosity of the cloud remained almost
%constant in both $L'$ band and Br-$\gamma{}$ line for all observations
%(2003-2013).

The orbit of G2 was traced back to $\sim{}2003$, thanks to archive VLT data, and showed a
three-dimensional velocity increase from 1200 km s$^{-1}$ (in 2004,
\citealt{Gillessen12}) to 2200 km s$^{-1}$ (in 2013,
\citealt{Gillessen13b}), consistent with a pure Keplerian motion. The internal velocity dispersion of $\approx{}100\kms$ (\citealt{Gillessen13a}) 
is another peculiar feature of the velocity field of G2\index{G2 cloud}: 
this corresponds to the sound speed of gas with temperatures of the order of a few million Kelvin.

%The
%most recent estimates of the orbital parameters
%(\citealt{Gillessen13b} and references therein) indicate an expected
%pericentre distance of $\sim{}2000$ $r_{\rm g}$ (where $r_{\rm g}$ is
%the Schwarzschild radius of the SMBH) from the position of Sgr~A$^\ast{}$. The
%bulk of the cloud will reach pericentre in the first months of
%2014, but the observation of a new blue-shifted ($-3000$ km s$^{-1}$)
%component in April 2013 indicates that a part of G2\index{G2 cloud} has already passed
%pericentre (see Fig.~\ref{fig:fig2gillessen13b}). Even in the most
%recent observations (\citealt{Gillessen13b}), the main features of G2\index{G2 cloud}
%in the position-velocity plane do not deviate from a ballistic orbit
%significantly: the effects due to hydrodynamical interactions with the
%hot gas are still negligible. These should become visible after
%pericentre passage. The internal velocity dispersion of $\approx{}100\kms$ (\citealt{Gillessen13a}) 
%is another peculiar feature of the velocity field of G2\index{G2 cloud}: 
%this corresponds to the sound speed of gas with temperatures of the order of a few million Kelvin.
%%%%%%%%%%%%%%%%%%%%%%%%%%%%%%%%FIGURE %%%%%%%%%%%%%%%%%%%%%%%%%%%%%%%%%%%%%%%%%
%%%%%%%%%%%%%%%%%%%%%%%%%%%%%%%%FIGURE %%%%%%%%%%%%%%%%%%%%%%%%%%%%%%%%%%%%%%%%%

The best-matching orbital parameters indicate that G2\index{G2 cloud} is almost
coplanar with the early-type CW disc (\citealt{Gillessen13b}):
the orbit of the cloud is $\sim{}20^\circ{}$ tilted with respect to
the most recent estimates of the CW disc\index{Clockwise disc} orientation. Finally, one of
the most peculiar features of G2\index{G2 cloud}'s orbit is its very high eccentricity
($e\sim{}0.98$, \citealt{Phifer13}; \citealt{Gillessen13a}; while the
average eccentricity of stellar orbits in the CW disc\index{Clockwise disc} is $\sim{}0.3$).

\cite{Pfuhl14} reported the analogy of G2 with another dusty object, the G1 cloud (already observed by \citealt{Clenet04} and \citealt{Ghez05b}). G1 transited at periapsis in 2001-2002, was observed in Br-$\gamma{}$ line and in $L'$ (the same as G2), and has approximately the same dust mass as G2.  The eccentricity of G1 is lower ($\sim{}0.86$) and the semimajor axis smaller ($\sim{}0.36$ arcsec instead of $\sim{}1.05$ arcsec), but its appearance and behaviour are very similar to the ones of G2.

\subsection{Does the Galactic Centre host an intermediate-mass black hole?}
\label{subsec:2.5}
In this subsection, we discuss the possibility that the GC hosts one
or more intermediate-mass black holes (IMBHs\index{Intermediate-mass black hole|textbf}, 
i.e. black holes with mass in the $10^2-10^5\msun$ range). 
IMBHs have been invoked to explain various phenomena that take
place in the GC (such as the ejection of hypervelocity stars\index{Hypervelocity stars}).

The presence of an IMBH in the GC might be detected in a
number of ways:
\begin{enumerate}
\item Reflex motion of the SMBH. %Upper limits on the astrometric
  %wobble of the radio source SgrA$^\ast{}$ are so far consistent with
  %the presence of an IMBH with mass and semi-major axis in the range
  %considered by \citet{MGM2009}.
\item A IMBH-SMBH binary might be revealed by emission of gravitational waves (GWs)\index{Gravitational waves|textbf}.
\item Stars can remain bound to the IMBH if its Hill sphere is larger
  than its tidal disruption\index{Tidal disruption} sphere; this condition is satisfied for
  SMBH--IMBH separations greater than $\sim 0.05\mpc$. The motion of a star
  bound to the IMBH would be the superposition of a Keplerian ellipse
  around the SMBH and an additional periodic component due to its
  motion around the IMBH; the latter would have a velocity amplitude
  $∼0.1–10$ times the IMBH orbital velocity and an orbital frequency
  from several hours to a few years, potentially accessible to
  astrometric monitoring.
\item In favorable circumstances, a near encounter of the IMBH with a
  star unbound to it could produce observable changes in the star’s
  orbit over month- or year-long timescales.
\item Interactions with an IMBH may result in ejections of stars to
  unbound orbits. A star ejected at $\sim 1000\kms$ requires about 100
  yr to move beyond 0.1 pc implying a probability $\sim 0.2 (N/10^4)$
  of observing an escaping star at any given time in the GC region, 
  where $N$ is the number of ejected
  stars. Interestingly, at least one S-star (S111) in the sample of
  \cite{Gillessen09a} appears to be on an unbound trajectory due to its large radial velocity.
\end{enumerate}

In the following two subsections, we focus on points 1. and 2. of the above enumeration. 

\subsubsection{Constraints on the presence of IMBHs in the GC from radio measurements}
At present, the measurements of the proper motion of the radio source
associated with SgrA$^\ast$ are the strongest constraints about the
presence of IMBHs\index{Intermediate-mass black hole} in the GC (\citealt{Hansen2003};
\citealt{ReidBru2004}). In fact, the perturbations induced onto the
SMBH by the nearly Keplerian motion of an IMBH orbiting around it are
expected to affect the proper motion of SgrA$^\ast$. In particular,
\cite{Hansen2003} showed that the perturbations induced on the proper
motion of SgrA$^\ast$ by an IMBH with mass $10^3\le{}m_{\rm IMBH}/{\rm
  M}_\odot{}\le{}10^4$, moving in a circular orbit with semi-major
axis $10^3\le{}a/{\rm AU}\le{}10^4$, can be detected if the proper
motion of SgrA$^\ast$ is measured with an accuracy higher than
$\approx{}0.1$ mas. On this basis, measurements of SgrA$^\ast$
proper motion, derived from Very Long Baseline Array (VLBA) data
(\citealt{ReidBru2004}), exclude the presence of IMBHs more massive
than $\sim{}10^4\msun$ with $10^3\le{}a/{\rm AU}\le{}10^5$.

\section{The formation of the early-type stars\index{Early-type stars}}
\label{sec:3}
A molecular cloud\index{Molecular cloud} close to the Galactic centre (GC) is tidally
disrupted if its number density does not exceed the Roche density
\begin{equation}\label{eq:roche}
%n_{\rm RL} \sim{}10^7\,{}{\rm cm}^{-3}\,{}\left(\frac{\sigma{}}{80\,{}{\rm km}\,{}{\rm s}^{-1}}\right)^2\,{}\left(\frac{{\rm pc}}{d}\right)^2,
n_{\rm RL} \sim{}10^7\,{}{\rm cm}^{-3}\,{}\left(\frac{m_{\rm
    BH}}{3\times{}10^6\,{}{\rm M}_\odot{}}\right)\,{}\left(\frac{{\rm
    pc}}{r}\right)^3,
\end{equation}
where $m_{\rm BH}$ is the mass of the SMBH and $r$ the distance of the
molecular cloud\index{Molecular cloud} from the SMBH. Since the density of molecular cloud\index{Molecular cloud}
cores is generally much lower ($\sim{}10^{4-6}$ cm$^{-3}$), molecular
clouds\index{Molecular cloud} are expected to be quickly disrupted when approaching the
central SMBH by less than a few parsecs. Thus, the early-type stars\index{Early-type stars}
inside the central parsec cannot have formed {\it in situ} from a
`typical' molecular cloud\index{Molecular cloud} (\citealt{Phinney89};
\citealt{Sanders98}). And yet, given their young age, they cannot have
migrated from larger distances by standard dynamical friction. %This is generally referred to as the `paradox of youth\index{Paradox of youth}' (\citealt{Ghez03}).

Various scenarios have been proposed to solve the `paradox of youth'
and to explain the formation of the early-type stars\index{Early-type stars} that orbit within
the central parsec. These scenarios can be divided in the following
two families: (i) `{\it in situ}' formation models, which assume local
star formation by some non standard process, and the (ii) migration
models, which assume formation at larger distances from the SMBH
followed by fast migration to their current location.

The inspiral and destruction of a star cluster
(Sect.~\ref{subsec:cluster}) belongs to the latter family, together
with the tidal breakup of stellar binaries (Sect.~\ref{subsec:3.4}),
while the fragmentation\index{Fragmentation} of the outer regions of an accretion disc
(Sect.~\ref{subsec:3.1}) and the disruption\index{Tidal disruption} of a molecular cloud\index{Molecular cloud}
(Sect.~\ref{subsec:3.2}) are the most likely `{\it in situ}'
formation pathways.

\subsection{Fragmentation\index{Fragmentation} of the accretion disc\index{Accretion disc|textbf}}
\label{subsec:3.1}
Keplerian accretion discs\index{Accretion disc} around SMBHs may become gravitationally
unstable to fragmentation\index{Fragmentation} and collapse to form stars
(\citealt{Paczynski78}; \citealt{Kolykhalov80}; \citealt{Lin87};
\citealt{Shlosman89}; \citealt{Hure98}; \citealt{Collin99};
\citealt{Collin99b}; \citealt{Gammie01}; \citealt{Goodman03};
\citealt{Nayakshin05a}; \citealt{Thompson05}; \citealt{Nayakshin06};
\citealt{Nayakshin07}; \citealt{Collin08}).

In particular, the Toomre stability parameter for Keplerian rotation (\citealt{Toomre64}) is
\begin{equation}\label{eq:toomre}
Q=\frac{c_{\rm s}\,{}\Omega{}}{\pi{}\,{}G\,{}\Sigma{}}=\frac{\Omega{}^2}{2\,{}\pi{}G\,{}\rho{}}\,{}\sqrt{(1+\zeta{})},
\end{equation}
where $c_{\rm s}$ is the sound speed, $\Omega{}$ is the angular
frequency, $G$ is the gravity constant, $\Sigma{}$ and $\rho{}$ are
the surface density and the volume density of the disc,
respectively. In equation~\ref{eq:toomre}, we have taken
$\Sigma{}=2\,{}H\,{}\rho{}$, where $H$ is the half-disc thickness. The
equation of hydrostatic equilibrium writes as $c_{\rm
  s}=\Omega{}\,{}H\,{}\sqrt{(1+\zeta{})}$, where
$\zeta\equiv{}4\,{}\pi{}G\,{}\rho{}\,{}\Omega{}^{-2}$.

The disc becomes unstable to fragmentation\index{Fragmentation} when $Q\le{}1$. This is expected to occur at a radius
\begin{equation}\label{eq:radius}
r_{\rm Q=1}\sim{}1.2\,{}{\rm pc}\,{}\left(\frac{m_{\rm
    BH}}{3\times{}10^6\,{}{\rm
    M}_\odot}\right)^{1/3}\,{}\left(\frac{\rho{}}{2\times{}10^{-17}\,{}{\rm
    g}\,{}{\rm cm}^{-3}}\right)^{-1/3},
\end{equation}
where we approximated $\Omega{}^2=G\,{}m_{\rm BH}\,{}r^{-3}$ and $\zeta{}=0$.

Following \cite{Collin99b}, the condition necessary for the collapse
of a fragment is that the time scale for star formation ($t_{\rm SF}$)
and the cooling time ($t_{\rm cool}$) be shorter than the
characteristic mass transport time in the disc ($t_{\rm trans}$).

According to \cite{Wang94}, $t_{\rm
  SF}=\Omega{}^{-1}\,{}Q/\sqrt{1-Q^2}$. Provided that $Q$ is not too
close to 1,
\begin{equation}\label{eq:tSF}
t_{\rm SF}\sim{}\Omega{}^{-1}=3\times{}10^{11}\,{}{\rm s}\,{}(m_{\rm BH}/3\times{}10^6\,{}{\rm M}_\odot)^{-1/2}\,{}(r/{\rm pc})^{3/2}. 
\end{equation}

For a gravitationally heated disc at nearly solar metallicity, 
\begin{equation}\label{eq:tcool}
t_{\rm cool}\sim{}\frac{8\,{}\pi{}\,{}\rho{}\,{}H^3}{3\,{}\dot{M}}=8\times{}10^{9}\,{}{\rm s}\,{}\left(\frac{\rho{}}{2\times{}10^{-17}\,{}{\rm g}\,{}{\rm cm}^{-3}}\right)\,{}\left(\frac{H}{0.01\,{}{\rm pc}}\right)^3\,{}\left(\frac{10^{-2}{\rm M}_{\odot}\,{}{\rm yr}^{-1}}{\dot{M}}\right)
\end{equation}

Finally, the mass transport time is
\begin{equation}\label{eq:ttrans}
t_{\rm trans}\sim{}\frac{2\,{}\pi{}\,{}r^2\,{}\rho{}\,{}H}{\dot{M}}=6\times{}10^{13}\,{}{\rm s}\,{}\left(\frac{r}{\rm pc}\right)^2\,{}\left(\frac{\rho{}}{2\times{}10^{-17}\,{}{\rm g}\,{}{\rm cm}^{-3}}\right)\,{}\left(\frac{H}{0.01\,{}{\rm pc}}\right)\,{}\left(\frac{10^{-2}{\rm M}_{\odot}\,{}{\rm yr}^{-1}}{\dot{M}}\right).
\end{equation}

For a wide range of accretion disc\index{Accretion disc}  parameters, $t_{\rm SF}$ and
$t_{\rm cool}$ are shorter than $t_{\rm trans}$. Thus, not only stars
are expected to form in the outer parts of accretion discs\index{Accretion disc}, but star
formation may be sufficiently vigorous to quench accretion and destroy
the accretion disc\index{Accretion disc}. Thus, recent studies searched for mechanisms that
can efficiently transfer angular momentum\index{Angular momentum} in the accretion disc\index{Accretion disc}, to
keep feeding the SMBH. \cite{Collin08} find that gas accretion onto
the SMBH is still possible (even if moderate star formation takes
place in the accretion disc\index{Accretion disc}), provided that supernovae and/or clump
collisions enhance the angular momentum\index{Angular momentum} transfer.

\begin{figure}[t]
%\sidecaption[t]
\includegraphics[width=6.5cm]{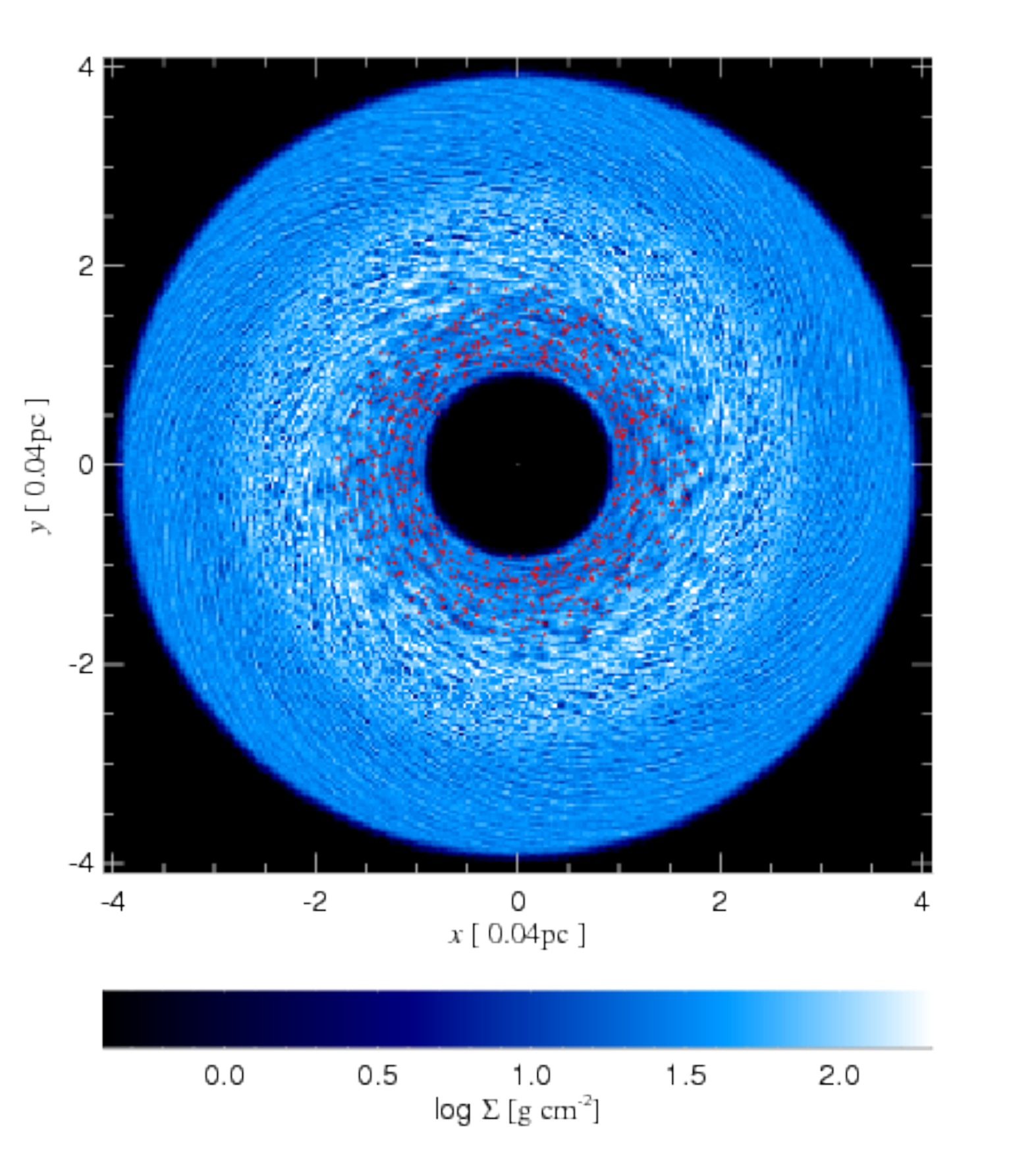}
\includegraphics[width=6.5cm]{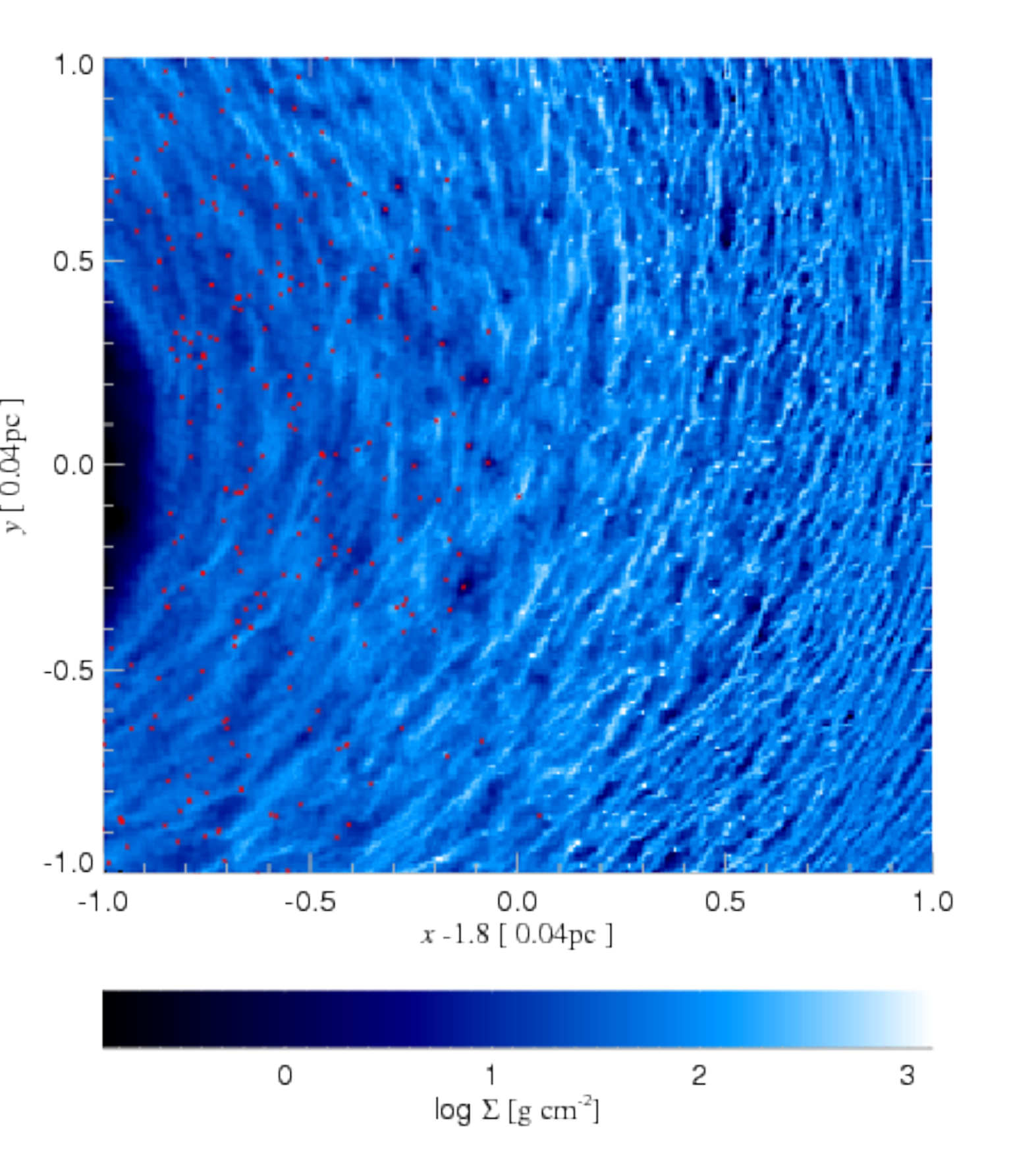}
\caption{Snapshot of the disc column density at time $t = 75$ (in
  $N$-body units) for run S2 of \cite{Nayakshin07}. In this $N$-body/SPH
  simulation\index{N-body simulations}, stars form as sink particles from an accretion disc\index{Accretion disc}.
  The left-hand panel shows the full simulation domain, whereas the
  right-hand one zooms in on a region of the disc centred at $x =
  1.8$. Stars with masses greater than $3\msun$ are plotted as the red
  asterisks. From Fig.~2 of \cite{Nayakshin07}.}
\label{fig:nayakshin07}
\end{figure}

\citet{Nayakshin06} find that, if star formation takes place in a
marginally stable accretion disc\index{Accretion disc}, the protostars heat up and thicken
the accretion disc\index{Accretion disc}, preventing further fragmentation\index{Fragmentation}. This occurs
because the accretion luminosity of the protostars exceeds the disc
radiative cooling, heating and puffing the disc up. While stellar
feedback stops further fragmentation\index{Fragmentation}, mass accretion on the already
formed protostars continues very efficiently, producing a top-heavy
MF\index{Mass function}. \cite{Nayakshin07} confirm these findings by means of
$N$-body/smoothed particle hydrodynamics (SPH) simulations\index{N-body simulations} of an
accretion disc\index{Accretion disc} (see Fig.~\ref{fig:nayakshin07}). Despite a number of
severe approximations (e.g. a constant cooling time, and the usage of
sink particles to model star formation without resolving gas
fragmentation\index{Fragmentation} directly), this is the first self-consistent simulation
of an accretion disc\index{Accretion disc} showing that (i) the thermal feedback associated
with gas accretion on to protostars slows down disc fragmentation\index{Fragmentation},
(ii) the initial MF (IMF)\index{Mass function} of the stars may be considerably
top-heavy with respect to Salpeter IMF\index{Mass function} (\citealt{Salpeter55}).
%%%%%%%%%%%%%%%%%%%%%%%%%%%%%%%%FIGURE %%%%%%%%%%%%%%%%%%%%%%%%%%%%%%%%%%%%%%%%%
%%%%%%%%%%%%%%%%%%%%%%%%%%%%%%%%FIGURE %%%%%%%%%%%%%%%%%%%%%%%%%%%%%%%%%%%%%%%%%

So far, the main issues of the accretion-disc-fragmentation\index{Fragmentation} scenario
are (i) if the orbits of gas particles in the accretion disc\index{Accretion disc} are
tidally circularised by viscosity, the orbits of the newly born stars
are circular too, and cannot reproduce the observed eccentricity
distribution in the GC (\citealt{Cuadra08}; on the other hand, this
issue may be overcome by starting with an eccentric accretion disc\index{Accretion disc} and
by imposing that it forms stars before tidal circularisation,
\citealt{Nayakshin07}; \citealt{Alexander08}); (ii) the newly born
stellar disc is expected to be very thin, much thinner (and with
smaller individual inclinations) than the observed disc of early-type
stars\index{Early-type stars} in the GC (e.g. \citealt{Cuadra08}); (iii) the MW SMBH is
currently quiescent and there is no evidence of an accretion disc\index{Accretion disc}:
which mechanisms induced the formation of an accretion disc\index{Accretion disc} and then
destroyed it a few Myrs since the formation of the stellar disc
(e.g. \citealt{Alexander12})?

\subsection{Molecular cloud\index{Molecular cloud|textbf} disruption\index{Tidal disruption|textbf}}
\label{subsec:3.2}

A molecular cloud\index{Molecular cloud} is disrupted well before reaching the inner
parsec. This was the main argument against the {\it in situ} formation
of the early-type stars\index{Early-type stars} in the GC. On the other hand, star formation
may take place even within a disrupted molecular cloud\index{Molecular cloud}. The two
necessary requirements for a disrupted molecular cloud\index{Molecular cloud} to form stars
in the central parsec of the MW are (i) that the molecular cloud\index{Molecular cloud} orbit
has very low angular momentum\index{Angular momentum}; (ii) that the streamers of the
disrupted cloud collide with each other and are shocked.

The former requirement is necessary for the streamers to settle on a
sufficiently tight orbit (i.e. the initial pericentre of the cloud
orbit must be $\lesssim{}1$ pc). The latter, i.e. collisions
between the filaments, is requested because collisions produce
shocks, which induce fast cooling and enhance the gas density by
orders of magnitude. In this way, the density of the post-shock
streamers can overcome the threshold for tidal disruption\index{Tidal disruption} (equation~\ref{eq:roche}), and the densest gas clumps collapse into
protostars. This is the basic motivation of the pioneering study by
\cite{Sanders98} and of a number of recent papers studying the
disruption\index{Tidal disruption} of a molecular cloud\index{Molecular cloud} in the surroundings of the GC, by
means of $N$-body/SPH simulations\index{N-body simulations} (\citealt{Bonnell08}; \citealt{Mapelli08}; \citealt{Hobbs09}; \citealt{Alig11};
\citealt{Mapelli12}; \citealt{Lucas13};
\citealt{Alig13}).

 The aforementioned papers describe simulations of the infall of one or more molecular clouds\index{Molecular cloud} toward
Sgr~A$^\ast{}$. 
 %{\bf For example, \cite{Mapelli12} simulate the infall of a molecular cloud\index{Molecular cloud} toward
 %Sgr~A$^\ast{}$. 
They consider different cloud masses (ranging from $\sim{}10^4$ M$_\odot{}$ to $\sim{}10^6$ M$_\odot{}$),  
%, %($4\times{}10^4$, $1.3\times{}10^5\msun$)
 temperatures\footnote{Many temperature components have been observed in the GC, ranging from $\sim{}20$ K to $\sim{}200$ K, and the temperature distribution is highly non-uniform (e.g. \citealt{McGary02}; \citealt{Herrnstein02}; \citealt{Herrnstein05}; \citealt{Montero09}; \citealt{baobab13}).} (ranging from $\sim{}10$ K to $\sim{}500$ K) and thermodynamics (adiabatic gas, isothermal gas or radiative cooling). 

\begin{figure}[t]
%\sidecaption[t]
\includegraphics[width=13cm]{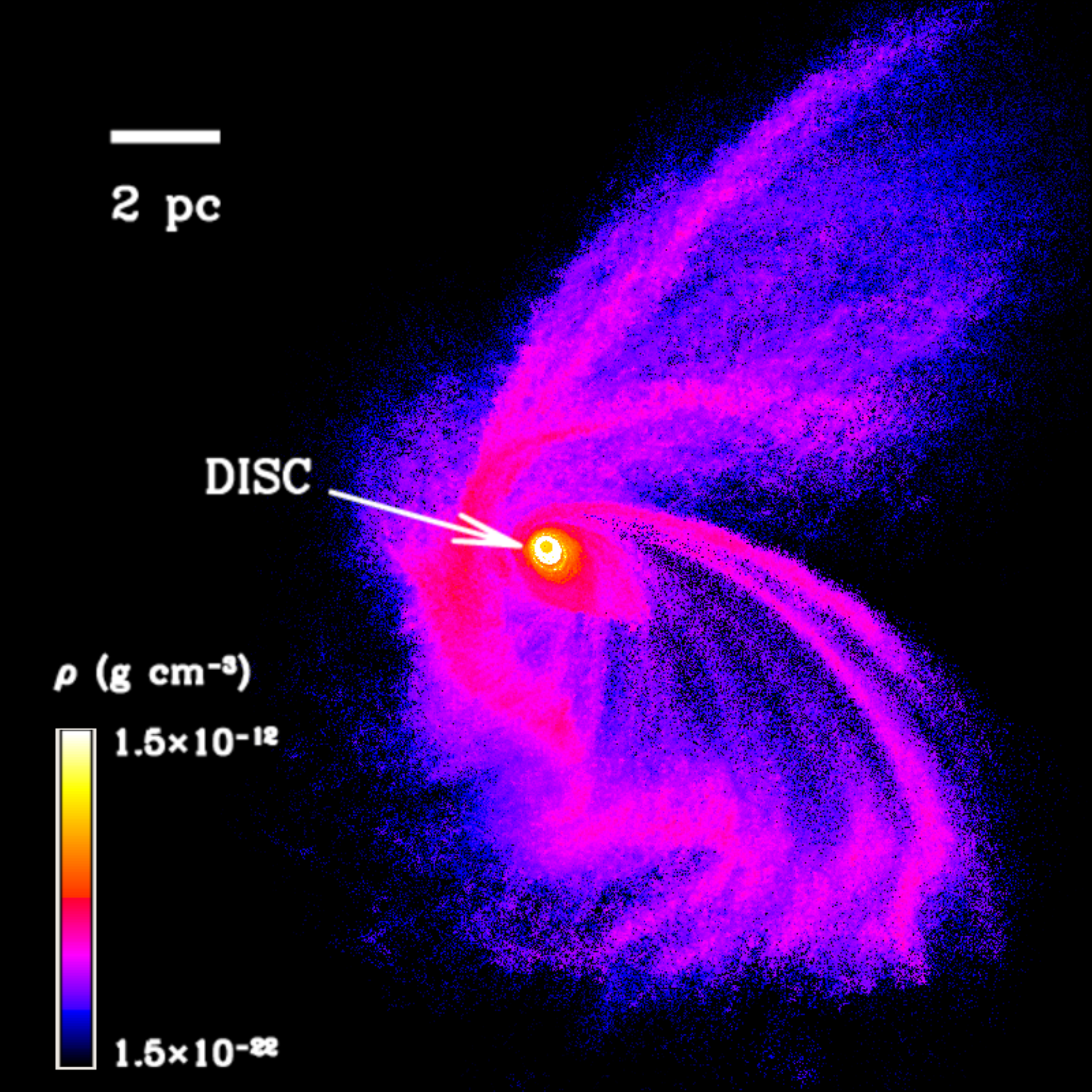}
\caption{Density map of the gas in run~E of \cite{Mapelli12} at $t=4.8\times{}10^5$
  yr. Simulation of a $1.3\times{}10^5\msun$ molecular cloud\index{Molecular cloud},
  disrupted by a $3.5\times{}10^6\msun$ SMBH. The simulation is
  projected in the plane where the gaseous disc (at the centre) is
  seen face-on. The box measures 20 pc per edge. The simulation has
  been run with the $N$-body/SPH\index{N-body simulations} code GASOLINE (\citealt{Wadsley04}) and
  includes radiative cooling (\citealt{Boley09};
  \citealt{Boley10}). The colour-coded map is logarithmic and ranges from $1.5\times{}10^{-22}$ to $1.5\times{}10^{-12}$ g cm$^{-3}$. From \cite{Gualandris12}.}
\label{fig:mapelli12}
\end{figure}

In all the simulations, the cloud is disrupted by the tidal forces of the
SMBH and spirals towards it. In less than $10^5$ yr, more than one
tenth of the gas in the parent cloud ends up in a dense and distorted disc around the SMBH, with a small outer radius
($\sim{}0.5$ pc, see e.g. Fig.~\ref{fig:mapelli12}). 
If the angular momentum\index{Angular momentum} of the cloud orbit is low, the
resulting gaseous disc is eccentric, consistently with the
observations of the stellar orbits in the CW disc. 
Locally, the surface
density of the gaseous disc may overcome the tidal shear from the SMBH
and fragmentation\index{Fragmentation} may take place.

%All the considered papers indicate that a dense gaseous disc may form
%around SgrA$^\ast$ as a consequence of the infall and of the disruption\index{Tidal disruption} of
%a massive molecular cloud\index{Molecular cloud}. 

%The eccentricity
%and the size of the final gaseous disc are similar in all the
%considered papers. 

%%%%%%%%%%%%%%%%%%%%%%%%%%%%%%%%FIGURE %%%%%%%%%%%%%%%%%%%%%%%%%%%%%%%%%%%%%%%%%
%%%%%%%%%%%%%%%%%%%%%%%%%%%%%%%%FIGURE %%%%%%%%%%%%%%%%%%%%%%%%%%%%%%%%%%%%%%%%%

\begin{figure}[t]
%\sidecaption[t]
\includegraphics[width=6.5cm]{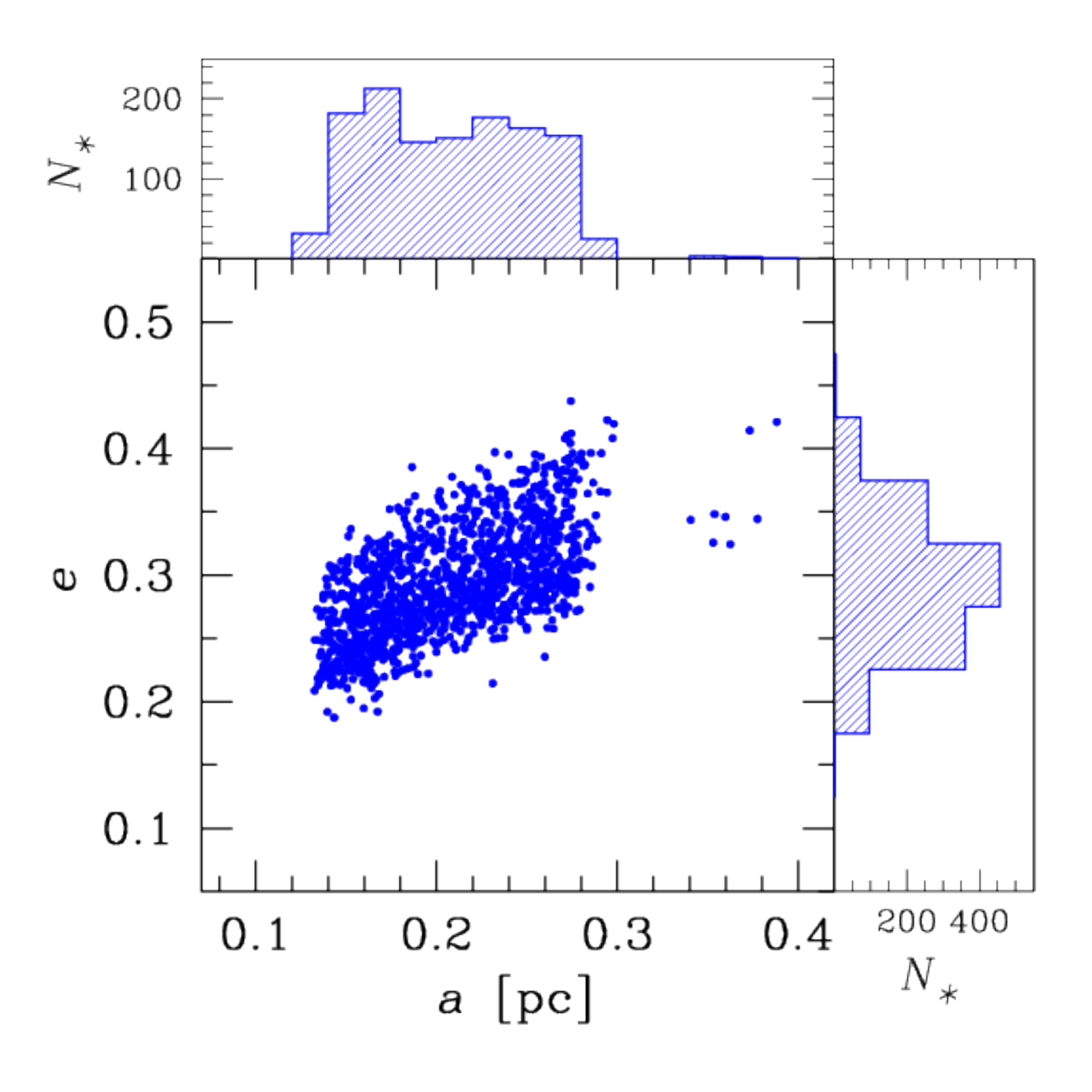}
\includegraphics[width=6.5cm]{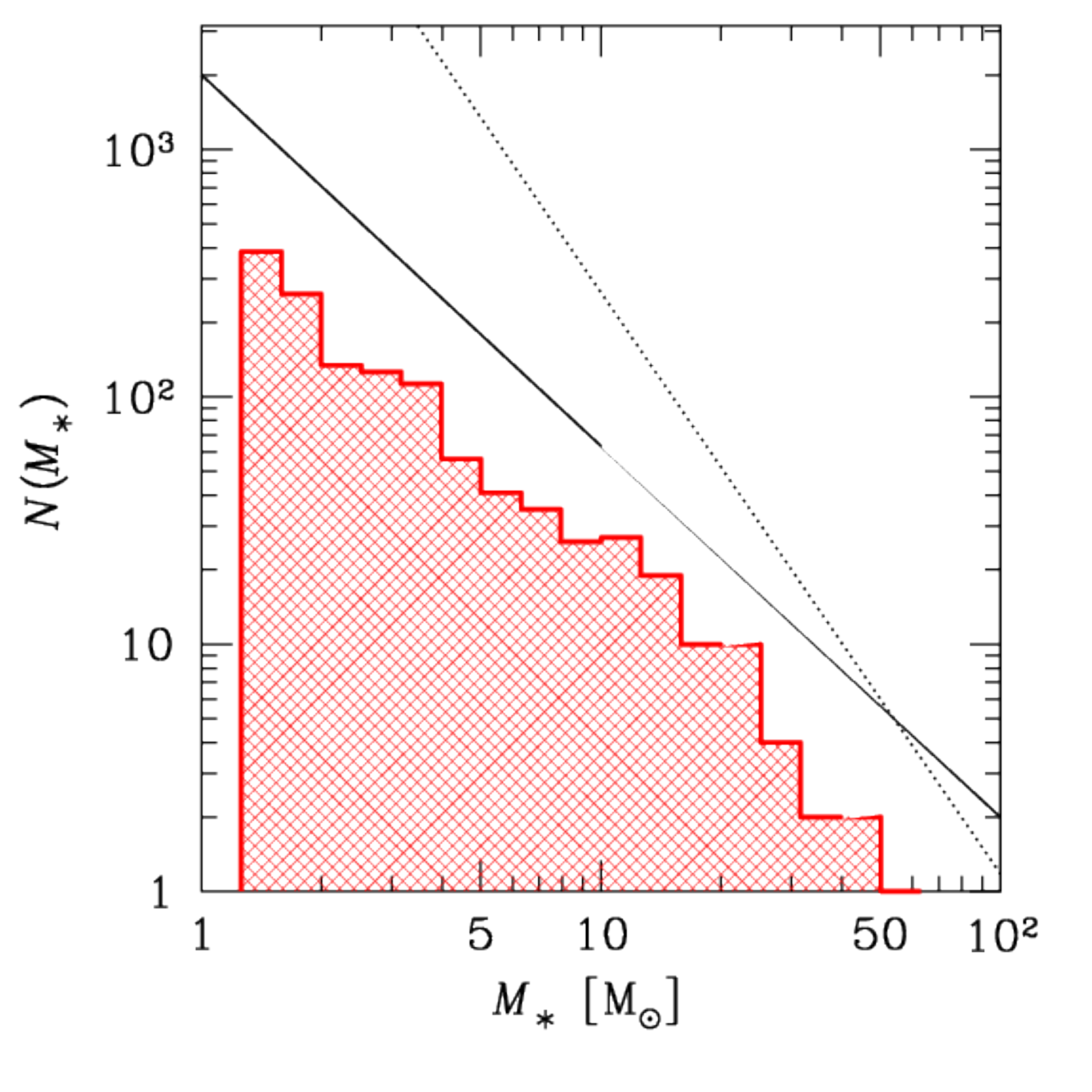}
\caption{Left-hand panel: Eccentricity $e$ versus semi-major axis $a$ at $t=4.8\times{}10^5$ yr in run~E of \cite{Mapelli12} (the same as in Fig.~\ref{fig:mapelli12}). The marginal histograms show the distribution of $a$ (top histogram) and $e$ (right-hand histogram). Right-hand panel: Stellar MF\index{Mass function} in run~E of \cite{Mapelli12} at $t=4.8\times{}10^5$ yr (hatched red histogram). $x-$axis: star mass $M_\ast{}$. $y-$axis: number of stars per mass bin $N(M_\ast{})$.  Solid (dotted) black thin line: MF\index{Mass function} $dN/dm\propto{}m^{-\alpha{}}$ with $\alpha{}=1.5$ ($\alpha{}=2.35$). From \cite{Mapelli12}.} 
\label{fig:mapelli12b}
\end{figure}

Among the aforementioned papers, the simulations presented in \cite{Mapelli12} are the first attempt to trace the fragmentation\index{Fragmentation} of the gas disc, without adopting the sink particle technique. The star candidates formed in these simulations are distributed in a thin ring at a distance of
$\sim{}0.1-0.4$ pc from the SMBH. They have eccentric orbits
($0.2\le{}e\le{}0.4$), with average eccentricity
$\langle{}e\rangle{}=0.29\pm{}0.04$ (Fig.~\ref{fig:mapelli12b}). Both the semi-major axis and the eccentricity distribution are in agreement with the properties
of the observed CW disc\index{Clockwise disc} (e.g. Fig.~\ref{fig:yelda12}). 

Both \cite{Bonnell08} and \cite{Mapelli12} agree that, if the parent molecular cloud\index{Molecular cloud} is
sufficiently massive ($1.3\times{}10^5\msun$), the total mass
of simulated star candidates ($2-5\times{}10^3\msun$) is consistent with
the estimated mass of the CW disc\index{Clockwise disc} (e.g.~\citealt{Paumard06};
\citealt{Bartko09}; but see \citealt{Lu13} for a slightly different estimate). 

Furthermore, if the minimum temperature (i.e. the temperature floor due to diffuse radiation in the GC) is sufficiently high (T$\sim{}100$ K), the MF\index{Mass function} of stellar candidates is top-heavy (fitted by a single power-law with $\alpha{}\sim{}1.5$ in the case of \citealt{Mapelli12}, Fig.~\ref{fig:mapelli12b}), in good agreement with the recent measurements by Lu et al. (2013, see Fig.~\ref{fig:lu2013_10}). The main reason is that a higher gas temperature corresponds to a higher Jeans mass ($m_{\rm J}\propto{}T^{3/2}$, \citealt{Jeans19}).

%%%%%%%%%%%%%%%%%%%%%%%%%%%%%%%%FIGURE %%%%%%%%%%%%%%%%%%%%%%%%%%%%%%%%%%%%%%%%%
%%%%%%%%%%%%%%%%%%%%%%%%%%%%%%%%FIGURE %%%%%%%%%%%%%%%%%%%%%%%%%%%%%%%%%%%%%%%%%

%These results are consistent with those of the other papers (e.g. \citealt{Mapelli08}; \citealt{Bonnell08}; \citealt{Hobbs09}; \citealt{Alig11}) 

All the papers that simulate the infall of a
molecular cloud\index{Molecular cloud} towards the GC (e.g. \citealt{Mapelli08}; \citealt{Bonnell08}; \citealt{Hobbs09};
\citealt{Alig11}) agree on the general picture.  However, there are significant differences between these papers, both in the initial conditions and in some of the results.

As to the initial conditions, \cite{Mapelli08}, \cite{Bonnell08} and \cite{Mapelli12} adopt models of gas clouds that are turbulently supported, while \cite{Hobbs09} and \cite{Alig11} consider a simplified model of spherical and homogeneous cloud. 
The simulations reported in \cite{Mapelli08} are isothermal, with
T$_{\rm MC}=10$~K (likely too low, if compared to the background radiation field in the GC).  The simulations in
\cite{Bonnell08} include an approximate radiative transfer formalism,
with compressional heating balanced by cooling rates derived from
estimated optical depths. In \cite{Hobbs09}, the simulations include a
very simplified model of cooling, and the initial temperature of the
cloud is low (T$_{\rm MC}=20$ K).
In \cite{Alig11} and \cite{Mapelli12}, the simulations include
different thermodynamical treatments for the gas, considering both
isothermal and radiative cooling cases. The floor temperature for the
simulations with radiative cooling is set to be 50 K in \cite{Alig11}
and 100 K in both \cite{Bonnell08} and \cite{Mapelli12}. \cite{Alig11}
stop their simulation before fragmentation\index{Fragmentation} takes place in the disc,
whereas the other considered papers study the formation of star
candidates in the disc. \cite{Mapelli08}, \cite{Bonnell08} and
\cite{Hobbs09} adopt the sink particle technique, to model SF. Only
\cite{Mapelli12} follow the initial fragmentation\index{Fragmentation} of the disc. Table~1 is a summary of the differences in the initial conditions of the aforementioned simulations.

%%%%%%%%%%%%%%%%%%%%%%%%%%%%%%%%TABLE %%%%%%%%%%%%%%%%%%%%%%%%%%%%%%%%%%%%%%%%%
\begin{table}
\begin{center}
\caption{Main differences in the initial conditions of simulations of molecular cloud disruption.} \leavevmode
\begin{tabular}[!h]{ccccc}
\hline
Paper & Cloud model & $T_{\rm MC}$ (K) & gas treatment & sink particles\\
\hline
\cite{Bonnell08} & turbulently supported & 100  & radiative cooling  & yes\vspace{0.2cm}\\
\cite{Mapelli08} & turbulently supported & 10   & isothermal         & yes\vspace{0.2cm}\\
\cite{Hobbs09}  & homogeneous  sphere    & 20   & simplified cooling & yes\vspace{0.2cm}\\
\cite{Alig11}   & homogeneous   sphere    & 50   & both isothermal and & no \\
                &                        &      & radiative cooling & \vspace{0.2cm}\\
\cite{Mapelli12} & turbulently supported & 100, 500 & both isothermal and & no\\ 
                &                        &      & radiative cooling & \vspace{0.2cm}\\
\cite{Lucas13} & turbulently supported   & 100  &  radiative cooling    & yes\vspace{0.2cm}\\
\hline
\end{tabular}
\end{center}
\end{table}
%%%%%%%%%%%%%%%%%%%%%%%%%%%%%%%%FIGURE %%%%%%%%%%%%%%%%%%%%%%%%%%%%%%%%%%%%%%%%%

The main differences among the results of these papers %various simulations
are about the formation of star candidates, and especially about the
MF\index{Mass function}. The MF\index{Mass function} in \cite{Hobbs09} is quite bottom-heavy, because of the
approximations in the cooling recipes and because of the absence of
opacity prescriptions.  

\cite{Bonnell08} adopt a very conservative
value of the critical density for converting gaseous particles into
sink particles ($=10^{14}$ M$_\odot{}$ pc$^{-3}=1.6\times{}10^{15}$
cm$^{-3}$, assuming molecular weight $\mu{}=2.46$), well above the
critical tidal density. Therefore, their MF\index{Mass function} is consistent with that
predicted by the Jeans mass for the local density and temperature of
the clouds. Similarly, the MFs\index{Mass function} derived in \cite{Mapelli12} are
consistent with the predictions from Jeans mass and Toomre
instability.

On the other hand, \cite{Mapelli12} do not observe the formation of
the very massive stars ($>60\msun$) that were found in the
massive cloud simulated by \cite{Bonnell08}. The MF\index{Mass function} in
\cite{Mapelli12} is consistent with a single power-law with index
$\alpha{}\sim{}1.5$, whereas that in \cite{Bonnell08} is clearly
bimodal, showing two distinct stellar populations (see fig.~4 of
\citealt{Bonnell08}). The very massive stars in \cite{Bonnell08} are
all formed at $r\sim0.02$ pc, where massive stars have not been
observed in the MW (the observed ring of young stars having an inner
radius of $\sim{}0.04$ pc). In \cite{Mapelli12}, star candidates do
not form at $r<0.05$ pc, because the shear from the SMBH prevents
local collapse.  This difference is likely due to the
different orbits of the parent clouds, to the different initial
densities and to the different recipes for opacity.

\begin{figure}[t]
%\sidecaption[t]
\includegraphics[width=6.5cm]{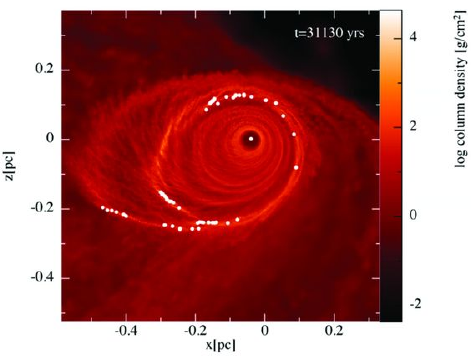}
\includegraphics[width=6.5cm]{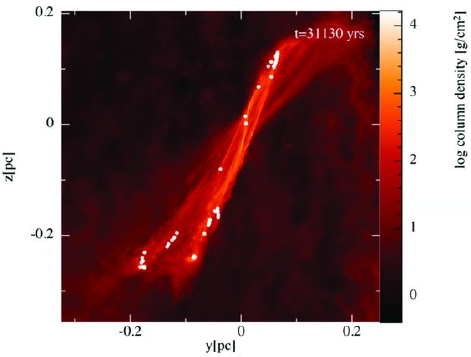}
\caption{Column densities for run I10 of \cite{Lucas13} in the $xz$
  and $yz$ planes at the simulation's end at $t =
  3.113\times{}10^4$ yr. In this simulation, a prolate molecular
  cloud\index{Molecular cloud}, oriented perpendicular to its orbital plane, is disrupted by
  a SMBH. The white circles mark the position of sink particles. The
  streamers are oriented 17$^\circ$ out of plane from the disc,
  resulting in two stellar systems separated by this angle. From
  Fig.~13 of \cite{Lucas13}.}
\label{fig:lucas13}
\end{figure}

Recently, \cite{Lucas13} showed that the disruption\index{Tidal disruption} of a single
prolate cloud, oriented perpendicular to its orbital plane, produces a
spread in angular momenta\index{Angular momentum} of gas particles, and leads to the formation
of stars with slightly misaligned orbital planes (see
Fig.~\ref{fig:lucas13}). This matches the observations, which indicate
that the early-type stars\index{Early-type stars} in the CW disc\index{Clockwise disc} have different orbital
inclinations. On the other hand, we will show in Sect.~\ref{sec:4}
that the misalignment of orbits can be the result of various dynamical
processes, taking place after the formation of the first disc.
%%%%%%%%%%%%%%%%%%%%%%%%%%%%%%%%FIGURE %%%%%%%%%%%%%%%%%%%%%%%%%%%%%%%%%%%%%%%%%
%%%%%%%%%%%%%%%%%%%%%%%%%%%%%%%%FIGURE %%%%%%%%%%%%%%%%%%%%%%%%%%%%%%%%%%%%%%%%%
\vspace{1cm}

The main problem with the disruption\index{Tidal disruption} scenario is the fact that the
molecular cloud\index{Molecular cloud} must have a very low angular momentum\index{Angular momentum}. In fact,
shock-induced radiative cooling reduces the orbital energy of the
cloud rather than its angular momentum\index{Angular momentum}. Therefore, the initial mean
angular momentum\index{Angular momentum} per unit mass of the cloud ($\sim{}b\,{}v$, where $b$
is the impact parameter and $v$ the initial velocity of the cloud
centre of mass) is nearly preserved during the disruption\index{Tidal disruption} of the cloud
and the settling of the disc. Thus, the radius of the formed gas disc
$R_{\rm d}$ will be approximately (\citealt{Wardle08}):
\begin{equation}\label{eq:wardle}
R_{\rm d}\,{}\left(\frac{G\,{}m_{\rm BH}}{R_{\rm d}}\right)^{1/2}\sim{}b\,{}v.
\end{equation}
Adopting $m_{\rm BH}=3.5\times{}10^6\msun$ and $R_{\rm d}\sim{}1$ pc, we obtain $b\lesssim{}1\,{}{\rm pc}\,{}v_{100}^{-1}$ (with $v=v_{100}\,{}100$ km s$^{-1}$).

\begin{figure}[t]
%\sidecaption[t]
\includegraphics[width=6.0cm]{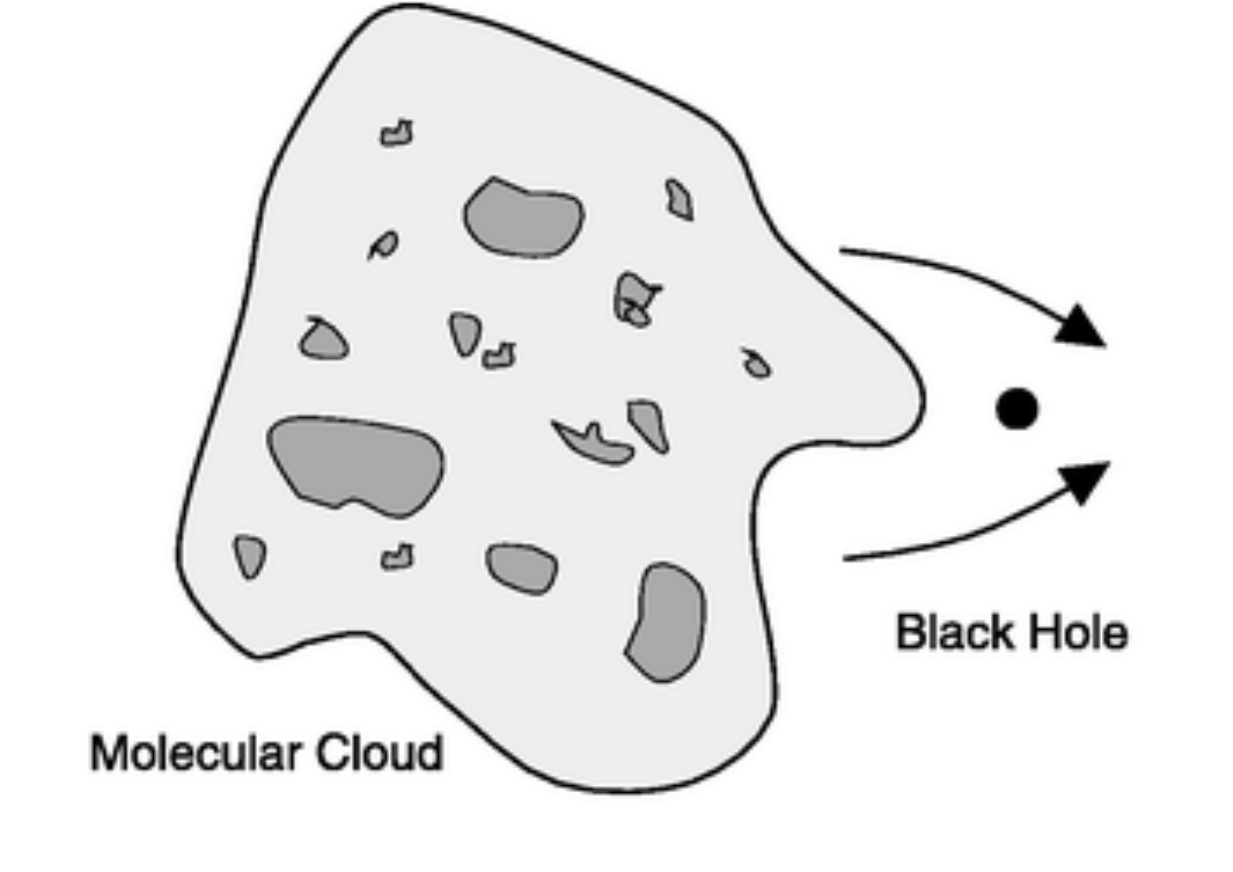}
\includegraphics[width=6.0cm]{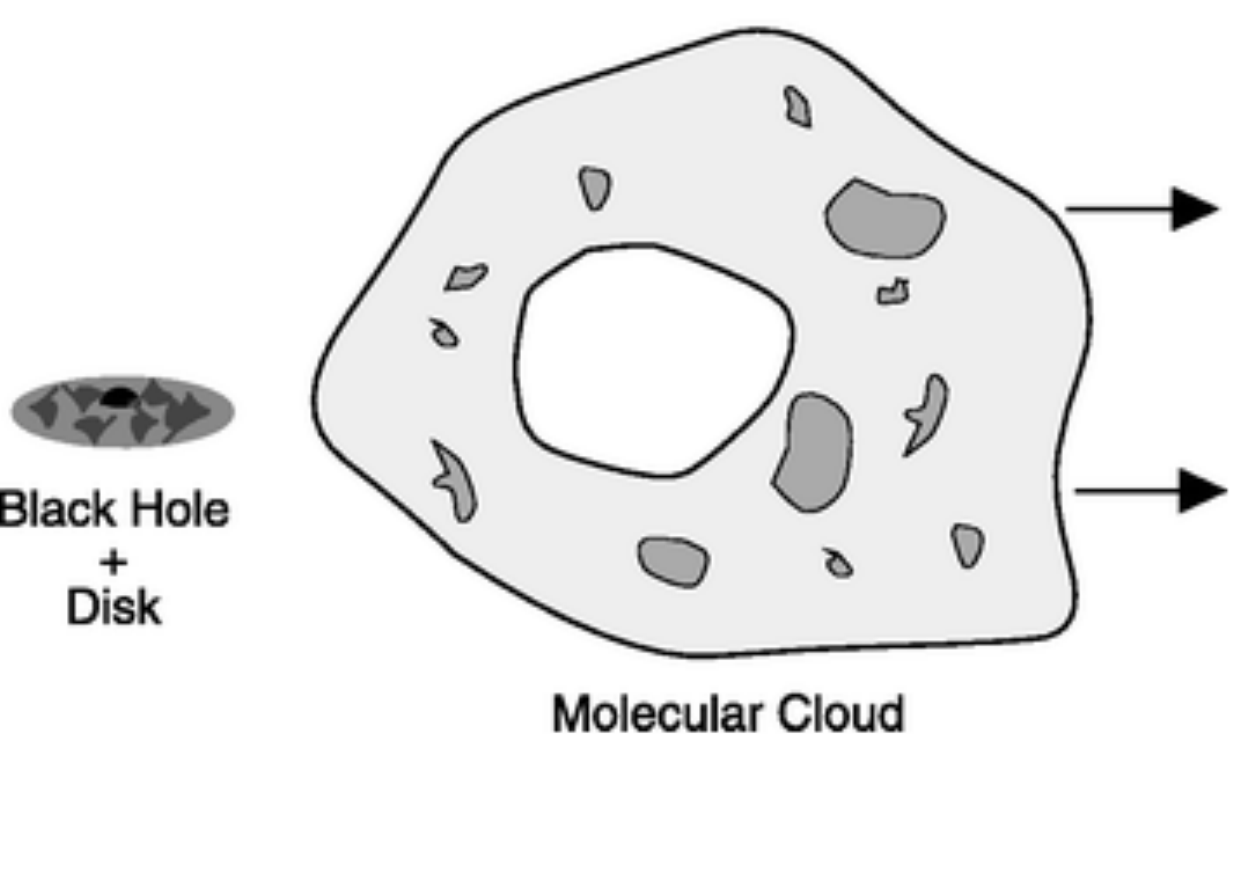}
\caption{Schematic diagram of a cloud engulfing Sgr~A$^\ast$, from
  Fig.~1 of \cite{Wardle08}. The left-hand panel indicates the
  gravitational focusing of incoming molecular cloud\index{Molecular cloud} material
  (incident from the left). The right-hand panel shows the carved-out
  inner region of the cloud that has been captured by Sgr~A$^\ast$ and
  circularised to form a disc. The outer region of the cloud continues
  its motion in the direction away from Sgr~A$^\ast$.}
\label{fig:wardle}
\end{figure}

This argument does not hold 
%(i) if the turbulent velocity of gas
%clumps inside the cloud is important with respect to the orbital
%angular momentum; (ii) 
if a molecular cloud\index{Molecular cloud} engulfs Sgr~A$^\ast{}$
during its passage through the GC, and it is partially captured by the
SMBH (see Fig.~\ref{fig:wardle}). The partial capture of a portion of
the molecular cloud\index{Molecular cloud} is enhanced by gravitational focusing. Fluid
elements passing on opposite sides of Sgr~A$^\ast{}$ have oppositely
directed orbital angular momenta\index{Angular momentum}, so that the collision between them
leads to a partial cancellation of the specific angular momentum\index{Angular momentum}. The
efficiency of angular momentum\index{Angular momentum} cancellation depends on the density and
velocity inhomogeneities in the gas.

%%%%%%%%%%%%%%%%%%%%%%%%%%%%%%%%FIGURE %%%%%%%%%%%%%%%%%%%%%%%%%%%%%%%%%%%%%%%%%
%%%%%%%%%%%%%%%%%%%%%%%%%%%%%%%%FIGURE %%%%%%%%%%%%%%%%%%%%%%%%%%%%%%%%%%%%%%%%%

\begin{figure}[t]
%\sidecaption[t]
\includegraphics[width=12.5cm]{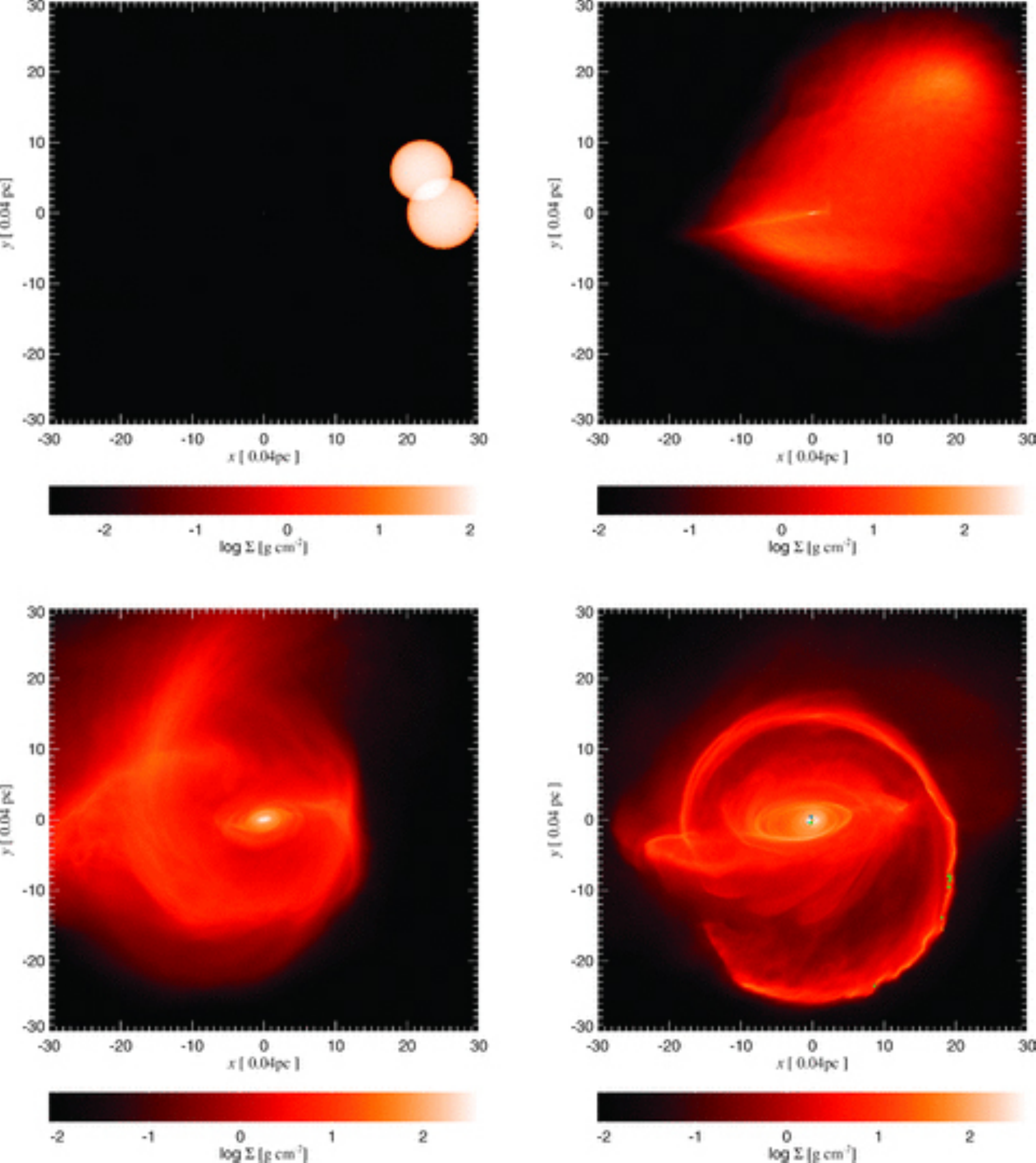}
\caption{Gas surface density and star locations (in the bottom
  right-hand panel) for snapshots from simulation S1 of \cite{Hobbs09}
  taken at times $t= 0, 100, 250$ and $1000$ (in $N$-body units), left
  to right and top to bottom, respectively. Sgr A$^\ast{}$ is located
  at $(0,0)$, and the line of sight is along the $z-$direction. This
  simulation follows the collision of two molecular clouds\index{Molecular cloud} and the
  infall of the collision product toward the GC. Fig.~1 of
  \cite{Hobbs09}.}
\label{fig:hobbs}
\end{figure}

An alternative solution to the angular momentum\index{Angular momentum} problem is to assume
that two molecular clouds\index{Molecular cloud} collided a few parsecs away from
Sgr~A$^\ast{}$, lost part of their angular momentum\index{Angular momentum} during the
collision, and fell towards the SMBH with a very small impact
parameter. According to this scenario, \cite{Hobbs09} simulate the
collision between two spherical clouds and the disruption\index{Tidal disruption} of the
collision product by a SMBH. Their $N$-body/SPH\index{N-body simulations} simulations reproduce many
interesting features of the observed early-type stars\index{Early-type stars} in the GC, such
as the presence of stars with high orbital inclinations with respect
to the main disc (see Fig.~\ref{fig:hobbs}). Other studies
(e.g. \citealt{Mapelli12}) simulate clouds with very low initial
angular momentum\index{Angular momentum}, assuming that this was the result of a previous
collision between two different clouds.
%%%%%%%%%%%%%%%%%%%%%%%%%%%%%%%%FIGURE %%%%%%%%%%%%%%%%%%%%%%%%%%%%%%%%%%%%%%%%%
%%%%%%%%%%%%%%%%%%%%%%%%%%%%%%%%FIGURE %%%%%%%%%%%%%%%%%%%%%%%%%%%%%%%%%%%%%%%%%
%-problem of low angular momentum orbit
%--> Hobbs \& Nayakshin
%Wardle \& Yusef-Zadeh (equations)

Finally, \cite{Alig13} propose that the young stellar disc is the
result of the collision between a molecular cloud\index{Molecular cloud} and the CNR\index{Circumnuclear ring} (see
Sect.~\ref{subsec:2.3} for a description of the observed properties
of the CNR\index{Circumnuclear ring}). The $N$-body/SPH simulations\index{N-body simulations} %(run with the Gadget3 code, \citealt{Springel01}; \citealt{Springel05})
 described in \cite{Alig13} show that the collision between a molecular cloud\index{Molecular cloud} and the CNR\index{Circumnuclear ring} leads to
multiple streams of gas flowing toward the SMBH. This simulation shows
that more than a single disc can be formed through this pathway.

\subsection{Star cluster disruption{\index{Cluster disruption|textbf}} }
\label{subsec:cluster}

Star formation from standard collapse can proceed outside the central
parsec and lead to the formation of young star clusters\index{Star cluster} like the
Arches and the Quintuplet, which then inspiral due to dynamical
friction and deposit stars while being tidally disrupted
\citep{Gerhard2001}. A dissolving cluster would lead to the formation
of a stellar disc similar to the CW disc\index{Clockwise disc}, possibly accompanied by a
number of isolated outliers. It would also preferentially deposit
massive stars close to the SMBH, as these would form a compact core
and survive tidal effects down to smaller separations than low-mass
stars. However, a cluster would need to be very dense and massive to
be able to inspiral within the lifetime of its massive stars
\citep{KimMorris2003}.

The inspiral\index{Cluster inspiral|textbf} would be accelerated by the presence of an
IMBH\index{Intermediate-mass black hole} in the centre of the cluster, if
this was as massive as 10\% of the cluster mass
\citep{Kim2004}. Formation of IMBHs has been predicted from a number of
$N$-body simulations\index{N-body simulations}. The simulations indicate that a runaway sequence of mergers of massive stars
leads to the formation of a very massive object which is assumed to
eventually collapse \citep{SPZ2004}. The evolution of such a very
massive star, however, is not well known, and it has been argued that
collapse to an IMBH might be prevented by severe mass-loss in the form
of stellar winds \citep{Yungelson2008}.

Another potential difficulty of this model is that an inspiralling
cluster would deposit stars all along its orbit while being stretched
and tidally disrupted. While young stars have been observed outside
the central $0.5$ pc \citep{Buchholz2009}, the required number of
young stars is much larger than what currently inferred from
observations \citep{PeretsGualandris2010}.

%%%%%%%%%%%%%%%%%%%%%%%%%%%%%%%%FIGURE  %%%%%%%%%%%%%%%%%%%%%%%%%%%%%%%%%%%%%%%%%
\begin{figure}[t]
%\sidecaption[t]
\includegraphics[width=0.97\textwidth]{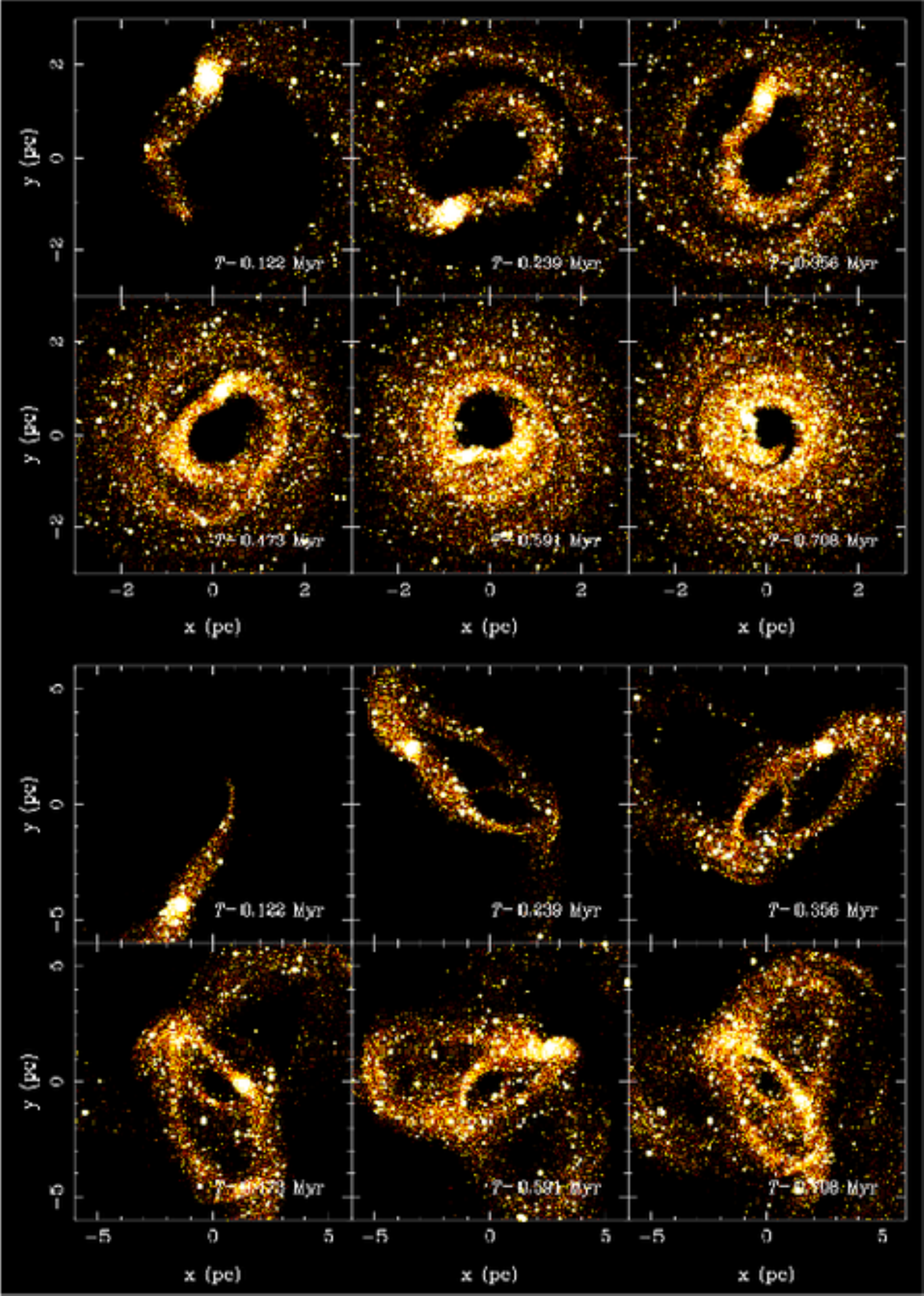}
\caption{Snapshots from the simulations of Fujii et al. (2008, Fig.2) of the
  inspiral of a star cluster in the case of a circular orbit (top
  panels) and an eccentric orbit (bottom panel). }
\label{fig:fujii2008}
\end{figure}
%%%%%%%%%%%%%%%%%%%%%%%%%%%%%%%%FIGURE %%%%%%%%%%%%%%%%%%%%%%%%%%%%%%%%%%%%%%%%%
\citet{Fujii2008} perform self-consistent simulations of the inspiral
of a star cluster\index{Star cluster!inspiral} in its parent galaxy and find that the inspiral time
is somewhat shorter than expected by simple application of
Chandraskhar's dynamical friction formula \citep{Ch43}, especially if
the cluster undergoes core collapse. In addition, an eccentric orbit
for the star cluster leads to a faster inspiral than a circular orbit,
mitigating the requirements on cluster density for survival down to
small separations. Snapshots from simulations of both a circular and
an eccentric orbit are shown in Fig.~\ref{fig:fujii2008}.  In these
simulations, the clusters are positioned at an initial distance of
either 2 or 5 parsecs, and no stars are found at distances smaller
than 0.5 pc at the end of the integration.

A further speed up of the inspiral is found in simulations in which
the star cluster\index{Star cluster} forms an IMBH in its centre in the early stages of
evolution \citep{Fujii2010}. In this case, the cluster can deposit
massive stars in a disc configuration around the IMBH.  Further
evolution can quickly randomise the orbital configuration and lead to
an isotropic distribution (\citealt{MGM2009}, see
Sect.~\ref{subsec:IMBHpert} for details), in agreement with
observations of the S-stars\index{S-stars}.  However, for this to
happen within the lifetime of the young stars, the cluster needs to be
dense and massive and on a very eccentric orbit.  While an IMBH of a
few thousand solar masses is sufficient for the purpose of randomising
the orbits of the bound stars, the simulated cluster in the models of
\citet{Fujii2010} forms an IMBH which is more massive than the
currently accepted upper limit for a secondary black hole in the GC
\citep{ReidBru2004}.

\subsection{The binary breakup scenario}
\label{subsec:3.4}
\begin{figure}[t]
\sidecaption[t]
\includegraphics[width=7.5cm]{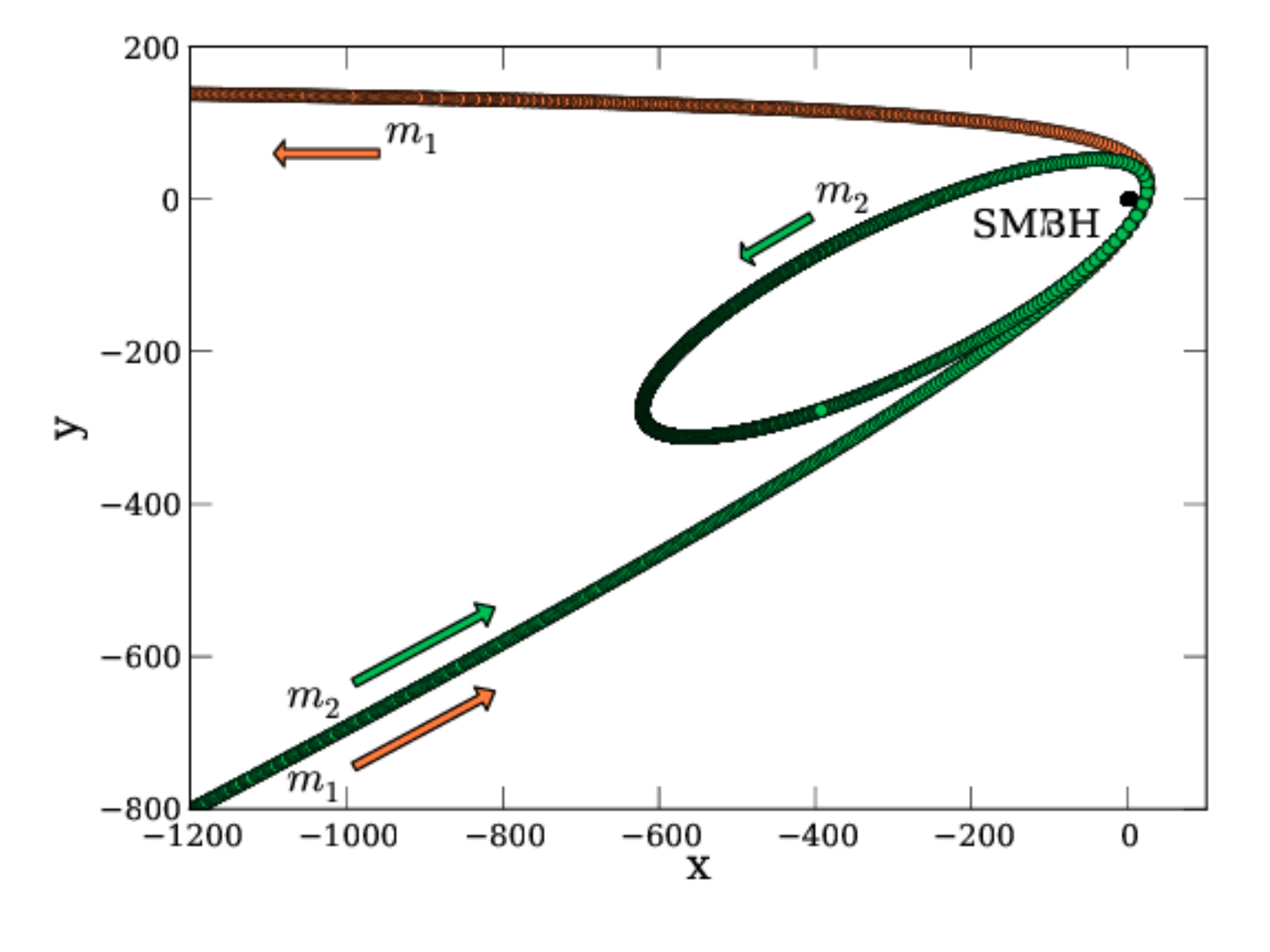}
\caption{Example of an encounter between a stellar binary and the SMBH
  in which the binary is broken apart; one star is ejected while the
  other is captured by the SMBH into a bound orbit. Distances are
  expressed in units of the initial binary separation.}
\label{fig:exchange}
\end{figure}

Another formation scenario that involves migration from outside the
central parsec is the breakup of stellar binaries scattered onto low
angular momentum\index{Angular momentum} orbits by relaxation processes.  A binary scattered
to pass very close to the SMBH\index{Supermassive black hole} is
likely to undergo an exchange interaction in which one of the stars is
ejected to large distances while the other is captured by the SMBH in
a wide and eccentric orbit (see Fig.~\ref{fig:exchange}).
%%%%%%%%%%%%%%%%%%%%%%%%%%%%%%%%FIGURE  %%%%%%%%%%%%%%%%%%%%%%%%%%%%%%%%%%%%%%%%%
%%%%%%%%%%%%%%%%%%%%%%%%%%%%%%%%FIGURE %%%%%%%%%%%%%%%%%%%%%%%%%%%%%%%%%%%%%%%%%

Theoretically, a binary is expected to be disrupted when it reaches a distance of the order of its
tidal radius\index{Tidal radius!binary}:
\begin{equation}
\label{eq:tidalradius}
r_{\rm t} \simeq \left(\frac{m_{\rm BH}}{M_{\rm b}}\right)^{1/3} \,a_{\rm b}\,,
\end{equation}
where $M_{\rm b}$ represents the binary mass and $a_{\rm b}$  its semi-major axis.
Simulations show that most binaries approaching the SMBH within the tidal radius
are actually disrupted \citep{Hills1991,Hills1992,Bromley2006}.
The mean semi-major axis of the captured star is \citep{Hills1991}
\begin{equation}
 \langle{} a_{\rm c} \rangle{} \simeq 0.56 \left(\frac{m_{\rm BH}}{M_{\rm b}}\right)^{2/3} a_{\rm b} \simeq 0.56 \left(\frac{m_{\rm BH}}{M_{\rm b}}\right)^{1/3} r_{\rm t}\,.
\end{equation}
This relation shows that harder stellar binaries (i.e. binaries with binding energy larger than average) tend to produce more tightly bound captured stars, and provides a direct mapping between the distribution of semi-major axes of incoming binaries and bound stars.
Because the pericentre distance of the captured stars is at the binary tidal radius, the eccentricity is quite high \citep{Hills1991, Miller2005}
\begin{equation}
e \simeq 1 - r_{\rm t} / a_{\rm c} \simeq 1 - 1.78 \left(\frac{M_{\rm b}}{m_{\rm BH}}\right)^{1/3} \simeq 0.97
\end{equation}
for stellar binaries interacting with the MW SMBH.  This is larger
than what is derived for any star in the S-cluster and may lead to the
conclusion that the binary breakup scenario is inconsistent with the
observed properties of the S-stars\index{S-stars}. Relaxation
processes, however, are able to alter the orbital eccentricities of
the bound population over the lifetime of the stars, bringing the
eccentricity distribution in agreement with the observed one (see
section \ref{subsec:rlx} for a discussion).

\citet{Antonini2010} study the dynamics of main-sequence binaries on
highly elliptical bound orbits with small pericentre distances.  They
find that bound stars can also be produced when the binary components
merge.  A coalescence remnant is not able to escape the SMBH
gravitational potential\index{Potential!gravitational} if the initial
binary is bound to the SMBH unless significant mass loss occurs.  The
probability for collisions between the components of the binary
increases with time, resulting in substantially larger numbers of
mergers when allowing for multiple pericentre passages.

Ejection velocities for the unbound star can be large enough to
explain the population of hypervelocity stars\index{Hypervelocity stars} 
(HVSs) \citep{Brown2005} in the halo of the Galaxy. In this
model, the HVSs are the former binary companions to the S-stars
\citep{GouldQuillen2003}.

In addition to an efficient mechanism to thermalise the eccentricities
within the lifetime of the stars, in order to be viable the binary
breakup model requires a continuous reservoir of binaries at large
radii, as well as a mechanism to scatter the binaries onto plunging
orbits. Scattering by massive perturbers\index{Massive perturbers}
like star clusters and molecular clouds\index{Molecular cloud} has been suggested to dominate
over two-body relaxation in the central 100 pc of the Galaxy
\citep{Perets2007}. Massive perturbers do not significantly contribute
to the disruption rate of single stars by the SMBH, but they may
enhance the tidal disruption\index{Tidal disruption} rate of binaries by a factor $10-1000$,
depending on their distribution.

\section{Evolution of the early-type stars\index{Early-type stars}}
\label{sec:4}
The different models described in Sect. 3 for the origin of the
young stars predict different distributions for
the orbital elements:
\begin{itemize}
\item Stars formed from the disruption\index{Tidal disruption} of a
molecular cloud\index{Molecular cloud} are expected to lie in a disc and have moderate
eccentricities ($0.2 \lesssim e \lesssim 0.5$).
\item Stars formed from the tidal breakup of stellar binaries are naturally
found in an isotropic configuration and with very large
eccentricities ($e \gtrsim 0.97$).
\item  Stars deposited by an inspiralling star
cluster with an IMBH in the centre will have distributions of
semi-major axes and eccentricities centred on the orbital elements of the
IMBH, and will also be orbiting in a plane. 
\end{itemize}

It is interesting to note that none of the suggested models predicts a
roughly thermal distribution of eccentricities $N(<e) \sim e^2$, as is
observed for the S-stars. The predicted distributions, however, cannot
be compared directly with observations, because they evolve during the
lifetime of the stars due to relaxation processes and external
perturbations. We review these processes in this section.

\subsection{Secular processes: precession and Kozai cycles}
\label{subsec:precession}
The motion of a star inside the SMBH sphere of influence can be
described as the motion of a test particle in a Keplerian
potential\index{Potential!Keplerian} (due to the SMBH), perturbed by
an external potential\index{Potential!external}. The sources of the
external potential may be the spherical cusp of old stars, a stellar
disc, a gaseous disc, the circumnuclear ring, a molecular cloud\index{Molecular cloud}, an
IMBH or whatever other perturber is sufficiently massive and
sufficiently close to the GC.
 
Precession\index{Precession|textbf} is one of the main effects that are induced by the external
potential. Precession effects have been invoked to explain the
formation of the S-star cluster (e.g. \citealt{Ivanov05};
\citealt{Lockmann08}; \citealt{Lockmann09}) and the broad distribution
of angular momentum\index{Angular momentum} vectors of the orbits of early-type stars\index{Early-type stars}
(\citealt{Subr09}; \citealt{Haas11a}; \citealt{Haas11b};
\citealt{Mapelli13}). The strength and the effects of precession
depend on the nature of the potential. In particular, the precession
induced by a spherically symmetric potential (e.g. the spherical
stellar cusp of old stars) is very different from that induced by
either an axisymmetric potential (e.g another stellar disc, a gaseous
disc/ring) or a single massive object (e.g. an IMBH\index{Intermediate-mass black hole}) orbiting the
SMBH. In this section, we briefly describe the precession effects that
may affect the early-type stars\index{Early-type stars} in the GC, and give an estimate of the
corresponding timescales.

%%%%%%%%%%% FIGURE  %%%%%%%%%%%%%%%%%%%% 
%\begin{figure}[t]
%\sidecaption[t]
%\includegraphics[width=11.5cm]{orbit_mic.pdf}%{Orbit1.pdf}
%\caption{Schematic description of a Keplerian orbit.}
%\label{fig:precession1}
%\end{figure}
%%%%%%%%%%%%%%%%%%%%%%%%%%%%%%% 
%%%%%%%%%%% FIGURE  %%%%%%%%%%%%%%%%%%%% 
\begin{figure}[t]
\sidecaption[t]
\includegraphics[width=7.5cm]{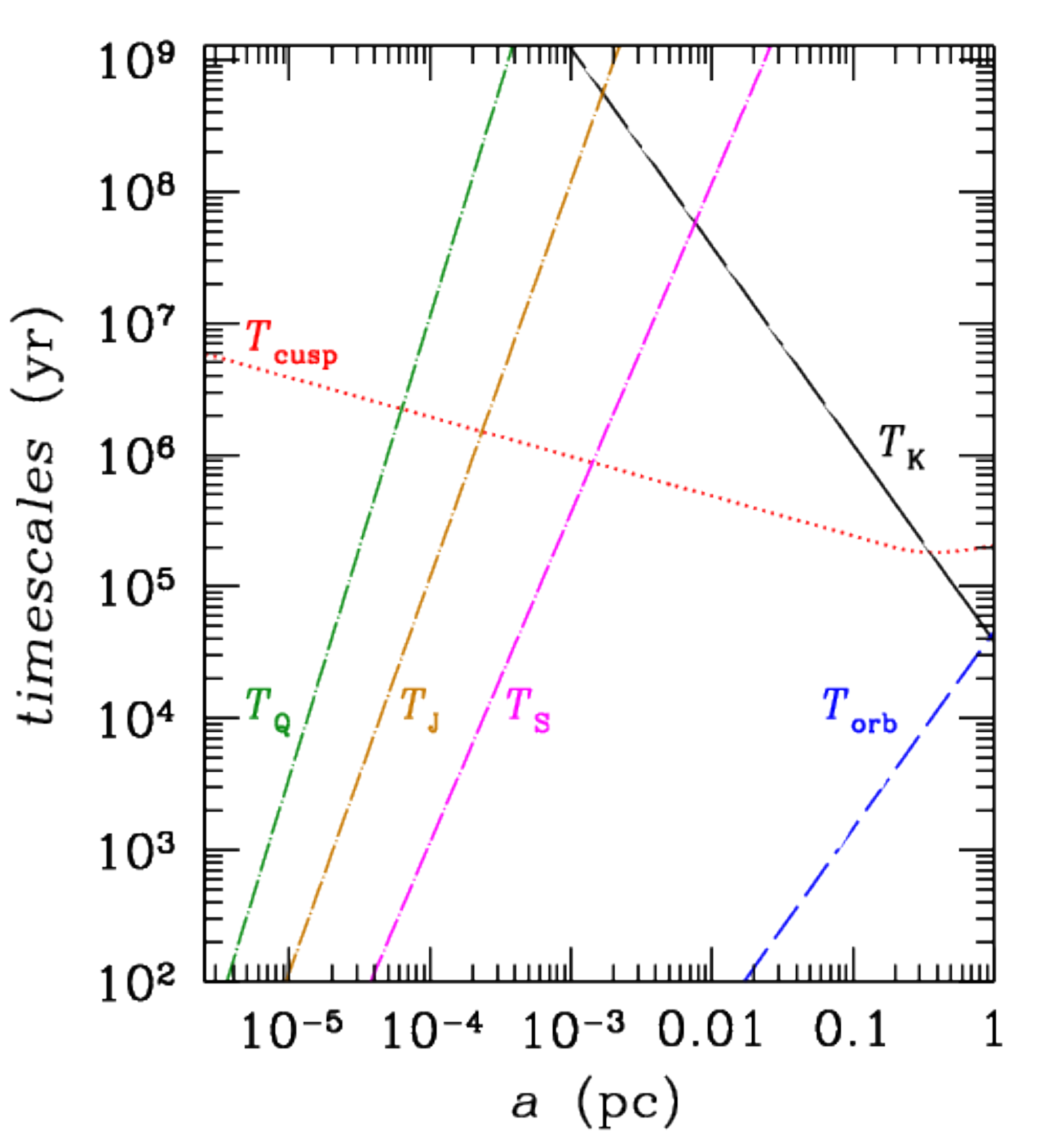}
\caption{Comparison of the relevant timescales as a function of the
  semi-major axis $a$ (see Sect.~\ref{subsec:precession} and
  Sect.~\ref{subsec:relativity} for details).  Solid black line:
  $T_{\rm K}$ (equation~\ref{eq:TK}, for $m_{\rm
    BH}=4.3\times{}10^6\msun$, $M_{\rm DISC}=10^5\msun$ and $R_{\rm
    DISC}=2$ pc); dotted red line: $T_{\rm cusp}$
  (equation~\ref{eq:Tcusp}, for $m_{\rm BH}=4.3\times{}10^6\msun$,
  $e=0$ and $M_{\rm cusp}$ derived from equation~\ref{eq:schoedelc});
  dashed blue line: $T_{\rm orb}$ (orbital period); dot-dashed green
  line: $T_{\rm Q}$ (equation~\ref{eq:TQ}); dot-dashed ochre line:
  $T_{\rm J}$ (equation~\ref{eq:TJ}); dot-dashed magenta line: $T_{\rm
    S}$ (equation~\ref{eq:TS}). $T_{\rm S}$, $T_{\rm J}$ and $T_{\rm
    Q}$ have been derived for $m_{\rm BH}=4.3\times{}10^6\msun$ and $e=0$. $\chi{}=1$ (i.e. a maximally rotating SMBH) has been assumed for $T_{\rm J}$ and $T_{\rm Q}$.}
\label{fig:precession2}
\end{figure}
%%%%%%%%%%%%%%%%%%%%%%%%%%%%%%% 

The orbit of a star in a Keplerian
potential\index{Potential!Keplerian} dominated by the SMBH mass is an
ellipse described by semi-major axis $a$ and eccentricity $e$.  The
orientation of the orbital plane in space is defined by two angles:
the inclination $i$ with respect to an (arbitrarily chosen) reference
plane and the longitude of the ascending node $\Omega{}$, with respect
to the same plane and to an arbitrarily selected direction in this
plane, called direction of the $\gamma$ point. 
%(Fig.~\ref{fig:precession1}).  
The argument of pericentre (angle
$\omega{}$) describes the orientation of the orbit within its
plane. Finally, the true anomaly $\psi{}$ describes the actual
position of the star on the orbit. Precession may affect both
$\Omega{}$ and $\omega{}$ or just one of them, depending on the nature
of the external potential\index{Potential!external}.

{\bf A spherical stellar cusp} induces a precession only on the
argument of pericentre ($\omega{}$), because the potential is
spherical and all non-spherical effects cancel out. The main effect of
this precession is pericentre advance (e.g. \citealt{Subrhaas12}).

The orbits of disc stars precess on a timescale (\citealt{Ivanov05};
\citealt{Lockmann08}; \citealt{Lockmann09}; \citealt{Gualandris12})
\begin{equation}\label{eq:Tcusp}
%T_{\rm cusp}=73\,{}T_{\rm orb}\,{}\left(\frac{a}{0.1\textrm{ pc}}\right)\,{}(1-e^2),
T_{\rm cusp}=\frac{m_{\rm BH}}{M_{\rm cusp}}\,{}T_{\rm orb}\,{}f(e),
\end{equation}
%where $T_{\rm orb}$, $a$ and $e$ are the orbital period, the semi-major axis and the eccentricity of a disc star, respectively. 
where $m_{\rm BH}$ is the mass of the SMBH, $T_{\rm orb}$ is the
orbital period of a disc star, $M_{\rm cusp}$ is the mass of the cusp
inside the stellar orbit, and
$f(e)=\frac{1+\sqrt{1-e^2}}{\sqrt{1-e^2}}$ is a function of the
eccentricity $e$ of a disc star.  Precession due to a spherical mass
distribution, also called {\it mass precession}\index{Precession!mass}, is retrograde,
i.e. in the opposite sense to orbital motion.
In the limit of $e \rightarrow 1$, mass precession becomes unimportant.

{\bf In an axisymmetric potential}, corresponding to a disc (e.g., a
stellar ring or a gas disc or the circumnuclear ring), a star orbiting
a SMBH of mass $m_{\rm BH}$ with semi-major axis $a$, at an
inclination $i$ relative to a disc of radius $R_{\rm DISC}$ and mass
$M_{\rm DISC}$ precesses on a timescale (\citealt{Nayakshin05b};
\citealt{Lockmann08}; \citealt{Subr09}):
\begin{equation}\label{eq:TK}
T_{\rm K}\equiv{}\frac{m_{\rm BH}}{M_{\rm DISC}}\,{}\frac{R_{\rm DISC}^3}{a^{3/2}\,{}\sqrt{G\,{}m_{\rm BH}}}.
%T_{\rm K}\equiv{}\frac{M_{\rm SMBH}}{M_{\rm DISC}}\sqrt{\frac{r^3}{G\,{}M_{\rm SMBH}}}\,{}\frac{(r^2+R_{\rm DISC}^2)^{5/2}}{r^3\,{}R_{\rm DISC}^2}
\end{equation}
In this case, the equations of motion for mean orbital elements read (\citealt{Subr09}, see also \citealt{Kozai62}, \citealt{Lidov62})
\begin{eqnarray}
T_{\rm K}\,{}\sqrt{1-e^2}\,{}\frac{{\rm d}e}{{\rm d}t}=\frac{15}{8}\,{}e\,{}(1-e^2)\,{}\sin{2\,{}\omega{}}\,{}\sin^2{i},\\
T_{\rm K}\,{}\sqrt{1-e^2}\,{}\frac{{\rm d}i}{{\rm d}t}=-\frac{15}{8}\,{}e^2\,{}\sin{2\,{}\omega{}}\,{}\sin{i}\,{}\cos{i},\\
T_{\rm K}\,{}\sqrt{1-e^2}\,{}\frac{{\rm d}\omega{}}{{\rm d}t}=\frac{3}{4}\,{}\left\{2\,{}-\,{}2\,{}e^2\,{}+\,{}5\,{}\sin^2{\omega{}}\left[e^2-\sin^2{i}\right]\right\},\label{eq:pericentre}\\
T_{\rm K}\,{}\sqrt{1-e^2}\,{}\frac{{\rm d}\Omega{}}{{\rm d}t}=-\frac{3}{4}\,{}\cos{i}\,{}\left[1\,{}+\,{}4\,{}e^2\,{}-\,{}5\,{}e^2\,{}\cos^2{\omega{}}\right].
\end{eqnarray}
In these equations, we chose the plane of the disc that perturbs the stellar orbits  as a reference plane. Energy conservation implies that $a$ is approximately constant.

 If $i=0$ (i.e. the star is coplanar with the disc generating the
 external potential), then only the longitude of the node $\Omega{}$
 and the argument of pericentre $\omega{}$ are affected, as all
 the terms $\propto{}\sin{i}$ cancel out. Furthermore, any precession of
 $\Omega{}$ does not affect the other properties of the orbit, as the
 plane of the stellar orbit and the plane of the perturbing disc coincide.

 If $0<i<90^\circ{}$, then all four quantities ($e$, $i$, $\omega{}$
 and $\Omega{}$) are expected to change. Finally, if $i=90^\circ{}$,
 only $e$ and $\omega{}$ are expected to change, as an effect of the
 perturbation.

It can be shown that the change of both eccentricity and inclination
with time is periodic, describing the so called `Kozai cycles\index{Kozai|textbf}'
(\citealt{Kozai62}). The change in eccentricity during each `Kozai
cycle' is particularly large if the initial inclination $i$ is high.

If the axisymmetric potential\index{Potential!axisymmetric} is not the
only potential that perturbs the stars, but it combines with a
spherical cusp, then things change significantly. The spherical cusp
enhances the change in the argument of pericentre. In presence of the
spherical cusp, equation~\ref{eq:pericentre} can be rewritten as
(\citealt{Ivanov05})
\begin{equation}\label{eq:pericentre2}
T_{\rm K}\,{}\sqrt{1-e^2}\,{}\frac{{\rm d}\omega{}}{{\rm d}t}=\frac{3}{4}\,{}\left\{2\,{}-\,{}2\,{}e^2\,{}+\,{}5\,{}\sin^2{\omega{}}\left[e^2-\sin^2{i}\right]\right\}\,{}(1-\kappa{})^{-1}.
\end{equation}
The term $\kappa{}$ is due to the spherical cusp, and can be expressed as (\citealt{Ivanov05})
\begin{equation}
\kappa{}=\tilde{\kappa{}}\,{}\left(\frac{T_{\rm K}}{T_{\rm orb}}\right)\,\left(\frac{M_{\rm cusp}(<a)}{m_{\rm BH}}\right),
\end{equation}
where $\tilde{\kappa}$ is a numeric constant (whose value depends on
the shape of the spherical cusp) and $M_{\rm cusp}(<a)$ is the total
mass of stars inside a sphere of radius $a$ (i.e. equal to the
semi-major axis of the orbit of the considered star).

It can be shown (\citealt{Ivanov05}) that if $\kappa{}$ is above a
certain threshold (i.e. if the spherical cusp is particularly massive
with respect to the other involved quantities), then Kozai
oscillations are dramatically damped. Thus, the value of the
eccentricity remains very close to the initial value.

If $M_{\rm cusp}>0.1\,{}M_{\rm DISC}$, the only remaining effect of
the gravitational influence of the perturbing disc on the stellar
orbits is the precession of the ascending node with frequency
(\citealt{Subr09})
\begin{equation}\label{eq:subr}
\frac{{\rm d}\Omega{}}{{\rm d}t}=-\frac{3}{4}\,{}\cos{i}\,{}\frac{1\,{}+\,{}\frac{3}{2}\,{}e^2}{\sqrt{\-e^2}}\,{}T_{\rm K}^{-1}.
\end{equation}

From equation~\ref{eq:subr}, it is apparent that the precession of the
ascending node depends on the semi-major axis of the stellar orbit
($\frac{{\rm d}\Omega{}}{{\rm d}t}\propto{}T_{\rm
  K}^{-1}\propto{}a^{3/2}$). In particular, stars with larger $a$ will
precess faster than stars with smaller $a$. This is very important for
the early-type stars\index{Early-type stars} that form the CW disc\index{Clockwise disc} around
Sgr~A$^\ast{}$, for the following reason. If $i=0$, this precession
has no effect on the inclination of the stellar orbits, as the plane
of the perturbing disc and the plane of the star orbit are the same. Instead, if
$i>0$, the orbits of the outer stars will become inclined with respect
to the orbits of the inner stars, producing a warp in the stellar
disc, and increasing its thickness.

Fig.~\ref{fig:precession2} shows a comparison of the relevant
precession timescales in the case of our GC.

\subsection{Relativistic effects}
\label{subsec:relativity}
According to general relativity, the SMBH itself is a source of
precession\index{Precession} for the stellar orbits. 
In the case of a non-rotating black hole, a star orbiting the SMBH
experiences an advance of the orbital periapse by an angle
\begin{equation}
\delta \omega_S = \frac{6 \pi}{c^2} \frac{G m_{\rm BH}}{a (1-e^2)}\,,
\end{equation}
which depends on the mass of the SMBH and the orbital elements of the star.
This apsidal precession
(also called geodetic, de Sitter, relativistic or Schwarzschild precession\index{Precession!relativistic})
is an in-plane prograde precession that operates on a time-scale \citep[see][]{GM2009, Merritt13}:
\begin{eqnarray}\label{eq:TS}
T_S & = & \frac{\pi T_{\rm orb}}{\delta \omega_S} = \frac{T_{\rm orb}\,c^2}{6} \frac{a (1-e^2)}{Gm_{\rm BH}} \nonumber \\
 %= \frac{\pi{}\,{}c^2\,{}(1-e^2)a^{5/2}}{3\,{}(G\,{}m_{\rm BH})^{3/2}} \\
 & = & 1.3\times10^3\,{}{\rm yr} \left(1-e^2\right) \left(\frac{a}{\rm mpc}\right)^{5/2} \left(\frac{4\times{}10^6\,{}{\rm M}_\odot}{m_{\rm BH}}\right)^{3/2}
\end{eqnarray}
%for $m_{\rm BH} = 4.0\times10^6\msun$.
In the case of our GC, Schwarzschild precession is large enough to
potentially be detectable via $\sim 10$ years’ monitoring of
identified stars at $\lesssim 10$ mpc separations from the SMBH
\citep{RubilarEckart2001}, see also Fig.~\ref{fig:precession2}.

In the case of a rotating black hole, the coupling between the spin of the SMBH\index{Supermassive black hole!spin} 
and the orbital angular momentum\index{Angular momentum} of the stars leads to additional sources of precession, both
in-plane and out-of plane.
The spin and quadrupole moment contributions to the in-plane precession are, respectively, \citep{MAMW2010}:
\begin{eqnarray}
%\delta \omega_J & = & -3\,{} \frac{4\pi}{c^3} \chi \left[\frac{G\,m_{\rm BH}}{a\,(1-e^2)}\right]^{3/2} \cos{i}\\
%\delta \omega_Q & = & \frac{1}{2} \frac{3\pi}{c^4} \chi^2 \left[\frac{G\,m_{\rm BH}}{a(1-e^2)}\right]^2 (1 - 5 {\rm cos^2} i)\,
\delta \omega_J & = & -\frac{8\pi}{c^3} \chi \left[\frac{G\,m_{\rm BH}}{a\,(1-e^2)}\right]^{3/2} \cos{i}\\
\delta \omega_Q & = & -\frac{3}{2} \frac{\pi}{c^4} \chi^2 \left[\frac{G\,m_{\rm BH}}{a(1-e^2)}\right]^2 (1 - 3 {\rm cos^2} i)\,
\end{eqnarray}
%where $\mathbf{\chi} = \mathbf{J} / (GM^2_{\rm BH}/c^2$ is the dimensionless spin angular momentum vector of the SMBH.
where $\chi = J / (Gm^2_{\rm BH}/c^2)$ is the dimensionless spin parameter of the SMBH.
The contributions to the precession of the orbital plane are \citep{MAMW2010}:
\begin{eqnarray}
\delta \Omega_J & = & \frac{4\pi}{c^3} \chi \left[\frac{G\,m_{\rm BH}}{a\,(1-e^2)}\right]^{3/2}\\
\delta \Omega_Q & = & - \frac{3\pi}{c^4} \chi^2 \left[\frac{G\,m_{\rm BH}}{a\,(1-e^2)}\right]^2 \cos{i}.
%\delta \Omega_J & = & \frac{4\pi}{c^3} \chi \left[\frac{G\,m_{\rm BH}}{a\,(1-e^2)}\right]^{3/2}\\
%\delta \Omega_Q & = & \frac{3\pi}{c^4} \chi^2 \left[\frac{G\,m_{\rm BH}}{a\,(1-e^2)}\right]^2 \cos{i}.
\end{eqnarray}
Of these terms, only the quadrupole term is dependent on inclination.
The associated timescales are:
\begin{eqnarray}\label{eq:TJ}
%T_J & = & \frac{T_{\rm orb}}{4\chi} \left[\frac{c^2\,a\,(1-e^2)}{Gm_{\rm BH}}\right]^{3/2} \nonumber \\
% & = & 1.4\times10^5\,{}{\rm yr}\left(1-e^2\right)^{3/2} \chi^{-1} \left(\frac{a}{\rm mpc}\right)^3\left(\frac{4\times{}10^6\,{}{\rm M}_\odot}{m_{\rm BH}}\right)^{2}
T_J & = & \frac{T_{\rm orb}}{4\chi} \left[\frac{c^2\,a\,(1-e^2)}{G\,m_{\rm BH}}\right]^{3/2} \nonumber \\
 & = & 1.4\times10^5\,{}{\rm yr}\left(1-e^2\right)^{3/2} \chi^{-1} \left(\frac{a}{\rm mpc}\right)^3\left(\frac{4\times{}10^6\,{}{\rm M}_\odot}{m_{\rm BH}}\right)^{2}
\end{eqnarray}
and
\begin{eqnarray}\label{eq:TQ}
%T_Q & = & \frac{T_{\rm orb}}{3\chi^3} \left[\frac{c^2\,a\,(1-e^2)}{Gm_{\rm BH}}\right]^{2} \nonumber \\
% & =  & 1.3\times10^7\,{}{\rm yr}\left(1-e^2\right)^{2} \chi^{-2} \left(\frac{a}{\rm mpc}\right)^{7/2}\left(\frac{4\times{}10^6\,{}{\rm M}_\odot}{m_{\rm BH}}\right)^{5/2}.
%T_Q & = & \frac{T_{\rm orb}}{3\chi^2} \left[\frac{c^2\,a\,(1-e^2)}{G\,m_{\rm BH}}\right]^{2}\,{}\cos^{-1}{i}\nonumber \\
 %& =  & 1.3\times10^7\,{}{\rm yr}\left(1-e^2\right)^{2} \chi^{-2} \left(\frac{a}{\rm mpc}\right)^{7/2}\left(\frac{4\times{}10^6\,{}{\rm M}_\odot}{m_{\rm BH}}\right)^{5/2}\left(\frac{0.25}{\cos{i}}\right).
T_Q & = & \frac{T_{\rm orb}}{3\chi^2} \left[\frac{c^2\,a\,(1-e^2)}{G\,m_{\rm BH}}\right]^{2}\nonumber \\
 & =  & 1.3\times10^7\,{}{\rm yr}\left(1-e^2\right)^{2} \chi^{-2} \left(\frac{a}{\rm mpc}\right)^{7/2}\left(\frac{4\times{}10^6\,{}{\rm M}_\odot}{m_{\rm BH}}\right)^{5/2}.
\end{eqnarray}
Detection of spin effects in the GC can in principle come from
observations of plane precession of stars in the inner mpc. However,
gravitational interactions\index{Interactions!gravitational} between
stars in this region are likely to induce orbital precession of the
same approximate amplitude as the precession due to frame dragging,
hampering detection.  Assuming near-maximal spin for the Milky Way
SMBH, detection of frame-dragging precession may be feasible after a
few years’ monitoring with an instrument like GRAVITY
\citep{Eisenhauer09} for orbits in the radial range $0.2-1$ mpc. At
smaller radii the number of stars is too small, while at larger radii
the star-star and star-remnant perturbations dominate over
relativistic effects \citep{MAMW2010}.

\subsection{Relaxation processes: two-body relaxation, resonant relaxation}
\label{subsec:rlx}
%%%%%%%%%%%%%%%%%%%%%%%%%%%%%%%%FIGURE  %%%%%%%%%%%%%%%%%%%%%%%%%%%%%%%%%%%%%%%%%
\begin{figure}[t]
\sidecaption[t]
\includegraphics[width=7.5cm]{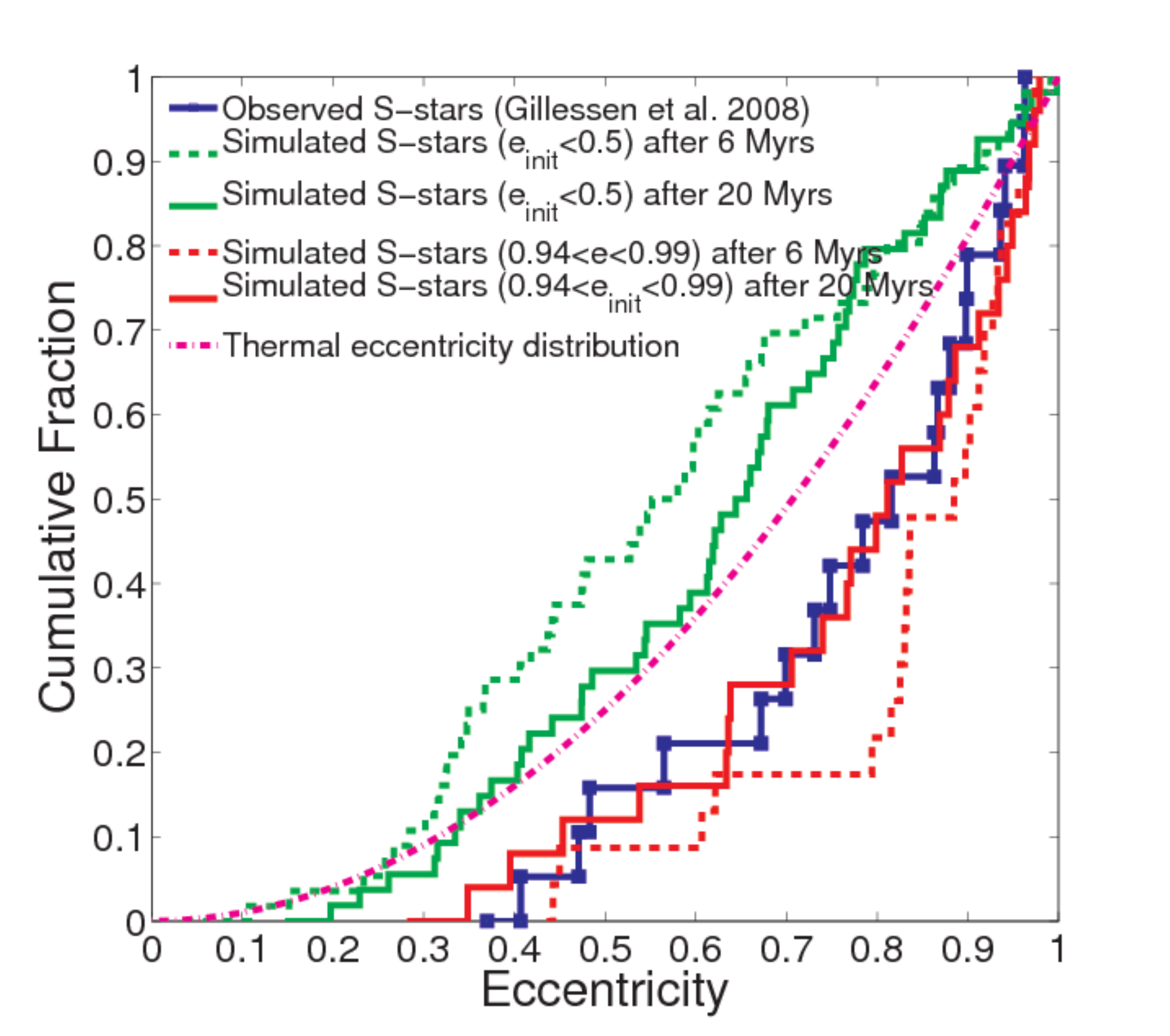}
\caption{Cumulative distribution of eccentricities for stars with
  initially low (green) and high (red) eccentricities, after 6 Myr
  (dashed) and 20 Myr (solid) of evolution in a relaxed cusp of stars
  and remnants. The distribution for the S-stars from the sample of
  \citet{Gillessen09a} is shown for comparison, as is a line giving
  the theoretical thermal distribution. The distribution predicted by
  the binary breakup model is the most consistent with the
  data. Fig.~2 of \citet{Perets2009}.}
\label{fig:perets2009fig2}
\end{figure}
%%%%%%%%%%%%%%%%%%%%%%%%%%%%%%%%FIGURE %%%%%%%%%%%%%%%%%%%%%%%%%%%%%%%%%%%%%%%%%
In an isotropic system, the angular momentum\index{Angular momentum} of the stars evolves both
due to the stochastic two-body relaxation \citep[e.g.][]{BT87}
and to the resonant relaxation \citep{RT1996}.  Nonresonant
two-body relaxation\index{Two-body relaxation|textbf} operates on a
timescale\index{Two-body relaxation!timescale} \citep{BT87}
\begin{eqnarray}
\label{eq:tnr}
T_{\rm NR} &=& \frac{0.34 \, \sigma^3}{G^2\,m\,\rho\,{\rm ln} \Lambda} \nonumber\\
 & \approx & 10^{10} \,{\rm yr} \left(\frac{\sigma}{200\kms}\right)^3 \left(\frac{10^6 \msun \pcc}{\rho}\right) \left(\frac{\msun}{m}\right)
\left(\frac{15}{{\rm ln} \Lambda}\right)\,
\end{eqnarray}
where $\rho$ is the stellar density, $\sigma$ is the one-dimensional
velocity dispersion of the stars, $m$ is the mass of a single star,
and ln$\Lambda$, the Coulomb logarithm, is a numerical factor that
corrects for the divergent force in a infinite homogeneous system.
Over a time $T_{\rm NR}$, gravitational encounters between stars act
to change orbital energies and angular momenta\index{Angular momentum}. In particular, angular
momentum\index{Angular momentum} changes with time in a random walk fashion.

Resonant relaxation\index{Resonant relaxation|textbf} occurs when the
symmetries of the potential\index{Potential} act to constrain the
stellar orbits (e.g. closed ellipses in a Kepler potential, or planar
rosettes in a spherical one). As long as the symmetry is approximately
maintained, gravitational
interactions\index{Interactions!gravitational} between stars are
highly correlated and stars experience coherent torques.  The
coherence timescale\index{Coherence timescale} $T_{\rm coh}$ (the time
over which orbits can be considered fixed), is the time associated
with the most rapid source of precession of the stellar
orbits. Sources of precession\index{Precession} (see also section
\ref{subsec:precession}) are: mass precession\index{Precession!mass},
due to the stellar mass distributed around the SMBH, relativistic
precession\index{Precession!relativistic} and precession due to
resonant relaxation itself.  The mass coherence time is always shorter
than the self-coherence time, but sufficiently close to the SMBH
relativistic precession must dominate.

For a time $\Delta\,t$ such that $T_{\rm
  orb} \ll \Delta\,t \ll T_{\rm coh}$, the so called {\it coherent
  resonant relaxation} is characterised by changes in the angular
momentum\index{Angular momentum} of a star at a roughly constant rate 
\begin{equation}
\frac{dJ}{dt} \sim \sqrt{N} \frac{G\,m}{a}, 
\end{equation}
where $a$ is the star's semi-major axis.
The angular momentum\index{Angular momentum} change is
\begin{equation}
(\Delta J/J_{c})_{\rm coh} \sim \sqrt{N} \frac{G\,m}{a} \frac{\Delta\,t}{\sqrt{Gm_{\rm BH} a}},
\end{equation}
where $J_c  = \sqrt{G\,m_{\rm BH} a}$ is the angular momentum\index{Angular momentum} of a circular orbit.
The coherent resonant relaxation timescale can be defined as the time for which $\Delta J = J_c$
\begin{eqnarray}
T_{\rm RR, coh} &=& \frac{T_{\rm orb}}{2\pi} \frac{m_{\rm BH}}{m} \frac{1}{\sqrt{N}} \nonumber\\
& \sim & 1.5\times10^4 {\rm yr} \left(\frac{a}{{\rm mpc}}\right)^{3/2} \left(\frac{10^6\msun}{m_{\rm BH}}\right)^{1/2} \left(\frac{10^{-6}}{m/m_{\rm BH}}\right) \left(\frac{10^3}{N}\right)^{1/2}.
\end{eqnarray}

On timescales longer than $T_{\rm coh}$, the field stars precess and
the direction of the torque exerted by the $N$ stars changes (while
its magnitude remains roughly unchanged).  Under the assumption that
the direction of the torque is essentially randomised after each
$T_{\rm coh}$, the angular momentum\index{Angular momentum} of a test star undergoes a random
walk, with step size given by the product of the torque and the
coherence time. The evolution of the angular momentum\index{Angular momentum} in this {\it
  incoherent resonant relaxation} regime is qualitatively similar to
the evolution under nonresonant two body relaxation, but can be
significantly faster. This is due to the fact that the mean free path
of the random walk in $\mathbf J$ is set by the (large) change
accumulated over $T_{\rm coh}$.  The incoherent resonant relaxation
timescale\index{Resonant relaxation!timescale} is then defined by
\citep[e.g.][]{Eilon2009}
\begin{eqnarray}
\Delta J/J_{c}  &=& \left(\Delta J/J_{c}\right)_{\rm coh} \sqrt{t/T_{\rm coh}} \equiv \sqrt{t/T_{\rm RR}},\\
T_{\rm RR} &=& \left(\frac{J_c}{\Delta J}\right)^2_{\rm coh} T_{\rm coh}.
\end{eqnarray}
If the coherence time is determined by mass precession\index{Precession!mass}, then
\begin{equation}
T_{\rm RR} \approx \left(\frac{m_{\rm BH}}{m}\right)\,T_{\rm orb}.
\end{equation}
If instead relativistic precession\index{Precession!relativistic} dominates,
\begin{equation}
T_{\rm RR} \approx \frac{3}{\pi^2} \frac{r_g}{a} \left(\frac{m_{\rm BH}}{m}\right)^2 \frac{T_{\rm orb}}{N},
\end{equation}
where 
\begin{equation}
r_g = \frac{G \,m_{\rm BH}}{c^2} \approx 2\times10^{-7} \left(\frac{m_{\rm BH}}{4.3\times10^6\msun}\right) {\rm pc}
\end{equation}
is the gravitational radius of the SMBH.

\citet{Merrittbook} estimates the distance from the SMBH at which
incoherent resonant relaxation becomes dominant over nonresonant two
body relaxation. In the case of a dynamically relaxed Bahcall-Wolf
cusp \citep{BW76} this distance is of about 0.06 pc, or $0.025 r_h$,
where $r_h$ represents the SMBH's influence radius.  In the case of a
low-density model for the innermost region of the NSC, 
resonant relaxation dominates inside $\sim 0.18 {\rm pc} \sim
0.1 r_h$, somewhat further out than in the relaxed model.

Simulations by \citet{Perets2009} show that perturbations from the
compact remnants tend to randomise stellar orbits in the GC, partially
erasing the dynamical signatures of their origin.  The simulations
follow the dynamical evolution of a population of stars in the inner
$\sim 0.3$\, pc of the Galaxy against a cusp of stars and
remnants. The initial conditions are based on the collisionally
relaxed cusp of stars and remnants by \citet{HA2006}, and intend to
represent products of both the {\it in situ} formation scenario and
the tidal breakup scenario. The former tends to produce stars with low
to moderate eccentricities, while the latter leaves stars bound to the
SMBH on highly eccentric orbits.  The eccentricities of the initially
highly eccentric stars evolve, in 20 Myr, to a distribution that is
consistent with the observed eccentricity distribution. In contrast,
the eccentricities of the initially more circular orbits fail to
evolve to the observed values in 20 Myr, arguing against the disc
migration scenario. Fig.~\ref{fig:perets2009fig2} shows the final
cumulative eccentricity distribution of the stars for the two models
under consideration and at two different times: 6 Myrs and 20 Myrs.
These times are chosen to represent the age of the current CW disc\index{Clockwise disc} and
the canonical S-star lifespan.  The binary breakup scenario after 20
Myr of evolution is found to be the preferred model for the origin of
the S-stars\index{S-stars}.  In contrast, the disc migration scenarios
seem to be excluded (for the given assumptions), since they have major
difficulties in explaining the large fraction of eccentric orbits
observed for the S-stars in the GC.

Resonant relaxation\index{Resonant relaxation} against the stellar
remnants acts to isotropise the inclination distribution of the
captured stars for all models, and can not therefore be used to
discriminate between them. However, randomisation of the inclinations
requires at least 4 Myr when starting from a single plane
configuration, and can be used to constrain the lifetime of the
S-stars in the {\it in situ} model.

Monte Carlo simulations by \citet{AM2013} of the orbital evolution of
the S-stars show that the distribution of the semi-major axis $a$ and
eccentricity $e$ of the S-stars predicted by the binary disruption\index{Tidal disruption} model is consistent with the observed orbits even when relativistic
effects are considered (see section \ref{subsec:relativity} for details).
Even though most of the orbits lie initially below the Schwarzschild
Barrier (i.e. the locus in the $(a,\,{}e)$ plane where resonant relaxation is
ineffective at changing eccentricities, \citealt{MAMW2010}), orbits
starting sufficiently close to the barrier are sometimes able to
penetrate it, diffusing above and reaching a nearly thermal
eccentricity distribution. After $\sim20$ Myr of evolution the
distributions are consistent with the observed ones, if a dynamically
relaxed model for the background stellar cusp is assumed.  This result
is particularly interesting given that relaxed models of the GC are
currently disfavored by observations \citep{Buchholz2009, Do2009} and
by some theoretical arguments \citep{Merritt2010,Antonini2011,
  GM2012}.

\subsection{Impact of relaxation and precession on the early-type stars\index{Early-type stars}}
\label{subsec:rlximpact}
Precession\index{Precession} of the stellar orbits, due to either stellar perturbations or relativistic effects,
has a number of implications for the evolution of the early-type stars\index{Early-type stars} in the GC.
We here discuss the most relevant to constrain the formation scenarios presented in the previous sections:

%%%%%%%%%% FIGURE  %%%%%%%%%%%%%%%%%%%% 
\begin{figure}[t]
\sidecaption[t]
\includegraphics[width=7.5cm]{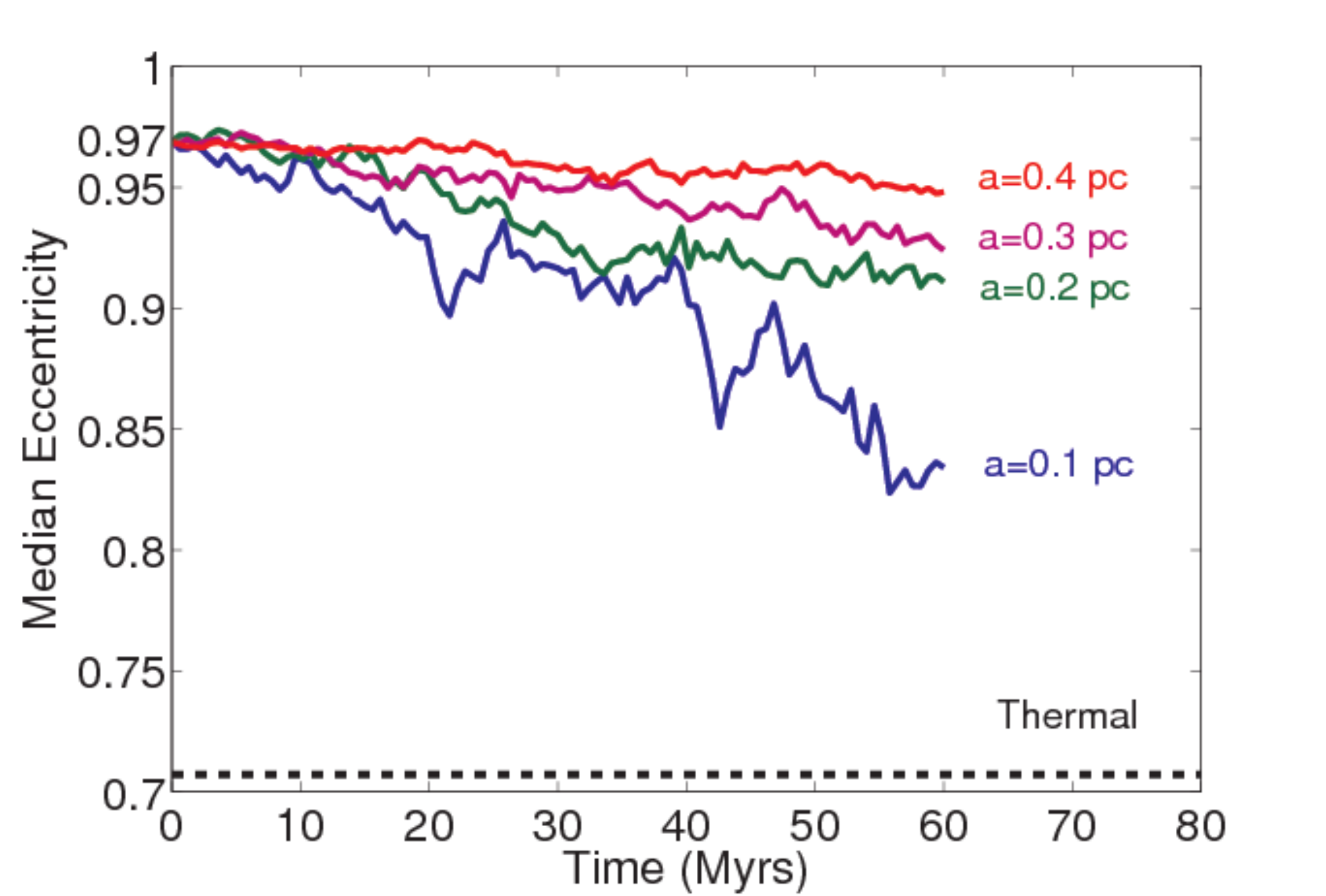}
\caption{Evolution of the median eccentricity of a population of
  B-stars in the central 0.5 pc of the GC, for different initial
  distances from the SMBH. Relaxation is driven by a cusp of remnants
  distributed between 0.04 and 0.8 pc. Only stars with initial
  eccentricity in the range $0.95-0.99$ are selected to represent
  captured stars. Fig.2 of \citet{PeretsGualandris2010}.}
\label{fig:PG2010fig2}
\end{figure}
%%%%%%%%%%%%%%%%%%%%%%%%%%%%%%% 
{\it Dependence of mean orbital eccentricity on distance}.  The
  resonant relaxation timescale increases with distance from the SMBH
  \citep[e.g.][]{HA2006}. Therefore, stars captured/formed further
  away from the SMBH are expected to have a less relaxed eccentricity
  distribution than stars closer to the black hole
  \citep{PeretsGualandris2010}.  As a result, relaxation processes
  will give rise, over time, to a correlation between the distance
  from the SMBH and the orbital eccentricity.  Far from the SMBH,
  where the resonant relaxation timescales are much longer than the
  typical lifetimes of the B-stars, stars should retain their original
  eccentricity distribution, i.e., highly eccentric orbits for
  captured stars after a binary disruption\index{Tidal disruption}, and likely low
  eccentricity orbits for stars formed in a stellar disc.  Closer to
  the SMBH, on the other hand, captured stars could have a relaxed
  (i.e. thermal) eccentricity distribution even after short times.
  These predictions have been confirmed by the simulations of
  \citet{PeretsGualandris2010}, who find an increase of the mean
  eccentricity of the stars with distance from the SMBH (see
  Fig.~\ref{fig:PG2010fig2}).  Therefore, the binary capture scenario
  provides a qualitatively unique signature, in which the typical
  eccentricity is an increasing function of distance, which can be
  tested against observations of the B-stars.  However, stars with
  large semi major axes have large orbital periods and it is difficult
  to determine their dynamical accelerations from astrometric data,
  from which orbital parameters are derived. \citet{Madigan2014}
  developed a statistical method which uses only sky positions and
  proper motions to infer the orbital eccentricities of a stellar
  population around the SMBH. They confirm the results by
  \citet{PeretsGualandris2010} regarding the binary disruption\index{Tidal disruption} 
  scenario that stellar orbits remain at very high eccentricities
  outside $\sim 0.1$ pc.  Similarly, stars formed with small
  eccentricities, as in the case of an in-situ formation from a
  dissolved disc, maintain small eccentricities at large
  distances. Applying the statistics to a sample of B-stars at
  projected radii $\sim 0.004-1$ pc from the SMBH they find that stars
  with $K$-band magnitudes $14 \lesssim m_{\rm K} \lesssim 15$ (i.e
  masses of $15-20\msun$ and ages of $8-13$ Myr) match well to an
  in-situ formation origin, while those with $m_{\rm K} \ge 15$
  (corresponding to masses $\le 15\msun$ and ages $\ge 13$ Myr), if
  isotropically distributed, form a population that is more eccentric
  than thermal, suggestive of a binary-disruption\index{Tidal disruption} origin.

{\it Thickness/warping of the CW disc\index{Clockwise disc}}. As discussed in
  Sect.~\ref{subsec:2.2}, recent observations (\citealt{Bartko09};
  \citealt{Lu09}; \citealt{Bartko10}; \citealt{Do13}; \citealt{Lu13};
  \citealt{Yelda2014}) show that the opening angle of the CW disc\index{Clockwise disc} is
  only $\sim{}10^\circ{}-14^\circ{}$, but about half (or even
  $\sim{}80$ per cent, \citealt{Yelda2014}) of the early-type stars\index{Early-type stars} in
  the inner $1-10$ arcsec ($0.04-0.4$ pc) do not belong to the CW
  disc\index{Clockwise disc}. Furthermore, the probability of early-type stars\index{Early-type stars} being members
  of the CW disc\index{Clockwise disc} decreases with increasing projected distance from
  Sgr~A$^\ast{}$ (\citealt{Bartko09}; \citealt{Lu09}). Finally, the CW
  disc\index{Clockwise disc} does not seem a flat structure, but rather a significantly
  warped ($\sim{}64^\circ{}$, \citealt{Bartko09}) and tilted object
  (but see \citealt{Yelda2014} for a different result). 
Recent studies (e.g. \citealt{Subr09}; \citealt{Haas11a};
\citealt{Haas11b}; \citealt{Mapelli13}) suggest a reasonable
interpretation for such observations: the precession exerted by a
slightly misaligned gas disc (or ring) enhances the inclinations of
the outer stellar orbits with respect to the inner stellar
orbits. Thus, while the inner disc remains quite coherent, the outer
stellar orbits change angular momentum\index{Angular momentum} orientation till they may even
lose memory of their initial belonging to the same disc. The result is
a tilted/warped disc, which is being dismembered in its outer parts.

%%%%%%%%%% FIGURE  %%%%%%%%%%%%%%%%%%%% 
\begin{figure}[t]
\sidecaption[t]
\includegraphics[width=6.5cm]{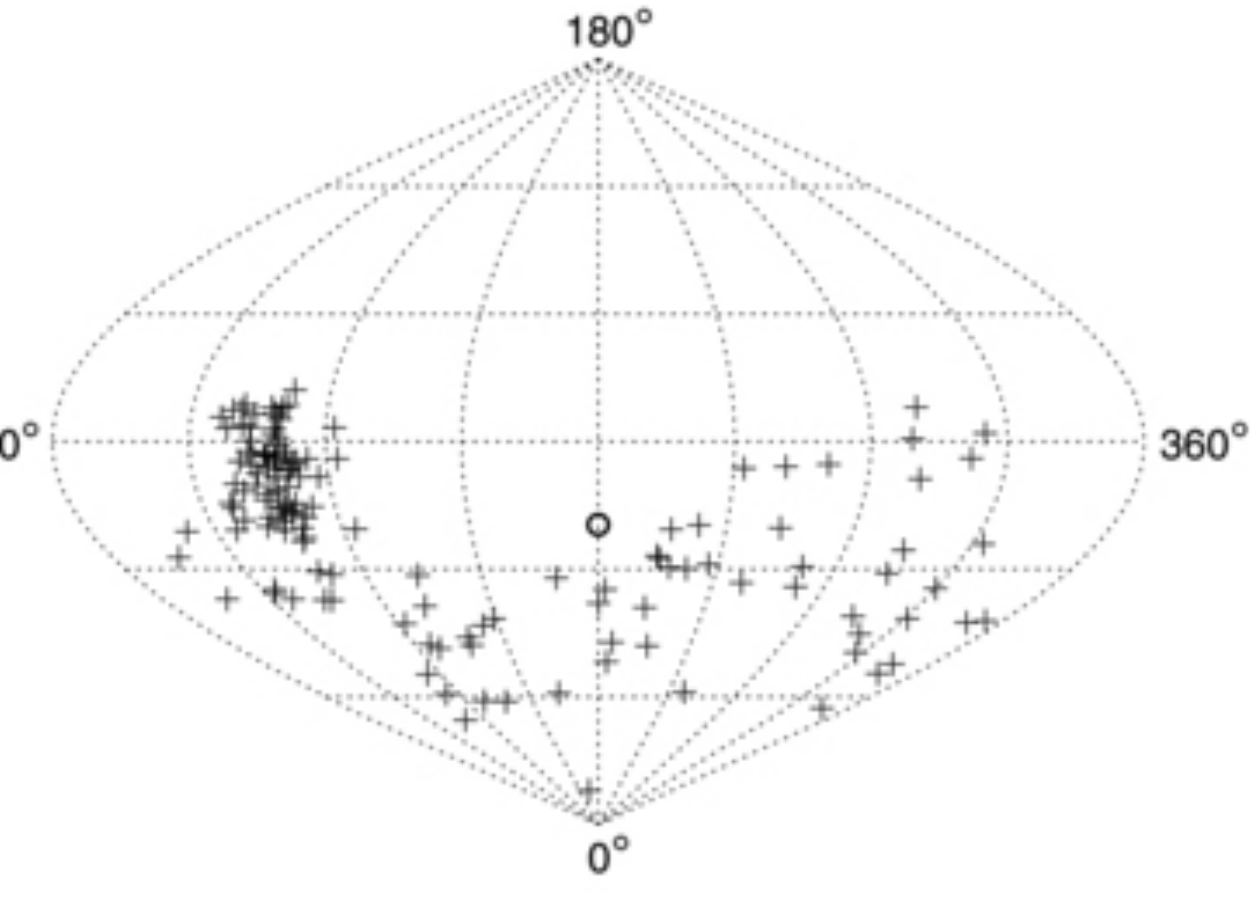}
\caption{Angular momenta\index{Angular momentum} of individual stars in the young stellar disc after 6 Myr of orbital evolution, integrated through $N$-body simulations\index{N-body simulations} (\citealt{Haas11a}).  The initial state is denoted by an empty circle. The plot is in sinusoidal projection. Latitude on the plots corresponds to $i$ while longitude is related to $\Omega{}$. Only stars with mass $m> 12\msun$ are displayed. From Fig.~2 of \cite{Haas11a}.} %For comparison, the bottom panel illustrates the situation with negligibly small mass of the stars in the disc (single mass, $m = 0.004$ M$_\odot$, the other parameters are the same as in the canonical model).}
\label{fig:haas}
\end{figure}
%%%%%%%%%%%%%%%%%%%%%%%%%%%%%%% 
The perturbing ring may be the CNR\index{Circumnuclear ring} (\citealt{Subr09};
\citealt{Haas11a}; \citealt{Haas11b}, see\footnote{The fiducial run
  reported in \cite{Haas11a} includes 200 early-type stars\index{Early-type stars} (modelled
  as $N$-body particles), a SMBH with mass $m_{\rm
    BH}=4\times{}10^6\msun$ (modelled as Keplerian
  potential\index{Potential!Keplerian}), a CNR\index{Circumnuclear ring} with mass
  $0.3\,{}m_{\rm BH}$ (modelled as a single particle), a stellar cusp
  with mass $M_{\rm cusp}=0.03\,{}m_{\rm BH}$ (modelled as a rigid
  potential).} Fig.~\ref{fig:haas}) or a transient gas ring that forms
from the disruption\index{Tidal disruption} of a low-angular momentum\index{Angular momentum} molecular cloud\index{Molecular cloud}
(\citealt{Mapelli13}).  %In particular, 

\begin{figure}[t]
\sidecaption[t]
\includegraphics[height=8.5cm]{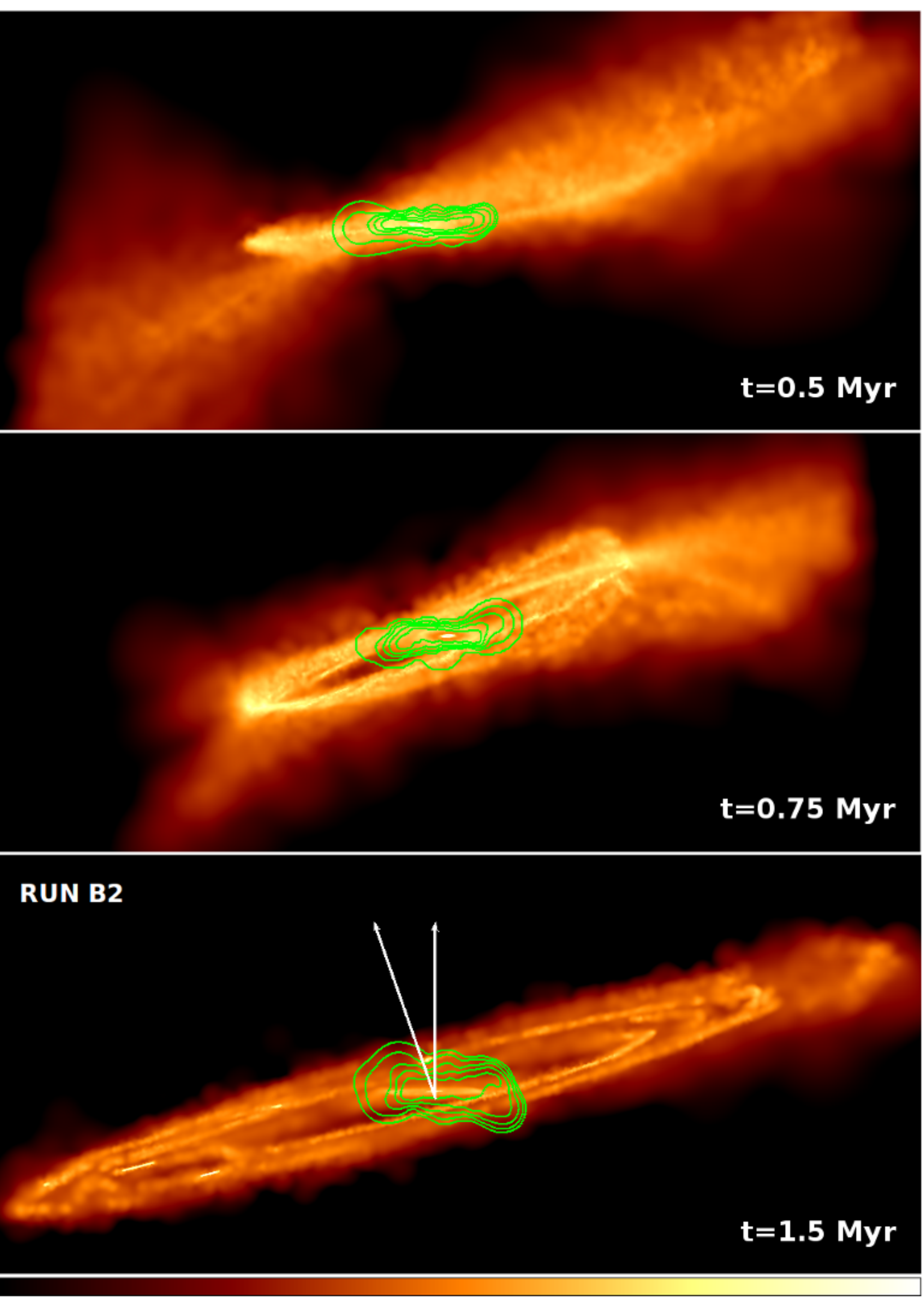}
\includegraphics[height=8.5cm]{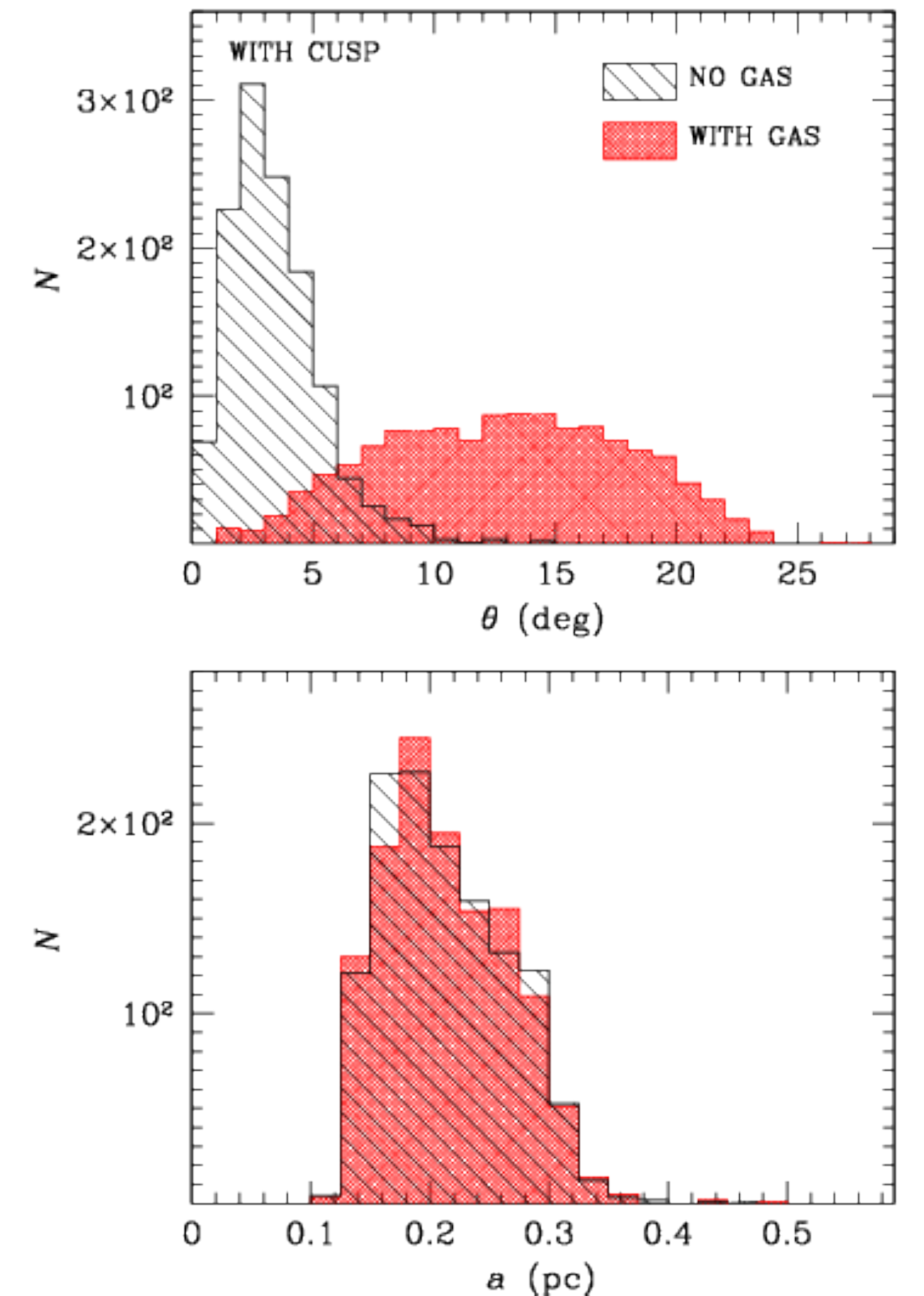}
\caption{Left-hand panel: Projected density of gas in run B2 of
  \cite{Mapelli13} at $t = 0.5, 0.75$ and 1.5 Myr in the top, central
  and bottom panel, respectively.  In this simulation a pre-existing
  stellar disc is perturbed by the joint effect of the stellar cusp
  (modelled as a rigid potential) and of a second molecular cloud\index{Molecular cloud}
  (modelled as SPH particles)disrupted by the SMBH. The color map is
  logarithmic, ranging from $2\times{} 10^{-2}$ to $2\times{}10^{10}$
  M$_\odot$ pc$^{-3}$. The contours show the projected density of
  stars in the stellar disc. The box size is $4.0 \times{} 1.8$
  pc. The projection was chosen so that the total angular momentum\index{Angular momentum} of
  the stellar disc is aligned to the vertical axis of the plot. The
  two white arrows in the bottom panel show the direction of the total
  angular momentum\index{Angular momentum} of the stellar disc and the total angular momentum\index{Angular momentum}
  of the outer gas disc. The length of the arrows is arbitrary. Fig.~4
  of \cite{Mapelli13}. Right-hand panel: distribution of inclinations
  ($\theta{}$) and semi-major axes ($a$) of the disc stars at $t=1.5$
  Myr in the top and bottom panel, respectively. From the $N$-body/SPH
  simulations\index{N-body simulations} of \cite{Mapelli13}. Cross-hatched red histogram:
  simulation including a spherical cusp and a perturbing gas disc
  (run~B2); hatched black histogram: simulation with only a spherical
  cusp (run~A2). From Fig.~9 of \cite{Mapelli13}.}
\label{fig:mapelli13}
\end{figure}

 \cite{Mapelli13} is the first study in which the gas perturber is represented by 'live' SPH particles, rather than by a rigid potential. In particular, \cite{Mapelli13} simulate the
 perturbations exerted on a thin stellar disc (with outer radius
$\sim{}0.4$ pc) by a molecular cloud\index{Molecular cloud} that falls towards the GC and is
disrupted by the SMBH. The initial conditions for the stellar disc
were drawn from the results of previous simulations
(\citealt{Mapelli12}) of molecular cloud\index{Molecular cloud} infall and disruption\index{Tidal disruption} in the
SMBH potential. \cite{Mapelli13} find that most of the gas from the
disrupted molecular cloud\index{Molecular cloud} settles into a dense and irregular disc
surrounding the SMBH (see Fig.~\ref{fig:mapelli13}).  If the gas disc
and the stellar disc are slightly misaligned ($\sim{}5-20^\circ{}$),
the precession of the stellar orbits induced by the gas disc
significantly increases the inclinations of the stellar orbits (by a
factor of $\sim{}3-5$ in 1.5 Myr) with respect to the normal vector to
the disc. Furthermore, the distribution of orbit inclinations becomes
significantly broader (see Fig.~\ref{fig:mapelli13}).
%%%%%%%%%%% FIGURE  %%%%%%%%%%%%%%%%%%%% 
%%%%%%%%%%%%%%%%%%%%%%%%%%%%%%% 

\begin{figure}[t]
%\sidecaption[t]
\includegraphics[width=10.5cm]{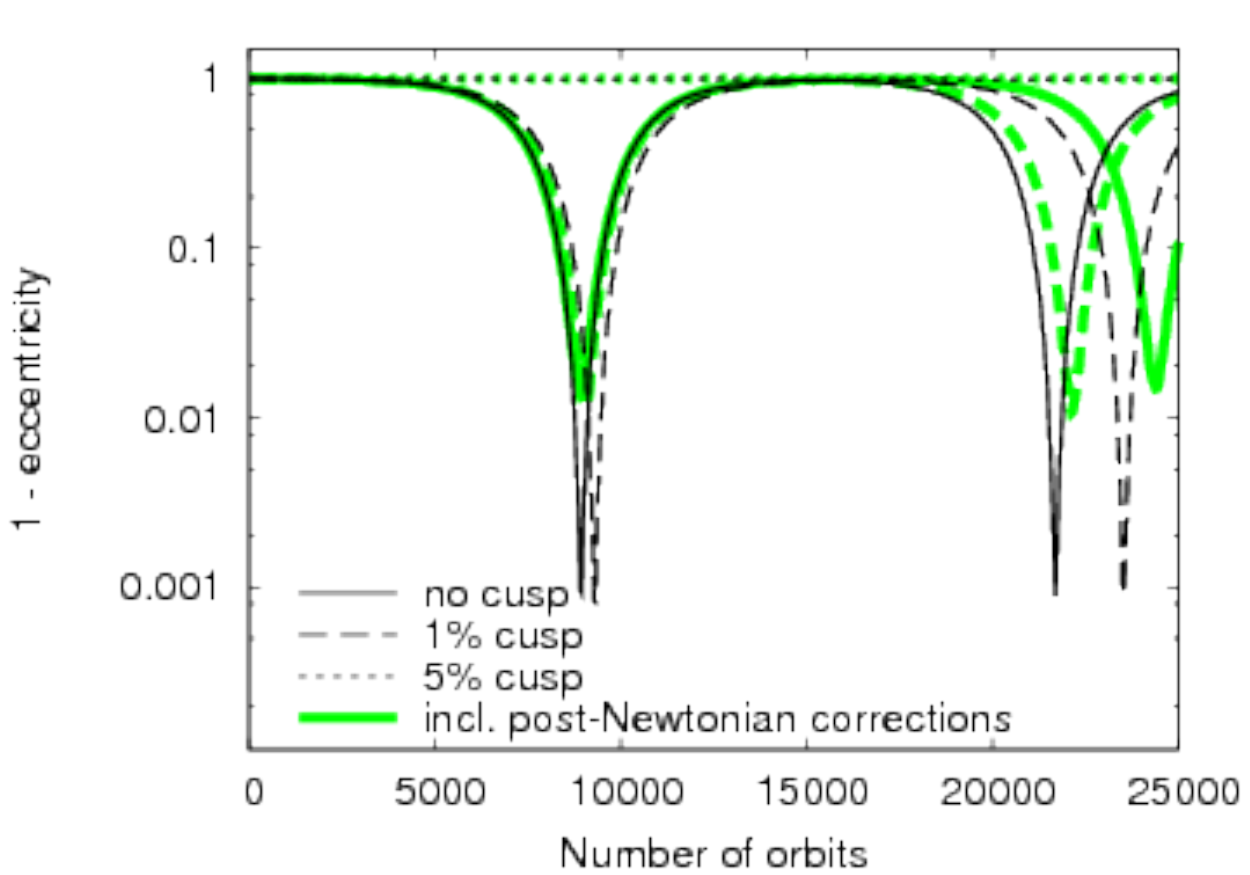}
\caption{Eccentricity evolution of a test star undergoing Kozai
  resonance driven by a fictitious $1.5\times{}10^4\msun$
  particle representing a disc potential. Both particles have
  initially circular orbits about the $3.5\times{}10^6\msun$ SMBH
  with semi-major axes of 0.04 and 0.16 pc, respectively. Simulations
  reported by \cite{Lockmann09}. Solid black line: models without
  cusp; dashed (dotted) line: models with 1 per cent (5 per cent) of
  the extended cusp mass observed in the Galactic Centre (modelled as
  a smooth potential). Each curve is accompanied by a corresponding
  thick green curve that represents a respective integration including
  post-Newtonian (PN) terms up to 2.5 PN to account for the effects of
  general relativity. While relativistic effects damp the Kozai effect
  at high eccentricities, a stellar cusp with mass of a few per cent of the
  observed value is sufficient to damp any eccentricity growth (dotted
  line). Fig.~6 of \cite{Lockmann09}.}
\label{fig:lockmann}
\end{figure}

{\it Origin of the S-cluster}. \cite{Lockmann08} propose that the
  orbits of the S-stars\index{S-stars} are the result of precession and Kozai
  resonance\index{Kozai!resonance} due to the interaction between two stellar discs.  In this
  scenario, binary stars in the young stellar disc are first moved to
  highly eccentric orbits by Kozai resonance with a second stellar
  disc and then disrupted by the SMBH at pericentre, as in the binary
  breakup model.  The inclusion of a stellar cusp, however, has been
  shown to damp Kozai oscillations in the disc
  (\citealt{Chang09,Lockmann09}, see Fig.~\ref{fig:lockmann}), which are
  a key factor in this scenario \citep[see also][for a discussion]{Gualandris12}.
%%%%%%%%%%% FIGURE  %%%%%%%%%%%%%%%%%%%% 
%%%%%%%%%%%%%%%%%%%%%%%%%%%%%%% 

\begin{figure}[t]
\sidecaption[t]
\includegraphics[width=7.5cm]{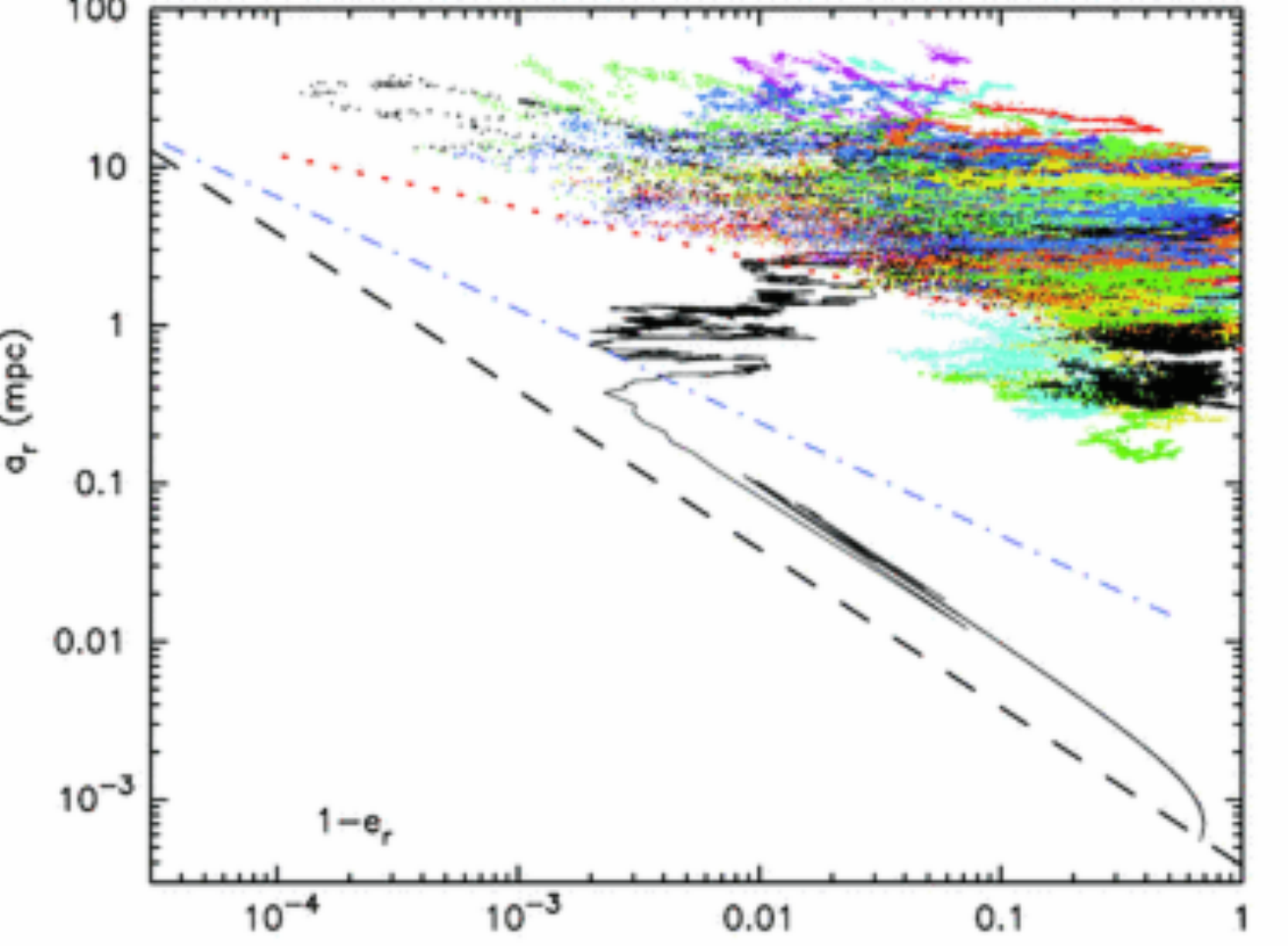}
\caption{Trajectories, over a time interval of 2 Myr, of stellar-mass
  black holes orbiting the SMBH as they undergo gravitational
  encounters with each other.  Dashed line: capture
  radius around the SMBH; dotted line: the Schwarzschild
  barrier; dot-dashed line: locus in the $a-e$
  plane where angular momentum\index{Angular momentum} loss due GW\index{Gravitational waves} emission
  dominates over gravitational encounters. From Fig.~5 of
  \citet{MAMW2011}.}
\label{fig:barrier}
\end{figure}

{\it Schwarzschild Barrier}. Relativistic precession\index{Precession!relativistic} limits the
  ability of torques from the stellar potential to modify orbital
  angular momenta\index{Angular momentum} via resonant relaxation.  This results in a sort of
  barrier \citep{MAMW2011} in the $(a,\,{}e)$ plane which sets an
  effectively maximum value of the eccentricity at each value of the
  semi-major axis (see Fig.~\ref{fig:barrier}). The Schwarzschild
  Barrier inhibits extreme-mass-ratio-inspirals and leads to capture
  rates that are $\sim 10-100$ times lower than in the
  non-relativistic case.
%%%%%%%%%%% FIGURE  %%%%%%%%%%%%%%%%%%%% 
%%%%%%%%%%%%%%%%%%%%%%%%%%%%%%% 

{\it Eccentric disc instability}\index{Eccentric disc instability}. 
An eccentric stellar disc around the SMBH is expected to exhibit an
instability as a result of the eccentricity dependence of the mass
precession timescale (Eq.~\ref{eq:Tcusp}). Retrograde precession due
to the presence of a stellar cusp induces coherent torques that
amplify deviations of individual orbital eccentricities from the
average, and thus drives all eccentricities away from their initial
value \citep{Madigan2009}, producing a bimodal eccentricity
distribution.
%%%%%%%%%%% FIGURE  %%%%%%%%%%%%%%%%%%%% 
\begin{figure}[t]
%\sidecaption[t]
\includegraphics[width=4.5cm, angle = -90]{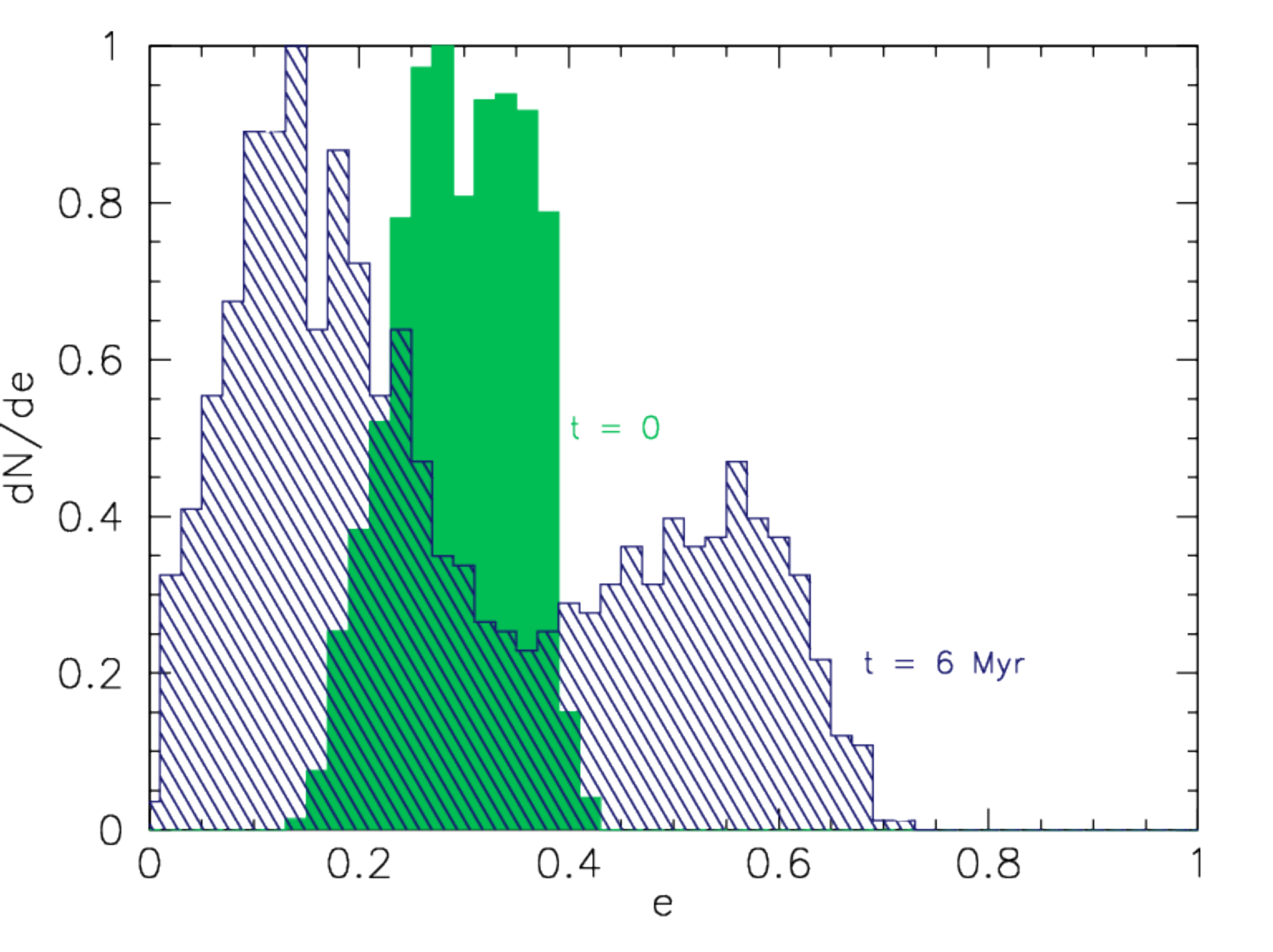}
\includegraphics[width=4.5cm, angle = -90]{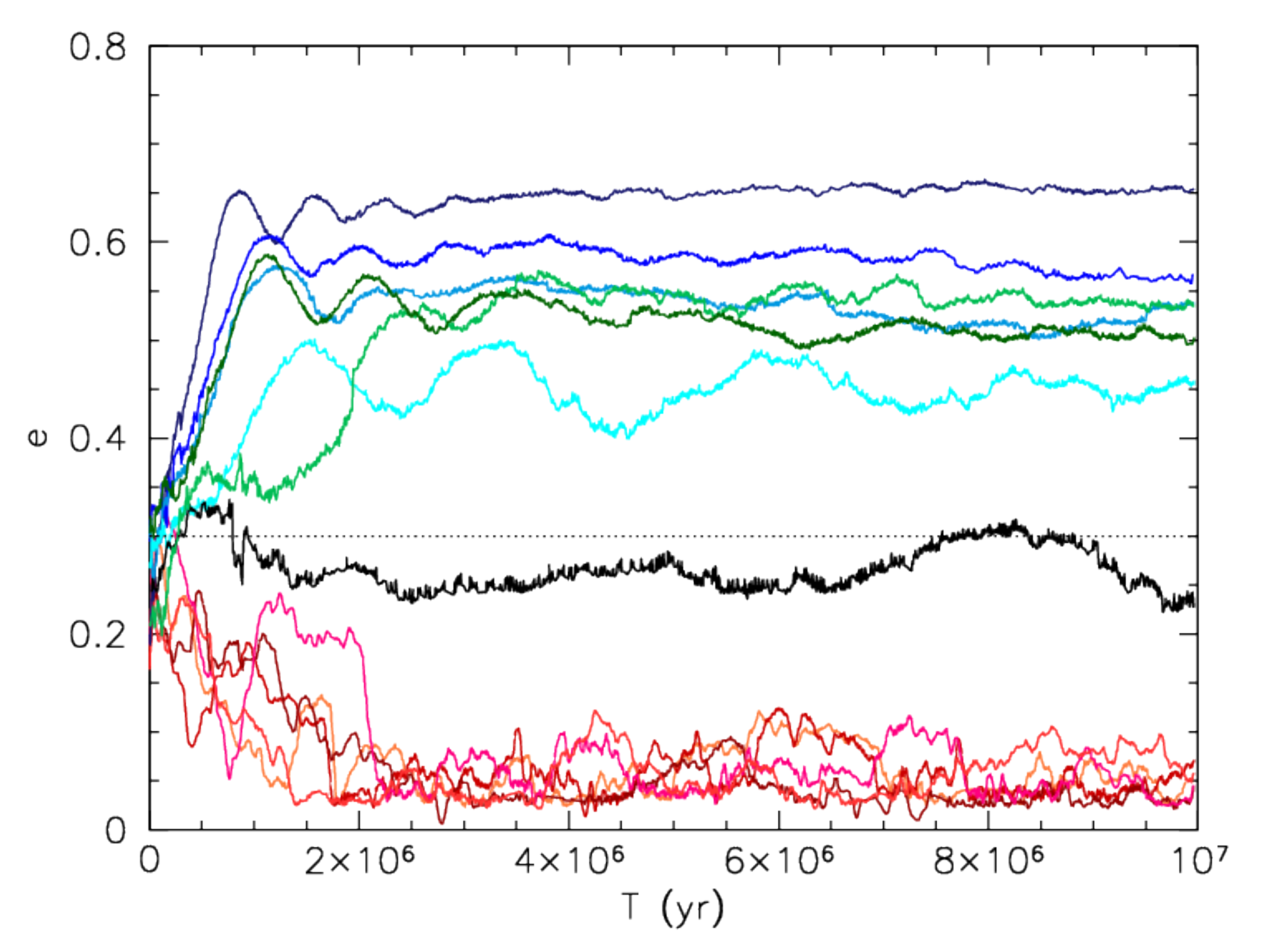}
\caption{(Left) Eccentricity distribution of a ring stars at the start
  of the integration (filled area) and after 6 Myr of evolution
  (hatched area), subject to the potential of the SMBH, a spherical
  stellar cusp and the parent gas disc.  Adapted from Fig.~5 of
  \citet{Gualandris12}. (Right) Evidence for the eccentric disc
  instability \citep{Madigan2009} in a random subset of stars in the
  simulation of \citet{Gualandris12}. The time-scale for the process
  is about 1 Myr. Fig.~6 of \citet{Gualandris12}.}
\label{fig:instability}
\end{figure}
%%%%%%%%%%%%%%%%%%%%%%%%%%%%%%% 
  \citet{Gualandris12} study the evolution of the ring of stars formed
  in the GC from fragmentation\index{Fragmentation} of the gas disc deposited by an
  inspiralling molecular cloud\index{Molecular cloud}. The stars are subject to the potential
  of the SMBH, a stellar cusp and the parent gas disc.  While the ring
  retains the original distribution of semi-major axes, and therefore
  also the initial inner and outer radius, the distribution of
  eccentricities evolves in time due to the onset of the eccentric
  disc instability. Torques exerted by other stars in the ring result
  in a change in the magnitude of the angular momentum\index{Angular momentum} and, as a
  consequence, in the eccentricity. As stars evolve away from the
  average eccentricity, a bimodal distribution is established, with a
  primary peak at $e\sim0.1$, a secondary peak at $e\sim0.5$ and a
  tail that extends to $e\sim0.7$ (see
  Fig.~\ref{fig:instability}). This is qualitatively consistent with
  the distribution found for the CW disc\index{Clockwise disc} stars \citep{Bartko09}.

\subsection{Perturbations from an intermediate-mass black hole}
\label{subsec:IMBHpert}
In the cluster inspiral scenario with an IMBH\index{Intermediate-mass black hole}, 
stars are naturally deposited in a disc structure in the same plane as the IMBH orbit and
with orbital elements similar to those of the IMBH itself.  The
inspiral of the IMBH is expected to slow down or stall completely at a
distance $\sim 10 (q / 10^{−3})\mpc$ from the SMBH, where $q$ is the
ratio of IMBH to SMBH masses \citep{Baumgardt06, MME2007,
  LockmannBaumgardt08}; this distance is comparable to the sizes of
the S-star orbits if $q \approx 10^{−3}$, i.e., if $M_{\rm IMBH} \sim10^{3.5}\msun$.  
At this separation, the total binding energy in
background stars within the IMBH orbit is comparable to that of the
IMBH itself and stars are easily ejected by the slingshot mechanism,
thereby causing the frictional force to drop.  

The orbit of the IMBH is likely to be quite eccentric at this stage,
depending on the initial orbit of the cluster and the detailed history
of interactions with the stars \citep{Baumgardt06, MME2007}.  If the
eccentricity is not so high ($e \lesssim 0.99$) that energy loss due
to emission of GWs\index{Gravitational waves} results in coalescence in less than
$10^8$ years, the semi-major axis of the IMBH orbit remains
essentially unchanged for times comparable to the
S-stars\index{S-stars} main-sequence lifetimes.  Prolonged
gravitational interactions\index{Interactions!gravitational} with the
IMBH can then scatter the young stars out of the thin disc into which
they were originally deposited \citep{MGM2009}.
%%%%%%%%%%% FIGURE  %%%%%%%%%%%%%%%%%%%% 
\begin{figure}[t]
%\sidecaption[t]
\includegraphics[width=12cm]{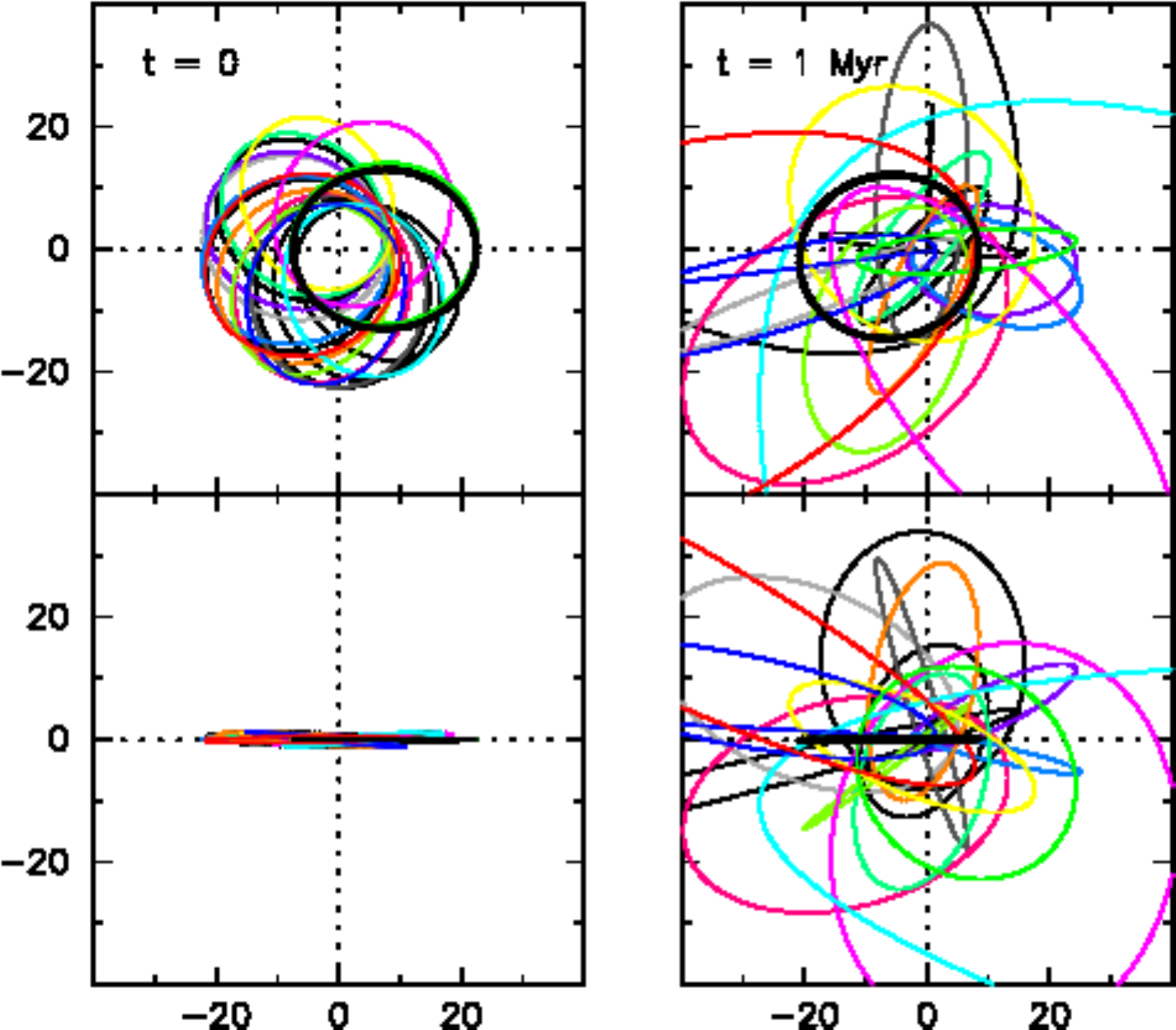}
\caption{Initial (left) and final (right, after 1 Myr) orbits of stars
  in a simulation with IMBH semi-major axis $a = 15\mpc$, eccentricity
  $e = 0.5$, mass ratio $q = 0.001$. Top panels show the view looking
  perpendicular to the IMBH orbital plane and bottom panels are from a
  vantage point lying in this plane. The IMBH orbit is the heavy black
  curve in all panels; the unit of length is milliparsecs. The
  initially disc-like, co-rotating distribution of stars is converted,
  after 1 Myr, into an approximately isotropic distribution of orbits
  with a range of eccentricities, similar to what is observed for the
  S-stars. Many of the orbits ``flip'' in response the perturbation
  from the IMBH, i.e. their angular momentum\index{Angular momentum} vector changes by
  $180^{\circ}$. Fig.~1 of \citet{MGM2009}.}
\label{fig:MGM2009fig1}
\end{figure}
%%%%%%%%%%%%%%%%%%%%%%%%%%%%%%% 
Fig.~\ref{fig:MGM2009fig1} shows the result of the simulations of
\citet{MGM2009} following the evolution of a disc of stars around an
IMBH with mass ratio of $q = 0.001$.  An initially planar
configuration for the stars is quickly ($\sim$ 1 Myr) turned into an
isotropic configuration by perturbations from the IMBH. An
eccentricity larger than $\sim 0.2$ is necessary for stellar
inclinations to be excited.

%%%%%%%%%%% FIGURE  %%%%%%%%%%%%%%%%%%%% 
\begin{figure}[t]
\sidecaption[t]
\includegraphics[width=7.5cm]{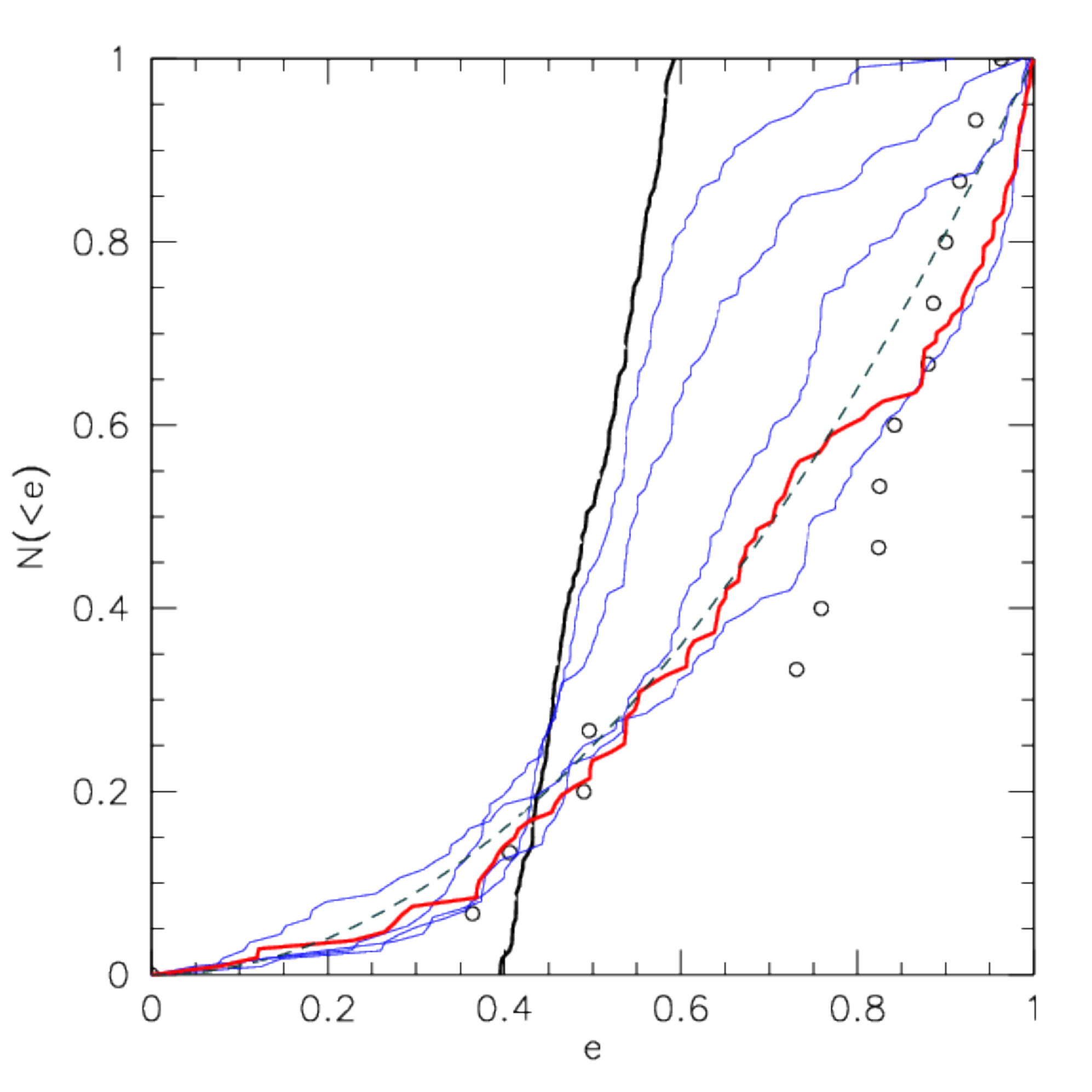}
\caption{Evolution of the distribution of stellar orbital
  eccentricities in a set of simulations with IMBH orbital parameters
  $q = 5\times10^{-4}$ , $a = 15\mpc$, $e = 0.5$. The initial
  distribution (thick black line) evolves in time (thin blue
  lines). After 1 Myr, the distribution (thick red line) is consistent
  with a thermal distribution (dashed line). Open circles represent
  the S-stars observed distribution \citep{Gillessen09a}. Fig.~3 of
  \citet{MGM2009}.}
\label{fig:MGM2009fig3}
\end{figure}
%%%%%%%%%%%%%%%%%%%%%%%%%%%%%%% 
The IMBH\index{Intermediate-mass black hole} also induces evolution in the eccentricities and energies
(semi-major axes) of the stars. Eccentricities were found to tend
toward a “thermal” distribution on a timescale of about 0.1 Myr for $q
\gtrsim 2.5\times10^{−4}$, as illustrated in
Fig.~\ref{fig:MGM2009fig3}.  The final distribution of stellar
semi-major axes depends on the assumed size of the IMBH orbit, but
stars with apastron distances as small as the periastron distance of
the IMBH are naturally produced.  Therefore, tightly bound orbits like
those of the innermost S-stars, e.g., S2, require an IMBH orbit with a
periastron distance of about $10\mpc$.

If the cluster inspiral scenario with an IMBH is deemed otherwise
viable, the results of \citet{MGM2009} show that the model can also
naturally reproduce the random and eccentric character of the stellar
orbits, and all in a time that is less than stellar evolutionary
timescales -- thus providing an essentially complete explanation for
the “paradox of youth\index{Paradox of youth}” of the S-stars\index{S-stars}.  

In order to avoid making the current epoch special, the IMBH inspiral
rate needs to be roughly equal to the inverse of the S-star lifetimes,
i.e $\sim 10^{−7} {\rm yr^{-1}}$. Such a rate has been proposed by
\citet{SPZ2006} based on a semi-analytic model of the formation and
evolution of star clusters in the galactic bulge.

%These results do not necessarily imply that an IMBH is currently
%present in the GC, as it may have been on a rapidly decaying orbit.

%%%%%%%%%%% FIGURE  %%%%%%%%%%%%%%%%%%%% 
\begin{figure}[t]
%\sidecaption[t]
\includegraphics[width=12cm]{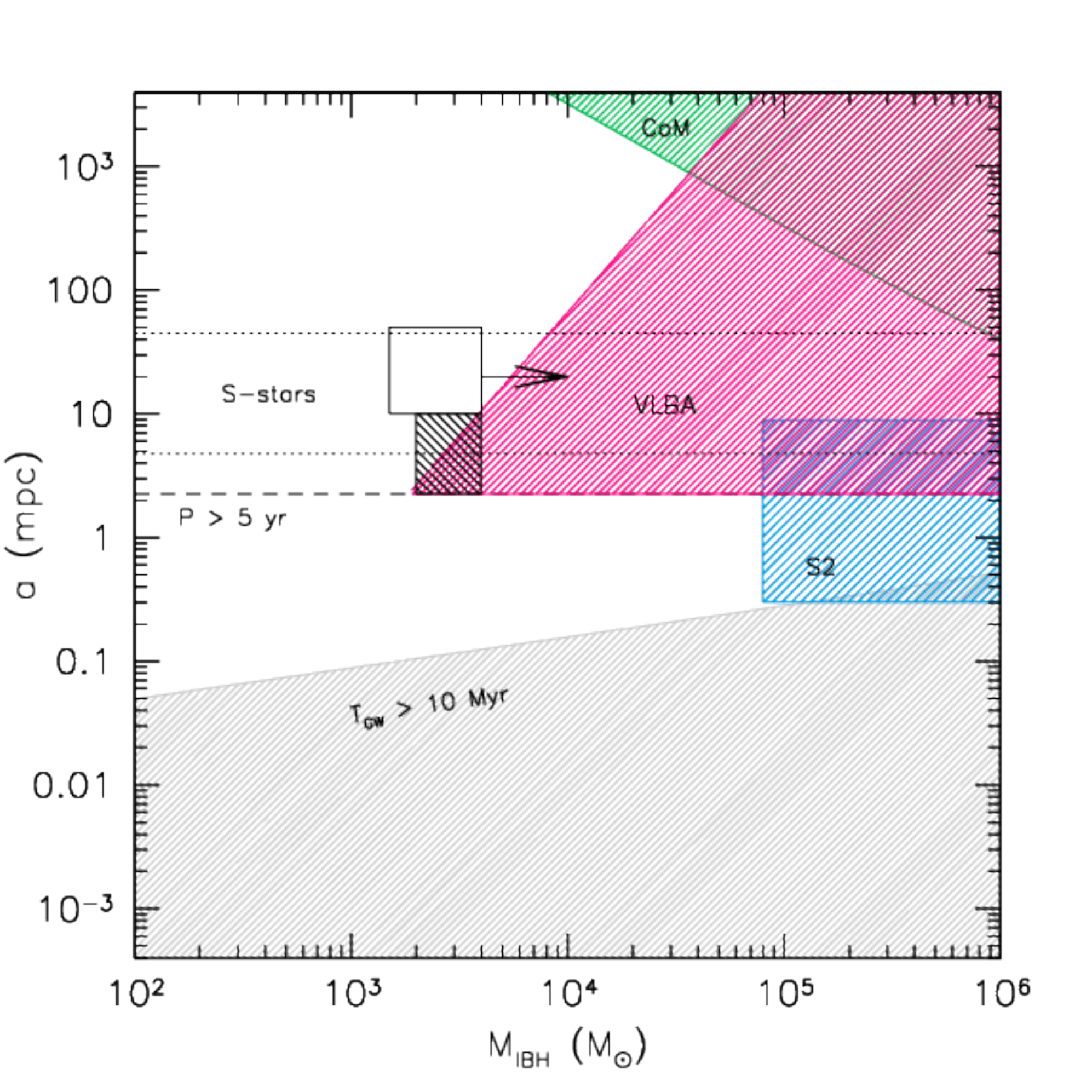}
\caption{Constraints on the orbital parameters of a hypothetical IMBH
  in the Galactic region. The shaded areas represent regions of
  parameter space that can be excluded based on observational or
  theoretical arguments. The dotted lines mark the distances at which
  the S-stars are currently observed. The dashed line represents the
  five year orbital period corresponding to discoverable systems. The
  parameters enclosed in the empty rectangular box are required for an
  efficient randomization of inclinations in the cluster infall
  scenario \citep{MGM2009}. The small rectangular region just below the
  empty box represents the parameter space excluded by
  \citet{GM2009}. Adapted from Fig.~13 of \citet{GM2009}.}
\label{fig:GM2009fig13}
\end{figure}
%%%%%%%%%%%%%%%%%%%%%%%%%%%%%%% 
\citet{GM2009} study the short- and long-term effects of an IMBH on
the orbits of the S-stars\index{S-stars}, for different choices of
IMBH parameters: mass, semi-major axis, eccentricity, spatial
orientation.  On long timescales, perturbations from an IMBH can
result in : (i) randomization of the inclinations of the stars; (ii)
ejection of stars from the region; (iii) scattering of stars onto
plunging orbits that result in tidal disruption\index{Tidal disruption} in the SMBH’s tidal
field; and (iv) secular effects like Kozai cycles\index{Kozai}.  When
considering individual stars, stars with initially large
eccentricities are the most susceptible to perturbations.

The result on the distribution of orbital
elements for the S-cluster depends on the IMBH\index{Intermediate-mass black hole} parameters.  The
distribution of S-star semi-major axes and eccentricities are
significantly altered from their currently observed form by IMBHs with
masses greater than $\sim 1000\msun$ if the IMBH--SMBH semi-major axis
lies in the range $3-10\mpc$.

These results can be used to constrain the allowed parameters of an
IMBH--SMBH binary at the Galactic centre.  The region of parameter
space corresponding to masses $\gtrsim 2000\msun$ and initial
semi-major axes $\sim 2-10\mpc$  can be excluded. Such region is represented
by the shaded box in Fig.~\ref{fig:GM2009fig13}.  All shaded areas in
the figure mark regions of parameter space that can be excluded based
on theoretical or observational arguments. Interestingly, the IMBH
parameters required for an efficient randomization of inclinations
\citep{MGM2009} in the cluster infall scenario ($M_{\rm IMBH}\gtrsim 1500\msun$
for the simulated range of separations $10-50\mpc$ - see rectangular
box in Fig.~\ref{fig:GM2009fig13}) are consistent with all the
constraints placed so far.

\citet{GGM2010} study short-term perturbations of an IMBH\index{Intermediate-mass black hole} on star S2,
\citep{Schodel02, Ghez03} whose short orbital period ($P\sim 16$\,yr)
and large eccentricity ($e \sim 0.88$) \citep{Gillessen09a} make it an
ideal candidate to detect small deviations from a purely Keplerian
orbit.  Deviations from a purely Keplerian orbit are expected for star
S2 due to relativistic and Newtonian precession (see
section~\ref{subsec:precession}).  Their only observable effect is an in-plane advance of the pericentre.
None of the other classical elements are affected by precession. 

In the absence of spin effects, which would not anyway manifest
themselves at the distance of S2 \citep{MAMW2010}, only non
spherically symmetric perturbations like those due to an IMBH are able
to produce changes in the angular momentum\index{Angular momentum} of S2's orbit, leading to
changes in eccentricity and the inclination of the orbital
plane. Perturbations due to the other S-stars\index{S-stars} have been
shown to be negligible.  Combining $N$-body simulations\index{N-body simulations} with
observational orbital fitting techniques, \citet{GGM2010} find that an
IMBH more massive than $\sim 1000\msun$ at a distance of $1-5\mpc$ is
potentially discoverable at S2's next pericentre passage in 2018.
Evidence for an IMBH would appear as significant deviations from the
assumed point mass relativistic potential in S2's orbital fit.

\section{Origin and evolution of the G2 cloud\index{G2 cloud|textbf}}
\label{sec:5}
In Sect.~\ref{subsec:2.4}, we discussed the orbital properties of the
dusty object G2. Several models were proposed to
explain the formation of G2.  Despite this, the nature of the G2 cloud
remains unclear, because none of the proposed models accounts for all
its properties in a satisfactory way.  The main open questions are
(see \citealt{Burkert12}): (i) is G2 only a cloud or is there a
compact source hidden inside the gas shell? (ii) where did G2 come
from? (iii) why is the orbit so eccentric? (iv) which are the
processes that affect G2 close to pericentre? (v) how many clouds like
G2 are currently orbiting Sgr~A$^\ast{}$?

In the following Sections, we review the most popular theoretical
scenarios proposed to explain the formation of G2, and
we highlight their major drawbacks.

\subsection{The pure gas cloud hypothesis}
\label{subsec:5.1}
The models proposed to explain the nature of G2 can be
grouped in two different families: (i) the `true' cloud scenarios, and
(ii) the `hidden' central object scenarios.  In the present Section we
consider the former models, while the latter will be discussed in the
next Section. The main difference between the two families of
scenarios is that the expected pericentre of the orbit is within
(outside) the tidal radius of a gas cloud (star).

According to the cloud scenario, G2 is a cold-ish gas clump, confined
by the hot gas surrounding SgrA$^\ast{}$. The gas temperature in the
inner arcsec is $\sim{}10^{7-8}$ K. The cooling timescale of such hot
gas is mush longer than the dynamical timescale
(\citealt{Cuadra05}). Thus, the cloud cannot have formed {\it in situ}
in the central arcsec, but must come from further out.

A possible scenario (e.g. \citealt{Burkert12}) is that the cloud
originated from the winds of the early-type stars\index{Early-type stars} 
in the CW disc\index{Clockwise disc}. Winds of a luminous
blue variable star can be as slow as $300-500\kms$. When shocked, they
are heated to $\sim{}10^6$ K and cool quickly to $\sim{}10^4$ K
(\citealt{Burkert12}), leading to the formation of cold cloudlets
embedded in the hot gas (\citealt{KooMcKee92}). The coincidence of the
orbital plane of G2 with the orientation of the CW
disc\index{Clockwise disc} encourages the `shocked wind debris'
hypothesis. While `upstream' winds (i.e. winds emitted in the
direction of motion of the parent star) have velocities in excess of
$1000\kms$ and are ejected from the GC, `downstream' winds (i.e. winds
emitted against the direction of motion of the parent star) have
velocities $<500\kms$ and may fall toward Sgr~A$^\ast$ on a very
eccentric orbit.

Alternatively, G2 might have originated from high-velocity stellar
winds that collided with each other, losing their initial angular
momentum\index{Angular momentum} (\citealt{Cuadra06}). Furthermore, G2 might have formed as a result of a cooling instability
in the accretion flow toward Sgr~A$^\ast$ (\citealt{Gillessen13a}). In
this case, the radial orbit is explained by the fact that the cloud
belongs to a gas inflow.

 Finally, \cite{Guillochon14} recently proposed that G2 formed out of the debris stream produced by the removal of mass from the outer envelope of a nearby giant star. Their adaptive mesh hydrodynamical simulations of the returning tidal debris stream show that the stream condenses into clumps that fall periodically onto Sgr A$^\ast{}$. G2 might be one of these clumps. Two intriguing results of this model are that (i) the orbits of several observed GC stars are consistent with the debris stream scenario, (ii) there might be several other G2-like clouds in the GC.

The cloud hypothesis (including the aforementioned `shocked wind
debris', 'stellar wind collision', `cooling instability' and `debris stream' scenarios) is consistent with existing observations
(\citealt{Gillessen13b}). The scenarios in which G2\index{G2 cloud} is
a collection of smaller droplets might even account for the observed
constant luminosity. In fact, G2\index{G2 cloud} is stretched by the
SMBH's tidal shear along its orbit, while it is compressed in the
transverse direction by the hot gas. This double effect is expected to
produce changes in the luminosity. On the other hand, if G2\index{G2 cloud} 
is composed of many little sub-clumps, the sub-clumps might
be less affected by the shear and compression internally. The
`collection of smaller droplets' would allow to explain even another
property: the large ($\approx{}100\kms$) internal velocity
dispersion. In fact, the cold droplets are embedded in diffuse hot gas
and might be pressure confined by this hot inter-droplet gas.

The main difficulty of the `true' cloud models \citet{Gillessen13b} is
their apparent inability to explain the `compactness' of G2\index{G2 cloud} 
found in the most recent data: the `head' of G2 (i.e. the
leading bulk of G2\index{G2 cloud} emission) is much more compact than
expected (from models and simulations, e.g. \citealt{Burkert12};
\citealt{Schartmann12}) for a gas cloud starting in pressure
equilibrium at the apocentre of the predicted orbit ($\approx{}0.041$
pc, i.e. the inner rim of the ring of early-type
stars\index{Early-type stars}). Furthermore, the head of G2 survived to the pericentre passage, without
undergoing significant disruption.

\begin{figure}[t]
%\sidecaption[t]
\includegraphics[width=12cm]{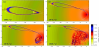}
\caption{Density evolution of a spherical shell matching the
  properties of GC (model SS01 of \citealt{Schartmann12}). The
  simulation was performed with the hydrodynamical code PLUTO
  (\citet{Mignone07}). Overlaid as dotted white lines are the positions
  of test particles initially located at the outer ring boundary. The
  axis labels are given in mpc. Fig.~5 of \citet{Schartmann12}.}
\label{fig:schartmann12}
\end{figure}

This issue might be overcome by
assuming that the cloud formed closer to the GC ($\approx{}0.0245$ pc,
\citealt{Burkert12}), or that it is `magnetically arrested' (\citealt{Shcherbakov14}).  Alternatively, the cloud might be a spherical
shell of gas (\citealt{Schartmann12}), or a nova ejecta (\citealt{Meyer12b}), rather than a compact cloud. Hydrodynamical simulations (see
Fig.~\ref{fig:schartmann12}) show that the spherical shell model is in
agreement with observations, even if the shell formed at apocentre, in
the ring of early-type stars. Finally, it may be that we observe only
the `tip of the iceberg', i.e. that the head of G2\index{G2 cloud} is
the very dense top of a much more massive (but less dense) gas inflow.

%%%%%%%%%%%%%%%%%%%%%%%%%%%%%%%%FIGURE %%%%%%%%%%%%%%%%%%%%%%%%%%%%%%%%%%%%%%%%%
%%%%%%%%%%%%%%%%%%%%%%%%%%%%%%%%FIGURE %%%%%%%%%%%%%%%%%%%%%%%%%%%%%%%%%%%%%%%%%

\subsection{The central object scenario}
\label{subsec:5.2}
In the central object scenario, G2\index{G2 cloud} is the atmosphere
of an unresolved central object that continuously loses gas. Ionizations and recombinations of this gas
would be responsible for the observed line emission (Br-$\gamma{}$ luminosity $\sim{}$ a few $\times{}10^{30}$ erg s$^{-1}$, corresponding to an emission measure $\sim{}$ 10$^{57}$ cm$^{-3}$). The object might
have formed in the disc of early-type stars\index{Early-type stars},
and was then scattered into a highly eccentric orbit due to a close
encounter with another stellar (or compact) object.

\begin{figure}[t]
%\sidecaption[t]
\includegraphics[width=12.0cm]{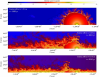}
\caption{Density maps of the stellar wind (around a T Tauri star)
  disrupted by the SMBH, in the fiducial run of \citet{Ballone13}. From
  top to bottom: source distance of 1''.21, 0''.43, and 0''.15 from
  Sgr~A$^\ast$. Fig.~2 of \citet{Ballone13}.}
\label{fig:ballone}
\end{figure}

As to the nature of the object,  a planetary nebula (\citealt{Gillessen12}),  a proto-planetary disc around a
low-mass star (\citealt{Murray12}),  a circumstellar gas disc around an old
low-mass star, disrupted by a stellar black hole (\citealt{Escude12}), 
the mass-loss envelope of a T Tauri star (\citealt{Scoville13}, see also
\citealt{Eckart13,Ballone13,Witzel14}, and Fig.~\ref{fig:ballone}), a merged star (\citealt{Prodan14}), and a giant gaseous proto-planet (i.e. a planetary embryo that formed from a gravitational instability in a protoplanetary disc, \citealt{Mapelli15}) have been proposed.

Both the compactness of G2's head and the survival of G2 to pericentre passage can be easily accounted for, in the frame of the compact source scenario, because of the small tidal radius of the central object ($\le{}10$ AU). In the hypothesis of a T Tauri star, the Br-$\gamma{}$ emission comes from
the inner cold bow shock, where the stellar wind is impacted by the
hot gas in proximity of Sgr~A$^\ast{}$ (\citealt{Scoville13}). In the scenarios of both a giant gaseous protoplanet and a protoplanetary disc, the  Br-$\gamma{}$ emission arises from photoevaporation due to the ultraviolet background of the nuclear star cluster, and is enhanced by partial tidal stripping (\citealt{Murray12,Mapelli15}). Finally, the scenario of a proto-planetary disc (\citealt{Murray12}) predicts
an increase in the luminosity of the Br$-\gamma{}$ line by a factor of $\approx{}5$ at pericentre passage, quite higher than
the observed value (which is only a factor of $\sim{}2$, \citealt{Pfuhl14}).  On the other hand, this mismatch could be due to an overestimate of the recombination rate (see e.g. \citealt{Mapelli15}). As recently highlighted by \citealt{Witzel14}, the high $L'$ continuum luminosity (corresponding to $\approx{}2\times{}10^{33}{\rm erg\,{}s}^{-1}$) can be easily explained by a dust-enshrouded $1-2$ M$_\odot$ star. If the central source is too weak (e.g. in the case of a protoplanet), the warm dust must be spread over a sufficiently large volume (radius $\gtrsim{}5\times{}10^{12}$ cm), to explain the $L'$ continuum luminosity.

In summary, most of the proposed central object scenarios and pure cloud scenarios are still viable to explain the dusty object G2: the nature of this object remains quite elusive.

%\citet{Murray-ClayLoeb13} do not balance ionizations and recombinations. In the case of disks with radius $\gtrsim 10\,{\rm AU}$ producing photoevaporative winds with $n\sim10^{7}\,{\rm cm^{-3}}$, they end up predicting a recombination rate much larger than the ionization rate, implying that the material would become neutral in $\sim5$ days.

%Furthermore, there is another possible issue for the
%compact-source scenario (both a planetary nebula, a proto-planetary
%disc, a circumstellar gas disc and a T Tauri star): the radial profile
%of gas surrounding a central object scales as $\sim{}r^{-2}$
%(\citealt{Scoville13}), which would produce a considerably steeper
%velocity gradient than observed in the data (\citealt{Gillessen13b}).

%%%%%%%%%%%%%%%%%%%%%%%%%%%%%%%%FIGURE %%%%%%%%%%%%%%%%%%%%%%%%%%%%%%%%%%%%%%%%%
%%%%%%%%%%%%%%%%%%%%%%%%%%%%%%%%FIGURE %%%%%%%%%%%%%%%%%%%%%%%%%%%%%%%%%%%%%%%%%

\section{Conclusions: open questions and future work}
\label{sec:6}
In this review, we have briefly summarized the most recent
observational results about the GC (Sect.~\ref{sec:2}), and we have
discussed the main theoretical scenarios for the formation of the
early-type stars\index{Early-type stars} (Sect.~\ref{sec:3}-\ref{sec:4}) and for the nature of 
the G2\index{G2 cloud} cloud, Sect.~\ref{sec:5}) in the
GC. In this Section, we would like to summarize the main scenarios for
the formation of the early-type stars\index{Early-type stars} in the GC and highlight the pros
and the cons of each of them.

The main scenarios for the formation of the {\bf CW disc of early-type stars}\index{Clockwise disc}\index{Early-type stars} are the following:
\begin{itemize}
\item{} Fragmentation\index{Fragmentation} of the outer parts of a past accretion disc\index{Accretion disc}
  (e.g. \citealt{levin2003}; \citealt{Nayakshin05a};
  \citealt{Nayakshin06}; \citealt{Nayakshin07};
  \citealt{Collin08}). This scenario appears promising when looking at
  the relevant timescales, but cannot easily explain (i) the non-zero
  eccentricity of the stellar orbits, (ii) the observed thickness of the
  disc, (iii) the absence of any remnant of a past accretion disc\index{Accretion disc}.
  The second issue can be circumvented by invoking some fast mechanism to
  increase inclinations (e.g. precession exerted by the CNR\index{Circumnuclear ring}, Kozai
  resonance), but the other two issues are more difficult to overcome.
\item{} Disruption\index{Tidal disruption} of a molecular cloud\index{Molecular cloud} which results in the formation
  of a gas disc sufficiently dense to fragment into stars
  (e.g. \citealt{levin2003}; \citealt{Bonnell08}; \citealt{Mapelli08};
  \citealt{Wardle08}; \citealt{Hobbs09}; \citealt{Alig11}; \citealt{Mapelli12};
  \citealt{Lucas13}; \citealt{Alig13}). Recent simulations show that
  this scenario can reproduce the observed distribution of
  eccentricities and semi-major axes of the CW stars, together with
  the thickness of the disc. A difficulty of this model is that the
  molecular cloud\index{Molecular cloud} must have been on a fine-tuned orbit (i.e. with
  sufficiently low angular momentum\index{Angular momentum}, or with nearly zero impact
  parameter, to engulf Sgr~A$^\ast$). This issue might be overcome by
  assuming either that a cloud-cloud collision reduced the angular
  momentum\index{Angular momentum} of the cloud or that the disc was formed by gaseous
  streamers (such as those observed in the region of the CNR\index{Circumnuclear ring}) rather
  than by a coherent molecular cloud\index{Molecular cloud}.
\item{} Inspiral and disruption\index{Tidal disruption} of a star cluster
  (e.g. \citealt{Gerhard2001}; \citealt{KimMorris2003}). This process
  appears to be too slow to be consistent with the age of the CW stars
  and unable to explain the top-heavy MF\index{Mass function}.  The presence of
  an IMBH at the centre of the cluster mitigates the requirements on
  the mass and density of the cluster for a fast inspiral but (i)
  there is so far no observational evidence for an IMBH in the GC,
  (ii) depending on the mass and eccentricity of the IMBH,
  interactions may act to randomise the inclinations and thermalise
  the eccentricities on timescales of 1 Myr or less, producing a
  system which is consistent with the properties of the S-cluster
  rather than those of the CW disc\index{Clockwise disc}.
\end{itemize}

The main scenarios for the formation of the {\bf early-type stars\index{Early-type stars} that
  do not lie in the CW disc\index{Clockwise disc}, including the (B-type) S-stars} can be
summarized as follows:
\begin{itemize}
\item{} Binary breakup scenario: the SMBH disrupts stellar binaries on
  eccentric orbits that take them within their tidal radius via the
  Hill's mechanism and captures one of the components on an eccentric
  bound orbit \citep[e.g.][]{Hills1991, Miller2005}. The large
  eccentricities of the stars are thermalised within the B-stars
  lifetime by resonant relaxation against the background cusp of stars
  and remnants.
\item{} The inspiral and disruption\index{Tidal disruption} of a star cluster with an IMBH can
  explain both the isotropic spatial distribution of the S-stars as
  well as the roughly thermal eccentricity distribution.  However,
  formation of an IMBH in a cluster has only been predicted from $N$-body
  simulations\index{N-body simulations}. In addition, tidal stripping of stars during the cluster inspiral
  predicts the deposition of a much larger number of stars outside the S-cluster
  than are actually observed.
\end{itemize}

Relaxation processes are necessary ingredients of all above models in
that they cause the orbital distributions of the young stars to evolve
in time.  In particular, resonant relaxation\index{Resonant relaxation} 
is required for the eccentricity distribution of stars
captured from disrupted binaries to be converted into a thermal
distribution.  In the same model, scattering off massive
perturbers\index{Massive perturbers} is necessary to ensure that a
sufficient number of stellar binaries born at large distances are
places onto highly eccentric orbits at any time.

Precession in an axisymmetric potential and Kozai-Lidov resonances
may explain the formation of the WR/O stars that do not lie in the CW
disc (e.g. \citealt{Lockmann08}; \citealt{Lockmann09};
\citealt{Subr09}; \citealt{Haas11a}; \citealt{Haas11b};
\citealt{Mapelli13}). According to these processes, the early-type
stars\index{Early-type stars} that do not belong to the CW disc\index{Clockwise disc} might be the former members of
a now dismembered disc and/or the former members of the outer parts of
the CW disc\index{Clockwise disc}.  This process can explain the outliers at $>0.04$ pc but
not the S-stars, unless the perturbing potential in the past was
different from the current one (e.g. \citealt{ChenAmaro2014} explain
the S-stars with Kozai-like resonance, by assuming that the inner edge
of the gas disc was $<<0.04$ pc in the past). Another intriguing idea is that the two-body relaxation time-scale in the inner parts of the disc ($\lesssim{}0.05$ pc) might be much shorter than previously thought (\citealt{Subr14}), leading to a fast relaxation of the innermost stellar orbits.

This short overview of formation scenarios for the young stars shows
that there are a number of open questions about the recent star
formation history and dynamical evolution of the Galactic Centre. The
scenario of molecular-cloud\index{Molecular cloud} disruption\index{Tidal disruption} has become increasingly popular
to explain the formation of the CW disc\index{Clockwise disc}, but current simulations are
far from realistically tracing the formation of stars in the gaseous
disc. The treatment of shocks in the gas is crucial in this context
but the SPH codes used so far to simulate the disruption\index{Tidal disruption} of the
molecular cloud\index{Molecular cloud} are not the most suitable to capture the physics of
shocks (e.g. \citealt{Agertz2007}). Simulations with different
techniques (e.g. the adaptive-mesh refinement, AMR, technique) are
absolutely needed to confirm these results. Radiative transfer from
the newly born stars has never been accounted for (even if this is a
likely minor effect with respect to SMBH heating). The explosion of
core-collapse supernovae (the stars in the CW disc\index{Clockwise disc} are $\gtrsim{}3$
Myr old) has never been considered: it might have a crucial impact on
the evaporation of the gas disc. The adopted cooling functions and
recipes for the chemical composition of gas in the GC are critical too
(see the discussion about opacity in \citealt{Mapelli12}). We also
know that the GC hosts strong magnetic fields: their effects on the
formation of the early-type stars\index{Early-type stars} have been neglected so far.

%{\bf MM ADD SOMETHING ON YELDA}

Furthermore, we find there has been a gap between $N$-body/SPH codes\index{N-body simulations},
used to simulate the evolution of gas, and a dissipationless
direct-summation $N$-body codes, used to probe the secular evolution
of stars. Only a few studies try to fill this gap \citep[e.g.][and
  references therein]{Mapelli13}. More accurate $N$-body\index{N-body simulations} integrators
need to be coupled to SPH or AMR codes, in order to have a global
picture of the interplay between gas physics and dynamics in the GC.
Finally, the formation of the S-stars is far from being understood, as all
the proposed mechanisms suffer potential difficulties and/or
substantial draw-backs. 

From an observational point of view, probably no other region in the
sky has been so thoroughly scanned and monitored as the GC, in the
last $\sim{}10$ years. %The recent results by \cite{Yelda2014} have revolutionised our knowledge of the CW disc\index{Clockwise disc} and of the early-type  stars\index{Early-type stars} in the GC. 
 ALMA is about to provide an exciting 
view of molecular gas and ongoing star formation in the GC \citep{Yusef-zadeh13}. Forthcoming observations with available
facilities (e.g. the 8-m class telescopes VLT and Keck II) will
provide more accurate measurements of the orbits of the S-stars, of
the mass of the SMBH, of the enclosed mass in the GC and of the main
properties of the early-type stars\index{Early-type stars}.  With an imaging resolution of a
few milliarcseconds and an astrometric accuracy of $10 \mu {\rm as}$,
the second generation instrument for the VLT Interferometer, GRAVITY,
will be able to measure the proper motion of matter (stars or hot
spots in the accretion disk) down to the event horizon of the black
hole, hereby probing spacetime in its immediate vicinity.  Future 30-m
class telescopes (the European Extremely Large Telescope, E-ELT, and
the Thirty Meter Telescope, TMT) will make a huge difference with
respect to the past:  a diffraction-limited resolution of $\sim{}12$
mas will be achieved, which will allow for unprecedented astrometric
precision ($\sim{}0.1$ mas, e.g. \citealt{Yelda2013}). This will offer a unique
laboratory to study the intriguing processes that take place in the
neighbourhood of a SMBH.  Of particular relevance is the potential
detection of relativistic effects, which can be accomplished by a
combination of new facilities, longer monitoring of the currently known
stars, and the detection of new stars at smaller distances from the
SMBH.

\begin{acknowledgement}
This review is not directly connected with the lectures held by
Prof. R. Genzel at the 10$^{\rm th}$ SIGRAV school on `Astrophysical
black holes'. We would like to thank Prof. R. Genzel,
Dr. S. Gillessen, the Organizers of the SIGRAV school and the Editors
of this Book for giving us the possibility to write this review. 
 We warmly thank S. Gillessen, H. B. Liu and P. Kroupa for their invaluable comments. 
MM acknowledges financial support from the Italian Ministry of
Education, University and Research (MIUR) through grant FIRB 2012
RBFR12PM1F (`New perspectives on the violent Universe: unveiling the
physics of compact objects with joint observations of gravitational
waves and electromagnetic radiation'), from INAF through grants 
PRIN-2011-1 (`Challenging Ultraluminous X-ray sources: chasing their
black holes and formation pathways') and PRIN-2014-14 (`Star formation and evolution in galactic nuclei'), and from CONACyT through grant
169554 (`Nearby and distant spheroids: cutting edge theoretical tools
for the analysis of stellar populations'). 
%If you want to include acknowledgments of assistance and the like at the end of an individual chapter please use the \verb|acknowledgement| environment -- it will automatically render Springer's preferred layout.
\end{acknowledgement}

%\bibliographystyle{aa} % style aa.bst
%\bibliography{referenc.bib} % your references Yourfile.bib

%%%%%%%%%%%%%%%%%%%%%%%% referenc.tex %%%%%%%%%%%%%%%%%%%%%%%%%%%%%%
% sample references
% %
% Use this file as a template for your own input.
%
%%%%%%%%%%%%%%%%%%%%%%%% Springer-Verlag %%%%%%%%%%%%%%%%%%%%%%%%%%
%
% BibTeX users please use
% \bibliographystyle{}
% \bibliography{}

\begin{thebibliography}{299.}%
\bibitem[\protect\citeauthoryear{Abadi et al.}{2009}]{Abadi2009} Abadi, M.G., Navarro, J.F., Steinmetz, M. 2009, ApJL, 691, L63
%\bibitem[\protect\citeauthoryear{Abbott et al.}{2009}]{Abbott2009} Abbott B. P., et al. 2009, Reports on Progress in Physics, 72, 076901
\bibitem[\protect\citeauthoryear{Accadia et al.}{2012}]{Accadia2012} Accadia, T., et al. 2012, Virgo Document VIR-0128A-12, {\tt https://tds.ego-gw.it/ql/?c=8940} %Acernese, F., et al. 2008, Classical and Quantum Gravity, 25, 184001
\bibitem[\protect\citeauthoryear{Agertz et al.}{2007}]{Agertz2007} Agertz, O., et al. 2007, MNRAS, 380, 963
\bibitem[\protect\citeauthoryear{Alexander}{1999}]{Alexander99} Alexander, T. 1999, ApJ, 527, 835
\bibitem[\protect\citeauthoryear{Alexander et al.}{2008}]{Alexander08} Alexander, R. D., Armitage, P. J., Cuadra, J., Begelman, M. C. 2008, ApJ, 674, 927
\bibitem[\protect\citeauthoryear{Alexander et al.}{2012}]{Alexander12} Alexander, R. D., Smedley, S. L., Nayakshin, S., King, A. R. 2012, MNRAS, 419, 1970
\bibitem[\protect\citeauthoryear{Alig et al.}{2011}]{Alig11} Alig, C., Burkert, A., Johansson, P. H., Schartmann, M. 2011, MNRAS, 412, 469
\bibitem[\protect\citeauthoryear{Alig et al.}{2013}]{Alig13} Alig, C., Schartmann, M., Burkert, A., Dolag, K. 2013, ApJ, 771, 119 
\bibitem[\protect\citeauthoryear{Allen}{1987}]{Allen87} Allen, D. A. 1987, AIP Conference Proceedings, 155, 1
\bibitem[\protect\citeauthoryear{Allen et al.}{1990}]{Allen90} Allen, D. A., Hyland, A. R., Hillier, D. J. 1990, MNRAS, 244, 706
\bibitem[\protect\citeauthoryear{Amaro-Seoane et al.}{2013}]{Amaro2013} Amaro-Seoane P., et al. 2013, GW Notes, 6, 4
\bibitem[\protect\citeauthoryear{Amo-Baladr\'on et al.}{2011}]{Amo-Baladron11} Amo-Baladr\'on, M. A., Mart\'in-Pintado, J., Mart\'in, S. 2011, A\&{}A, 526, 54
\bibitem[\protect\citeauthoryear{Antonini et al.}{2010}]{Antonini2010} Antonini, F., Faber, J., Gualandris, A., Merritt, D. 2010, ApJ, 713, 90
\bibitem[\protect\citeauthoryear{Antonini et al.}{2011}]{Antonini2011} Antonini, F.,  Capuzzo-Dolcetta, R., Mastrobuono-Battisti, A., Merritt, D. 2011, ApJ, 750, 111
\bibitem[\protect\citeauthoryear{Antonini \&{} Merritt}{2013}]{AM2013} Antonini, F., Merritt, D. 2013, ApJ, 763, L10
\bibitem[\protect\citeauthoryear{Ao et al.}{2013}]{Ao13} Ao, Y., Henkel, C., Menten, K. M., Requena-Torres, M. A., Stanke, T., Mauersberger, R., Aalto, S., M\"uhle, S., Mangum, J. 2013, A\&{}A, 550, 135
\bibitem[\protect\citeauthoryear{Arca Sedda \&{} Capuzzo-Dolcetta}{2014}]{ArcaSedda2014} Arca Sedda, M., Capuzzo-Dolcetta, R. 2014, ApJ, submitted, arXiv:1307.5717
\bibitem[\protect\citeauthoryear{Bahcall \&{} Tremaine}{1981}]{Bahcall81} Bahcall, J. N., Tremaine, S. 1981, ApJ, 244, 805
\bibitem[\protect\citeauthoryear{Bahcall \&{} Wolf}{1976}]{BW76} Bahcall, J. N., Wolf, R.A. 1976, ApJ, 209, 214
\bibitem[\protect\citeauthoryear{Balick \&{} Brown}{1974}]{balick74}Balick, B., Brown, R. L. 1974, Intense sub-arcsecond structure in the galactic center, published in `H II regions and the galactic centre : proceedings of the Eighth ESLAB Symposium', p. 261
\bibitem[\protect\citeauthoryear{Ballone et al.}{2013}]{Ballone13} Ballone, A., Schartmann, M., Burkert, A., Gillessen, S., Genzel, R., Fritz, T. K., Eisenhauer, F., Pfuhl, O., Ott, T. 2013, ApJ, 776, 13
\bibitem[\protect\citeauthoryear{Bartko et al.}{2009}]{Bartko09} Bartko, H., Martins, F., Fritz, T. K., Genzel, R., Levin, Y., Perets, H.B., Paumard, T., Nayakshin, S., Gerhard, O., Alexander, T., et al. 2009, ApJ, 697, 1741
\bibitem[\protect\citeauthoryear{Bartko et al.}{2010}]{Bartko10} Bartko, H., Martins, F., Trippe, S., Fritz, T. K., Genzel, R., Ott, T., Eisenhauer, F., Gillessen, S., Paumard, T., Alexander, T., et al. 2010, ApJ, 708,  834
\bibitem[\protect\citeauthoryear{Baumgardt et al.}{2006}]{Baumgardt06} Baumgardt, H., Gualandris, A., Portegies Zwart, S. 2006, MNRAS, 372, 174
\bibitem[\protect\citeauthoryear{Becklin et al.}{1982}]{Becklin82} Becklin, E. E., Gatley, I., Werner, M. W. 1982, ApJ, 258, 135
\bibitem[\protect\citeauthoryear{Binney \&{}Tremaine}{1987}]{BT87} Binney, J., Tremaine, S. 1987, Galactic Dynamics (Princeton, NJ: Princeton University Press)
\bibitem[\protect\citeauthoryear{Blum et al.}{1995a}]{Blum95a} Blum, R. D., Sellgren, K., Depoy, D. L. 1995a, ApJ, 440, L17
\bibitem[\protect\citeauthoryear{Blum et al.}{1995b}]{Blum95b} Blum, R. D., Depoy, D. L., Sellgren, K. 1995b, ApJ, 441, 603
\bibitem[\protect\citeauthoryear{Boley}{2009}]{Boley09} Boley, A. C. 2009, ApJ, 695, L53 
\bibitem[\protect\citeauthoryear{Boley et al.}{2010}]{Boley10} Boley, A. C., Hayfield, T., Mayer, L., Durisen, R. H. 2010, Icarus, 207, 509
\bibitem[\protect\citeauthoryear{Bonnell \&{} Rice}{2008}]{Bonnell08} Bonnell, I. A., Rice, W. K. M. 2008, Science, 321, 1060
\bibitem[\protect\citeauthoryear{Bromley et al.}{2006}]{Bromley2006} Bromley, B.~C.,Kenyon, S.~J.,Geller, M.~J., Barcikowski, E.,Brown, W.~R., Kurtz, M.~J. 2006, ApJ, 653, 1194
\bibitem[\protect\citeauthoryear{Brown et al.}{2005}]{Brown2005} Brown, W.R., Geller, M.J., Kenyon, S.J., Kurtz, M.J. 2005, ApJL, 622, L33
\bibitem[\protect\citeauthoryear{Buchholz et al.}{2009}]{Buchholz2009} Buchholz, R.~M., Sch{\"o}del, R., Eckart, A. 2009, A\&{}A, 499, 483
\bibitem[\protect\citeauthoryear{Burkert et al.}{2012}]{Burkert12} Burkert, A., Schartmann, M., Alig, C., Gillessen, S., Genzel, R., Fritz, T. K., Eisenhauer, F. 2012,  ApJ, 750, 58
\bibitem[\protect\citeauthoryear{Chan et al.}{1997}]{Chan97} Chan, K.-W., Moseley, S. H., Casey, S., Harrington, J. P., Dwek, E., Loewenstein, R., V\'arosi, F., Glaccum, W. 1997, ApJ, 483, 798
\bibitem[\protect\citeauthoryear{Chandrasekhar}{1943}]{Ch43} Chandrasekhar, S. 1943, ApJ, 97, 255  
\bibitem[\protect\citeauthoryear{Chang}{2009}]{Chang09} Chang, Ph. 2009, MNRAS, 393, 224
\bibitem[\protect\citeauthoryear{Chatzopoulos et al.}{2014}]{Chatzopoulos14}Chatzopoulos, S., Fritz, T., Gerhard, O., Gillessen, S., Wegg, C., Genzel, R., Pfuhl, O. 2014, MNRAS, submitted, arXiv:1403.5266
\bibitem[\protect\citeauthoryear{Chen \&{} Amaro-Seoane}{2014}]{ChenAmaro2014} Chen, X., Amaro-Seoane, P. 2014, ApJ, submitted, arXiv:1401.6456
\bibitem[\protect\citeauthoryear{Christopher et al.}{2005}]{Christopher05} Christopher, M. H., Scoville, N. Z., Stolovy, S. R., Yun, Min S. 2005, ApJ, 622, 346
\bibitem[\protect\citeauthoryear{Cl\'enet et al.}{2004}]{Clenet04}Cl\'enet, Y., Rouan, D., Gendron, E., Lacombe, F., Lagrange, A.-M., Mouillet, D., Magnard, Y., Rousset, G., Fusco, T., Montri, J., et a. 2004, A\&{}A, 417, L15
\bibitem[\protect\citeauthoryear{Coil \&{} Ho}{1999}]{Coil99} Coil, A. L., Ho, P. T. P. 1999, ApJ, 513, 752
\bibitem[\protect\citeauthoryear{Coil \&{} Ho}{2000}]{Coil00} Coil, A. L., Ho, P. T. P. 2000, ApJ, 533, 245
\bibitem[\protect\citeauthoryear{Collin \&{} Hur\'e}{1999}]{Collin99} Collin, S., Hur\'e, J.-M. 1999, A\&{}A, 341, 385
\bibitem[\protect\citeauthoryear{Collin \&{} Zahn}{1999}]{Collin99b} Collin, S., Zahn, J.-P. 1999,  A\&{}A, 344, 433
\bibitem[\protect\citeauthoryear{Collin \&{} Zahn}{2008}]{Collin08} Collin, S., Zahn, J.-P. 2008,  A\&{}A, 477, 419
\bibitem[\protect\citeauthoryear{C\^ot\'e et al.}{2006}]{Cote06} C\^ot\'e, P., Piatek, S., Ferrarese, L., Jord\'an, A., Merritt, D., Peng, E. W., Hasegan, M., Blakeslee, J. P., Mei, S., West, M. J., et al. 2006, ApJS, 165,  57
\bibitem[\protect\citeauthoryear{Cuadra et al.}{2005}]{Cuadra05} Cuadra, J., Nayakshin, S., Springel, V., Di Matteo, T. 2005, MNRAS, 360, L55
\bibitem[\protect\citeauthoryear{Cuadra et al.}{2006}]{Cuadra06} Cuadra, J., Nayakshin, S., Springel, V., Di Matteo, T. 2006,  MNRAS, 366, 358
\bibitem[\protect\citeauthoryear{Cuadra et al.}{2008}]{Cuadra08} Cuadra J., Armitage, Ph. J., Alexander, R. D. 2008, MNRAS, 388,  L64
\bibitem[\protect\citeauthoryear{Dent et al.}{1993}]{Dent93} Dent, W. R. F., Matthews, H. E., Wade, R., Duncan, W. D. 1993, ApJ, 410, 650
\bibitem[\protect\citeauthoryear{Depoyy et al.}{1989}]{DePoy89} Depoy, D. L., Gatley, I., McLean, I. S. 1989, Proceedings of the 136th Symposium of the International Astronomical Union, held in Los Angeles, U.S.A., July 25-29, 1988. Edited by Mark Morris. International Astronomical Union. Symposium no. 136, Kluwer Academic Publishers, Dordrecht, p.411
\bibitem[\protect\citeauthoryear{Detweiler}{1979}]{detweiler1979} Detweiler, S. 1979, ApJ, 234, 1100
\bibitem[\protect\citeauthoryear{Do et al.}{2009}]{Do2009} Do, T., Ghez, A. M., Morris, M. R., Lu, J. R., Matthews, K., Yelda, S., Larkin, J. 2009, ApJ, 703, 1323
\bibitem[\protect\citeauthoryear{Do et al.}{2013}]{Do13} Do, T., Lu, J. R., Ghez, A. M., Morris, M. R., Yelda, S., Martinez, G. D., Wright, S. A., Matthews, K. 2013, ApJ, 764, 154
\bibitem[\protect\citeauthoryear{Donovan et al.}{2006}]{Donovan06} Donovan, J. L., Herrnstein, R., M., Ho, P. T. P. 2006, ApJ, 647, 1159
\bibitem[\protect\citeauthoryear{Downes \&{} Martin}{1971}]{Downes71} Downes, D., Martin, A. H. M. 1971, Nature, 233, 112
\bibitem[\protect\citeauthoryear{Eckart et al.}{1993}]{Eckart93} Eckart, A., Genzel, R., Hofmann, R., Sams, B. J., Tacconi-Garman, L. E. 1993, ApJ, 407, L77
\bibitem[\protect\citeauthoryear{Eckart et al.}{1995}]{Eckart95} Eckart, A., Genzel, R., Hofmann, R., Sams, B. J., Tacconi-Garman, L. E. 1995, ApJ, 445, L23
\bibitem[\protect\citeauthoryear{Eckart et al.}{1997}]{Eckart97} Eckart, A., Genzel, R. 1997, MNRAS, 284, 576
\bibitem[\protect\citeauthoryear{Eckart et al.}{2013}]{Eckart13} Eckart, A., Muzi\'c, K., Yazici, S., Sabha, N., Shahzamanian, B., Witzel, G., Moser, L., Garcia-Marin, M., Valencia-S., M., Jalali, B., et al. 2013, A\&{}A, 551, 18
\bibitem[\protect\citeauthoryear{Eilon et al.}{2009}]{Eilon2009} Eilon, E., Kupi, G., Alexander, T. 2009, ApJ, 698, 641
\bibitem[\protect\citeauthoryear{Eisenhauer et al.}{2005}]{Eisenhauer05} Eisenhauer, F., Genzel, R., Alexander, T., Abuter, R., Paumard, T., Ott, T., Gilbert, A., Gillessen, S., Horrobin, M., Trippe, S., et al. 2005, ApJ, 628, 246
\bibitem[\protect\citeauthoryear{Eisenhauer et al.}{2009}]{Eisenhauer09} Eisenhauer, F. et al. 2009, Science with the VLT in the ELT Era, Astrophysics and Space Science Proceedings. ISBN 978-1-4020-9189-6. Springer Netherlands, 2009, p. 361
\bibitem[\protect\citeauthoryear{Ekers et al.}{1983}]{Ekers83} Ekers, R. D., van Gorkom, J. H., Schwarz, U. J., Goss, W. M. 1983, A\&{}A, 122, 143
\bibitem[\protect\citeauthoryear{Fritz et al.}{2014}]{fritz14} Fritz, T. K., Chatzpoulos, S., Gerhard, O., Gillessen, S., Dodd-Eden, K., Genzel, R., Ott, T., Pfuhl, O., Eisenhauer, F., 2014, submitted
\bibitem[\protect\citeauthoryear{Fujii et al.}{2008}]{Fujii2008} Fujii,M.,Iwasawa,M.,Funato,M.,Makino,J. 2008, ApJ, 686, 1082
\bibitem[\protect\citeauthoryear{Fujii et al.}{2010}]{Fujii2009} Fujii,M.,Iwasawa,M.,Funato,M.,Makino,J. 2009, ApJ, 695, 1421
\bibitem[\protect\citeauthoryear{Fujii et al.}{2010}]{Fujii2010} Fujii,M.,Iwasawa,M.,Funato,M.,Makino,J. 2010, ApJ, 716, L80
\bibitem[\protect\citeauthoryear{Gammie}{2001}]{Gammie01} Gammie, Ch. F. 2001, ApJ, 553, 174
\bibitem[\protect\citeauthoryear{Gatley et al.}{1986}]{Gatley86} Gatley, I., Jones, T. J., Hyland, A. R., Wade, R., Geballe, T. R., Krisciunas, K. 1986, MNRAS, 222, 299
\bibitem[\protect\citeauthoryear{Genzel et al.}{1990}]{Genzel90} Genzel, R., Stacey, G. J., Harris, A. I., Townes, C. H., Geis, N., Graf, U. U., Poglitsch, A., Stutzki, J. 1990, ApJ, 356, 160
\bibitem[\protect\citeauthoryear{Genzel et al.}{1994}]{Genzel94} Genzel, R., Hollenbach, D., Townes, C. H. 1994, 57, 417
\bibitem[\protect\citeauthoryear{Genzel et al.}{1996}]{Genzel96} Genzel, R., Thatte, N., Krabbe, A., Kroker, H., Tacconi-Garman, L. E. 1996, ApJ, 472, 153
\bibitem[\protect\citeauthoryear{Genzel et al.}{1997}]{Genzel97} Genzel, R., Eckart, A., Ott, T., Eisenhauer, F. 1997, MNRAS, 291, 219
\bibitem[\protect\citeauthoryear{Genzel et al.}{2003}]{Genzel03} Genzel, R., Sch\"odel, R., Ott, T., Eisenhauer, F., Hofmann, R., Lehnert, M., Eckart, A., Alexander, T., Sternberg, A., Lenzen, R., et al. 2003, ApJ, 594, 812
\bibitem[\protect\citeauthoryear{Genzel, Eisenhauer \&{} Gillessen}{2010}]{Genzel10} Genzel R., Eisenhauer F., Gillessen S. 2010, Reviews of Modern Physics, 82, 3121
\bibitem[\protect\citeauthoryear{Gerhard}{2001}]{Gerhard2001} Gerhard, O. 2001, ApJ, 546, L39
\bibitem[\protect\citeauthoryear{Ghez et al.}{1998}]{Ghez98} Ghez, A. M., Klein, B. L., Morris, M., Becklin, E. E. 1998, ApJ, 509, 678
\bibitem[\protect\citeauthoryear{Ghez et al.}{2003}]{Ghez03} Ghez, A. M., Duchene, G., Matthews, K., Hornstein, S. D., Tanner, A., Larkin, J., Morris, M., Becklin, E. E., Salim, S., Kremenek, T. et al. 2003, ApJ, 586, L127
\bibitem[\protect\citeauthoryear{Ghez et al.}{2005}]{Ghez05} Ghez, A. M., Salim, S., Hornstein, S. D., Tanner, A., Lu, J. R., Morris, M., Becklin, E. E., Duchene, G. 2005, ApJ, 620, 744
\bibitem[\protect\citeauthoryear{Ghez et al.}{2005}]{Ghez05b}Ghez, A. M., Hornstein, S. D., Lu, J. R., Bouchez, A., Le Mignant, D., van Dam, M. A., Wizinowich, P., Matthews, K., Morris, M., Becklin, E. E., et al. 2005b, ApJ, 635, 1087
\bibitem[\protect\citeauthoryear{Ghez et al.}{2008}]{Ghez08} Ghez, A. M., Salim, S., Weinberg, N. N., Lu, J. R., Do, T., Dunn, J. K., Matthews, K., Morris, M. R., Yelda, S., Becklin, E. E., et al. 2008, ApJ, 689, 1044
\bibitem[\protect\citeauthoryear{Gillessen et al.}{2009a}]{Gillessen09a} Gillessen, S., Eisenhauer, F., Trippe, S., Alexander, T., Genzel, R., Martins, F., Ott, T. 2009a, ApJ, 692, 1075
\bibitem[\protect\citeauthoryear{Gillessen et al.}{2009b}]{Gillessen09b} Gillessen, S., Eisenhauer, F., Fritz, T. K., Bartko, H., Dodds-Eden, K., Pfuhl, O., Ott, T., Genzel, R. 2009b, ApJ, 707, L114
\bibitem[\protect\citeauthoryear{Gillessen et al.}{2012}]{Gillessen12} Gillessen, S., Genzel, R., Fritz, T. K., Quataert, E., Alig, C., Burkert, A., Cuadra, J., Eisenhauer, F., Pfuhl, O., Dodds-Eden, K., Gammie, C. F., Ott, T. 2012, Nature, 481, 51
\bibitem[\protect\citeauthoryear{Gillessen et al.}{2013a}]{Gillessen13a} Gillessen, S., Genzel, R., Fritz, T. K., Eisenhauer, F., Pfuhl, O., Ott, T., Cuadra, J., Schartmann, M., Burkert, A. 2013a, ApJ, 763, 78
\bibitem[\protect\citeauthoryear{Gillessen et al.}{2013b}]{Gillessen13b} Gillessen, S., Genzel, R., Fritz, T. K., Eisenhauer, F., Pfuhl, O., Ott, T., Schartmann, M., Ballone, A., Burkert, A. 2013b, ApJ,  774, 44
\bibitem[\protect\citeauthoryear{Goicoechea et al.}{2013}]{Goicoechea13}Goicoechea, J. R., et al. 2013, ApJ, 769, L13
\bibitem[\protect\citeauthoryear{Goodman}{2003}]{Goodman03} Goodman, J. 2003, MNRAS, 339, 937
\bibitem[\protect\citeauthoryear{Gould \&{} Quillen}{2003}]{GouldQuillen2003} Gould, A,. Quillen, A. 2003, ApJ, 592, 935
\bibitem[\protect\citeauthoryear{Gualandris \&{} Merritt}{2009}]{GM2009} Gualandris, A., Merritt, D. 2009, ApJ, 705, 361
\bibitem[\protect\citeauthoryear{Gualandris et al.}{2010}]{GGM2010} Gualandris, A., Gillessen, S. , Merritt, D. 2010, MNRAS, 409, 1146
\bibitem[\protect\citeauthoryear{Gualandris et al.}{2012}]{Gualandris12} Gualandris, A., Mapelli, M., Perets, H.B. 2012, MNRAS, 427, 1793
\bibitem[\protect\citeauthoryear{Gualandris \&{} Merritt}{2012}]{GM2012} Gualandris, A., Merritt D. 2012, ApJ, 744, 74
\bibitem[\protect\citeauthoryear{Guesten \&{} Downes}{1980}]{Guesten80} Guesten, R., Downes, D. 1980, A\&{}A, 87, 6
\bibitem[\protect\citeauthoryear{Guesten et al.}{1987}]{Gusten87} Guesten, R., Genzel, R., Wright, M. C. H., Jaffe, D. T., Stutzki, J., Harris, A. I. 1987, ApJ, 318, 124
\bibitem[\protect\citeauthoryear{Guillochon et al.}{2014}]{Guillochon14}Guillochon, J., Loeb, A., MacLeod, M., Ramirez-Ruiz, E. 2014, ApJ,  786, L12
 \bibitem[\protect\citeauthoryear{Gvaramadze et~al.}{2009}]{Gvaramadze2009} Gvaramadze, V. V., Gualandris, A., Portegies Zwart, S. 2009, MNRAS,  396, 570
\bibitem[\protect\citeauthoryear{Gvaramadze \&{} Gualandris}{2011}]{GG2011} Gvaramadze, V.V., Gualandris, A. 2011, MNRAS, 410, 304
\bibitem[\protect\citeauthoryear{Haas et al.}{2011a}]{Haas11a} Haas, J., \v{S}ubr, L., Kroupa, P. 2011a, MNRAS, 412, 1905
\bibitem[\protect\citeauthoryear{Haas et al.}{2011b}]{Haas11b} Haas, J., \v{S}ubr, L., Vokrouhlick\'y, D. 2011b, MNRAS, 416, 1023
\bibitem[\protect\citeauthoryear{Haggard et al.}{2014}]{Haggard14} Haggard, D., Baganoff, F. K., Rea, N., Coti Zelati, F., Ponti, G., Heinke, C., Campana, S., Israel, G. L., Yusef-Zadeh, F., Roberts, D. 2014, The Astronomer's Telegram, \# 6242
\bibitem[\protect\citeauthoryear{Haller et al.}{1996}]{Haller96} Haller, Joseph W., Rieke, M. J., Rieke, G. H., Tamblyn, P., Close, L., Melia, F. 1996, ApJ, 456, 194
\bibitem[\protect\citeauthoryear{Hansen \&{} Milosavljevi\'c}{2003}]{Hansen2003} Hansen, B. M. S., Milosavljevi\'c, M. 2003, ApJ, 593, L77	
\bibitem[\protect\citeauthoryear{Harry et al.}{2010}]{Harry2010} Harry, G. M. and the LIGO Scientific Collaboration 2010, Classical and Quantum Gravity 27, 084006
\bibitem[\protect\citeauthoryear{Heber et al.}{2008}] {Heber2008} Heber, U., Edelmann, H., Napiwotzki, R., Altmann, M., Scholz, R.-D. 2008, A\&{}A, 483, L21
\bibitem[\protect\citeauthoryear{Herrnstein \&{} Ho}{2002}]{Herrnstein02} Herrnstein, R. M., Ho, P. T. P. 2002, ApJ, 579, L83
\bibitem[\protect\citeauthoryear{Herrnstein \&{} Ho}{2005}]{Herrnstein05} Herrnstein, R. M., Ho, P. T. P. 2005, ApJ, 620, 287
\bibitem[\protect\citeauthoryear{Hills}{1991}]{Hills1991} Hills, J.G. 1991, AJ, 102, 704
\bibitem[\protect\citeauthoryear{Hills}{1992}]{Hills1992} Hills, J.G. 1992, AJ, 103, 1955
\bibitem[\protect\citeauthoryear{Hills}{1998}]{Hills1998} Hills, J.G. 1992, Nature, 331, 687
\bibitem[\protect\citeauthoryear{Ho et al.}{1985}]{Ho85} Ho, P. T. P., Jackson, J. M., Barrett, A. H., Armstrong, J. T. 1985, ApJ, 288, 575
\bibitem[\protect\citeauthoryear{Ho et al.}{1991}]{Ho91} Ho, P. T. P., Ho, L. C., Szczepanski, J. C., Jackson, J. M., Armstrong, J. T. 1991, Nature, 350, 309
\bibitem[\protect\citeauthoryear{Hobbs \&{} Nayakshin}{2009}]{Hobbs09} Hobbs, A., Nayakshin, S. 2009, MNRAS, 394, 19
\bibitem[\protect\citeauthoryear{Hopman \&{} Alexander}{2006}]{HA2006} Hopman, C., Alexander, T. 2006, ApJ, 645, L133
\bibitem[\protect\citeauthoryear{Hur\'e}{1998}]{Hure98} Hur\'e, J.-M. 1998, A\&{}A, 290, 625
\bibitem[\protect\citeauthoryear{Irrgang et al.}{2010}]{Irrgang2010} Irrgang, A., Przybilla, N., Heber, U., Nieva, M.F., \&{} Schuh, S.  2010, ApJ, 711, 138
\bibitem[\protect\citeauthoryear{Ivanov et al.}{2005}]{Ivanov05} Ivanov, P. B., Polnarev, A. G., Saha, P. 2005, MNRAS, 358, 1361
\bibitem[\protect\citeauthoryear{Jackson et al.}{1993}]{Jackson93} Jackson, J. M., Geis, N., Genzel, R., Harris, A. I., Madden, S., Poglitsch, A., Stacey, G. J., Townes, C. H. 1993, ApJ, 402, 173
\bibitem[\protect\citeauthoryear{Jeans}{1919}]{Jeans19} Jeans, J. H.: Problems of cosmogony and stellar dynamics, 1919, Cambridge, University press, LCCN: 20-9684 (PREM), CALL NUMBER: QB981 .J4
\bibitem[\protect\citeauthoryear{Karlsson et al.}{2003}]{Karlsson03} Karlsson, R., Sjouwerman, L. O., Sandqvist, Aa., Whiteoak, J. B. 2003, A\&{}A, 403, 1011
\bibitem[\protect\citeauthoryear{Kim \&{} Morris}{2003}]{KimMorris2003} Kim, S. S., Morris, M. 2003, ApJ, 597, 312
\bibitem[\protect\citeauthoryear{Kim et al.}{2004}]{Kim2004} Kim, S. S., Figer,D.F., Morris, M. 2004, ApJL, 607, L123
\bibitem[\protect\citeauthoryear{Kocsis et al.}{2012}]{Kocsis2012} Kocsis, B., Ray, A., Portegies Zwart, S. 2012, ApJ, 752, 67
\bibitem[\protect\citeauthoryear{Kolykhalov \&{} Sunyaev}{1980}]{Kolykhalov80} Kolykhalov, P. I., Syunyaev, R. A. 1980, Soviet Astronomy Letters, 6, 357
\bibitem[\protect\citeauthoryear{Koo \&{} McKee}{1992}]{KooMcKee92} Koo, B.-C., McKee, C. F. 1992, ApJ, 388, 93
\bibitem[\protect\citeauthoryear{Koyama et al.}{1996}]{Koyama96} Koyama, K., Maeda, Y., Sonobe, T., Takeshima, T., Tanaka, Y., Yamauchi, S. 1996, PASJ, 48, 249
\bibitem[\protect\citeauthoryear{Kozai}{1962}]{Kozai62} Kozai, Y. 1962, AJ, 67, 591
\bibitem[\protect\citeauthoryear{Krabbe et al.}{1991}]{Krabbe91} Krabbe, A., Genzel, R., Drapatz, S., Rotaciuc, V. 1991, ApJ, 382, L19
\bibitem[\protect\citeauthoryear{Krabbe et al.}{1995}]{Krabbe95} Krabbe, A., Genzel, R., Eckart, A., Najarro, F., Lutz, D., Cameron, M., Kroker, H., Tacconi-Garman, L. E., Thatte, N., Weitzel, L., et al. 1995, ApJ, 447, L95
\bibitem[\protect\citeauthoryear{Kroupa}{2001}]{kroupa2001}Kroupa, P. 2001, MNRAS, 322, 231
\bibitem[\protect\citeauthoryear{Lacy et al.}{1980}]{Lacy80} Lacy, J. H., Townes, C. H., Geballe, T. R., Hollenbach, D. J. 1980, ApJ, 241, 132
\bibitem[\protect\citeauthoryear{Lacy et al.}{1982}]{Lacy82} Lacy, J. H., Townes, C. H., Hollenbach, D. J. 1982, ApJ, 262, 120
\bibitem[\protect\citeauthoryear{Lazio}{2013}]{Lazio2013} Lazio, T. J. W. 2013, Classical and Quantum Gravity, 30, 224011 
\bibitem[\protect\citeauthoryear{Levin \&{} Beloborodov}{2003}]{levin2003} Levin, Y., Beloborodov, A. M. 2003, ApJ, 590, L33
\bibitem[\protect\citeauthoryear{Libonate et al.}{1995}]{Libonate95} Libonate, S., Pipher, J. L., Forrest, W. J., Ashby, M. L. N. 1995, ApJ, 439, 202
\bibitem[\protect\citeauthoryear{Lidov}{1962}]{Lidov62} Lidov, M. L. 1962, Planetary and Space Science, 9, 719
\bibitem[\protect\citeauthoryear{Lin \&{} Pringle}{1987}]{Lin87} Lin, D. N. C., Pringle, J. E. 1987, MNRAS, 225, 607
\bibitem[\protect\citeauthoryear{Lis \&{} Carlstrom}{1994}]{Lis94} Lis, D. C., Carlstrom, J. E. 1994, ApJ, 424, 189
\bibitem[\protect\citeauthoryear{Liu et al.}{2012}]{baobab12} Liu, H. B., Hsieh, P.-Y., Ho, P. T. P., Su, Y.-N., Wright, M., Sun, A.-L., Minh, Y. C. 2012, ApJ, 756, 195
\bibitem[\protect\citeauthoryear{Liu et al.}{2013}]{baobab13} Liu, H. B., Ho, P. T. P., Wright, M. C. H., Su, Y.-N., Hsieh, P.-Y., Sun, A.-L., Kim, S. S., Minh, Y. C. 2013, ApJ, 770, 44
\bibitem[\protect\citeauthoryear{Lo \&{} Claussen}{1983}]{Lo83} Lo, K. Y., Claussen, M. J. 1983, Nature, 306, 647
\bibitem[\protect\citeauthoryear{L\"ockmann et al.}{2008}]{LockmannBaumgardt08} L\"ockmann, U., Baumgardt, H. 2008, MNRAS, 384, 323 
\bibitem[\protect\citeauthoryear{L\"ockmann et al.}{2008}]{Lockmann08} L\"ockmann, U., Baumgardt, H., Kroupa, P. 2008, ApJ, 683, L151
\bibitem[\protect\citeauthoryear{L\"ockmann  et al.}{2009}]{Lockmann09} L\"ockmann, U., Baumgardt, H., Kroupa, P. 2009, MNRAS, 398, 429
\bibitem[\protect\citeauthoryear{L\"ockmann  et al.}{2010}]{lockmann2010kroupa} L\"ockmann, U., Baumgardt, H., Kroupa, P. 2010, MNRAS, 402, 519
\bibitem[\protect\citeauthoryear{Lu et al.}{2006}]{Lu06} Lu, J. R., Ghez, A. M., Hornstein, S. D., Morris, M., Matthews, K., Thompson, D. J., Becklin, E. E. 2006, Journal of Physics: Conference Series, Volume 54, Proceedings of "The Universe Under the Microscope - Astrophysics at High Angular Resolution", held 21-25 April 2008, in Bad Honnef, Germany. Editors: Rainer Schoedel, Andreas Eckart, Susanne Pfalzner and Eduardo Ros, pp. 279-287
\bibitem[\protect\citeauthoryear{Lu et al.}{2009}]{Lu09} Lu, J. R., Ghez, A. M., Hornstein, S. D., Morris, M. R., Becklin, E. E., Matthews, K. 2009, ApJ, 690, 1463
\bibitem[\protect\citeauthoryear{Lu et al.}{2010}]{Lu2010} Lu, Y., Zhang, F., Yu, Q. 2010, ApJ, 709, 1356
\bibitem[\protect\citeauthoryear{Lu et al.}{2013}]{Lu13} Lu, J. R., Do, T., Ghez, A. M., Morris, M. R., Yelda, S., Matthews, K. 2013, ApJ, 764, 155
\bibitem[\protect\citeauthoryear{Lucas et al.}{2013}]{Lucas13} Lucas, W. E., Bonnell, I. A., Davies, M. B., Rice, W. K. M. 2013, MNRAS, 433, 353
\bibitem[\protect\citeauthoryear{Madigan et al.}{2009}]{Madigan2009} Madigan, A., Levin, Y., Hopman, C. 2009, ApJ, 697, L44
\bibitem[\protect\citeauthoryear{Madigan et al.}{2014}]{Madigan2014} Madigan, A., Pfuhl, O., Levin, Y., Gillessen, S., Genzel, R., Perets, H.B. 2014, ApJ, submitted,  arXiv:1305.1625
\bibitem[\protect\citeauthoryear{Maoz}{1998}]{Maoz98} Maoz, E. 1998, ApJ, 494, L181
\bibitem[\protect\citeauthoryear{Mapelli et al.}{2008}]{Mapelli08} Mapelli, M., Hayfield, T., Mayer, L., Wadsley, J. 2008, arXiv0805.0185
\bibitem[\protect\citeauthoryear{Mapelli et al.}{2010}]{Mapelli2010} Mapelli, M., Huwyler, C., Mayer, L., Jetzer, Ph., Vecchio, A. 2010, ApJ, 719, 987
\bibitem[\protect\citeauthoryear{Mapelli et al.}{2012}]{Mapelli12} Mapelli, M., Hayfield, T., Mayer, L., Wadsley, J. 2012, ApJ, 749, 168
\bibitem[\protect\citeauthoryear{Mapelli et al.}{2013}]{Mapelli13}  Mapelli, M., Gualandris, A., Hayfield, T. 2013, MNRAS, 436, 3809
\bibitem[\protect\citeauthoryear{Mapelli \&{} Ripamonti}{2015}]{Mapelli15} Mapelli, M., Ripamonti, E. 2015, ApJ, in press, arXiv:1504.04624
\bibitem[\protect\citeauthoryear{Marr et al.}{1993}]{Marr93} Marr, J. M., Wright, M. C. H., Backer, D. C. 1993, ApJ, 411, 667
\bibitem[\protect\citeauthoryear{Mart\'in et al.}{2012}]{Martin12} Mart\'in, S., Mart\'in-Pintado, J., Montero-Casta\~no, M., Ho, P. T. P., Blundell, R. 2012, A\&{}A, 539, 29
\bibitem[\protect\citeauthoryear{Martins et al.}{2008}]{Martins08} Martins, F., Gillessen, S., Eisenhauer, F., Genzel, R., Ott, T., Trippe, S. 2008, ApJ, 672, L119
\bibitem[\protect\citeauthoryear{Matsubayashi et al.}{2007}]{MME2007} Matsubayashi, T., Makino, J., Ebisuzaki, T. 2007, ApJ, 656, 897
\bibitem[\protect\citeauthoryear{McGary \&{} Ho}{2002}]{McGary02} McGary, R. S., Ho, P. T. P. 2002, ApJ, 577, 757
\bibitem[\protect\citeauthoryear{McGinn et al.}{1989}]{McGinn89} McGinn, M. T., Sellgren, K., Becklin, E. E., Hall, D. N. B. 1989, ApJ, 338,  824
\bibitem[\protect\citeauthoryear{Merritt et al.}{2009}]{MGM2009} Merritt, D., Gualandris, A., Mikkola, S. 2009, ApJL, 693, L35
\bibitem[\protect\citeauthoryear{Merritt}{2010}]{Merritt2010} Merritt, D. 2010, ApJ, 718, 739
\bibitem[\protect\citeauthoryear{Merritt et al.}{2010}]{MAMW2010} Merritt, D., Alexander, T., Mikkola, S., Will, C.M. 2010, Phys.Rev.D, 81, 6
\bibitem[\protect\citeauthoryear{Merritt et al.}{2011}]{MAMW2011} Merritt, D., Alexander, T., Mikkola, S., Will, C.M. 2011, Phys.Rev.D, 84, 044024
\bibitem[\protect\citeauthoryear{Merritt}{2013}]{Merritt13} Merritt, D. 2013, Loss Cone Dynamics, Invited article for the focus issue on ``Astrophysical Black Holes" in Classical and Quantum Gravity, guest editors: D. Merritt and L. Rezzolla
\bibitem[\protect\citeauthoryear{Merritt}{2013}]{Merrittbook} Merritt, D. 2013, Dynamics and Evolution of Galactic Nuclei (Princeton: Princeton University Press)
\bibitem[\protect\citeauthoryear{Meyer et al.}{2012}]{Meyer12} Meyer, L., Ghez, A. M., Sch\"odel, R., Yelda, S., Boehle, A., Lu, J. R., Do, T., Morris, M. R., Becklin, E. E., Matthews, K. 2012, Science, 338, 84
\bibitem[\protect\citeauthoryear{Meyer \&{} Meyer-Hofmeister}{2012}]{Meyer12b}Meyer, F., Meyer-Hofmeister, E. 2012,A\&{}A, 546, L2
\bibitem[\protect\citeauthoryear{Mezger et al.}{1989}]{Mezger89} Mezger, P. G., Zylka, R., Salter, C. J., Wink, J. E., Chini, R., Kreysa, E., Tuffs, R. 1989, A\&{}A, 209, 337
\bibitem[\protect\citeauthoryear{Mignone et al.}{2007}]{Mignone07} Mignone, A., Bodo, G., Massaglia, S., Matsakos, T., Tesileanu, O., Zanni, C., Ferrari, A. 2007, ApJS, 170, 228
\bibitem[\protect\citeauthoryear{Miller	et al.}{2005}]{Miller2005} Miller, M. C., Freitag, M., Hamilton, D. P., Lauburg, V. M. 2005, ApJ, 631, L117
\bibitem[\protect\citeauthoryear{Mills et al.}{2013}]{Mills13} Mills, E. A. C., G\"usten, R., Requena-Torres, M. A., Morris, M. R. 2013, ApJ, 779, 47
\bibitem[\protect\citeauthoryear{Minh et al.}{2013}]{Minh13} Minh, Y. C., Liu, H. B., Ho, P. T. P., Hsieh, P.-Y., Su, Y.-N., Kim, S. S., Wright, M. 2013, ApJ, 773, 31
\bibitem[\protect\citeauthoryear{Miralda-Escud\'e}{2012}]{Escude12} Miralda-Escud\'e, J. 2012, ApJ, 756, 86
\bibitem[\protect\citeauthoryear{Montero-Casta\~no et al.}{2009}]{Montero09} Montero-Casta\~no, M., Herrnstein, R. M., Ho, P. T. P. 2009, ApJ, 695, 1477
\bibitem[\protect\citeauthoryear{Morris}{1993}]{Morris93} Morris, M. 1993, ApJ, 408, 496
\bibitem[\protect\citeauthoryear{Morris \&{} Ghez}{2012}]{Morris12} Morris, M. R., Meyer, L., Ghez, A. M. 2012, Research in Astronomy and Astrophysics, 12, 995
\bibitem[\protect\citeauthoryear{Morris \&{} Serabyn}{1996}]{Morris96} Morris, M., Serabyn, E. 1996, Annual Review of Astronomy and Astrophysics, 34, 645
\bibitem[\protect\citeauthoryear{M\"uller S\'anchez et al.}{2009}]{Muller2009}M\"uller S\'anchez, F., Davies, R. I., Genzel, R., Tacconi, L. J., Eisenhauer, F., Hicks, E. K. S., Friedrich, S., Sternberg, A. 2009, ApJ, 691, 749
\bibitem[\protect\citeauthoryear{Murray-Clay \&{} Loeb}{2012}]{Murray12} Murray-Clay, R. A., Loeb, A. 2012, Nature Communications, 3
\bibitem[\protect\citeauthoryear{Nayakshin \&{} Cuadra}{2005}]{Nayakshin05a} Nayakshin, S., Cuadra, J. 2005, A\&{}A, 437, 437
\bibitem[\protect\citeauthoryear{Nayakshin}{2005}]{Nayakshin05b} Nayakshin, S. 2005, MNRAS, 359, 545
\bibitem[\protect\citeauthoryear{Nayakshin}{2006}]{Nayakshin06} Nayakshin, S. 2006, MNRAS, 372, 143
\bibitem[\protect\citeauthoryear{Nayakshin et al.}{2007}]{Nayakshin07} Nayakshin, S., Cuadra, J., Springel, V. 2007, MNRAS, 379, 21
\bibitem[\protect\citeauthoryear{Novak et al.}{2000}]{Novak00} Novak, G., Dotson, J. L., Dowell, C. D., Hildebrand, R. H., Renbarger, T., Schleuning, D. A. 2000, ApJ, 529, 241
\bibitem[\protect\citeauthoryear{Oka et al.}{2011}]{Oka11} Oka, T., Nagai, M., Kamegai, K., Tanaka, K. 2011, ApJ, 732, 120
\bibitem[\protect\citeauthoryear{Okumura et al.}{1991}]{Okumura91} Okumura, S. K., Ishiguro, M., Fomalont, E. B., Hasegawa, T., Kasuga, T., Morita, K.-I., Kawabe, R., Kobayashi, H. 1991, ApJ, 378, 127
\bibitem[\protect\citeauthoryear{Paczynski}{1978}]{Paczynski78} Paczynski, B. 1978, Acta Astronomica 28, 91
\bibitem[\protect\citeauthoryear{Paumard et al.}{2006}]{Paumard06} Paumard, T., Genzel, R., Martins, F., Nayakshin, S., Beloborodov, A. M., Levin, Y., Trippe, S., Eisenhauer, F., Ott, T., Gillessen, S., et al. 2006, ApJ, 643, 1011
\bibitem[\protect\citeauthoryear{Perets et al.}{2007}]{Perets2007} Perets, H.B., Hopman, C., Alexander, T. 2007, ApJ, 656, 709
\bibitem[\protect\citeauthoryear{Perets et al.}{2009}]{Perets2009} 	
	Perets, H.B, Gualandris, A., Kupi, G., Merritt, D., Alexander, T. 2009, ApJ, 702, 884
\bibitem[\protect\citeauthoryear{Perets \&{} Gualandris}{2010}]{PeretsGualandris2010} Perets, H.B., Gualandris, A. 2010, ApJ, 719, 220
\bibitem[\protect\citeauthoryear{Pfahl \&{} Loeb}{2004}]{pfahl2004} Pfahl, E., Loeb, A. 2004, ApJ, 615, 253
\bibitem[\protect\citeauthoryear{Pfuhl et al.}{2014}]{Pfuhl14}Pfuhl, O., Gillessen, S., Eisenhauer, F., Genzel, R., Plewa, P. M., Ott, T., Ballone, A., Schartmann, M., Burkert, A., Fritz, T. K. 2014, ApJ, accepted
\bibitem[\protect\citeauthoryear{Phifer et al.}{2013}]{Phifer13} Phifer, K., Do, T., Meyer, L., Ghez, A. M., Witzel, G., Yelda, S., Boehle, A., Lu, J. R., Morris, M. R., Becklin, E. E., Matthews, K. 2013, ApJ, 773, L13
\bibitem[\protect\citeauthoryear{Phinney}{1989}]{Phinney89} Phinney, E. S. 1989,  In: The Center of the Galaxy: Proceedings of the 136th Symposium of the International Astronomical Union, held in Los Angeles, U.S.A., July 25-29, 1988. Edited by Mark Morris. International Astronomical Union. Symposium no. 136, Kluwer Academic Publishers, Dordrecht, 543
\bibitem[\protect\citeauthoryear{Pierce-Price et al.}{2000}]{Pierce-Price00} Pierce-Price, D., et al. 2000, ApJ, 545, L121
\bibitem[\protect\citeauthoryear{Piffl et al.}{2011}]{Piffl2011} Piffl, T., Williams, M., Steinmetz, M. 2011, A\&{}A, 535, A70
\bibitem[\protect\citeauthoryear{Ponti et al.}{2010}]{Ponti10} Ponti, G., Terrier, R., Goldwurm, A., Belanger, G., Trap, G. 2010, ApJ, 714, 732
\bibitem[\protect\citeauthoryear{Prodan et al.}{2015}]{Prodan14}Prodan, S., Antonini, F., Perets, H. B. 2015, ApJ, accepted, arXiv1405.6029
\bibitem[\protect\citeauthoryear{Przybilla et al.}{2008}]{Przybilla2008} Przybilla, N., Nieva, M.F., Heber, U., \&{} Butler, K. 2008, ApJL, 684,
  L103
\bibitem[\protect\citeauthoryear{Portegies Zwart et al.}{2004}]{SPZ2004} Portegies Zwart, S., Baumgardt, H., Hut, P., Makino, J., McMillan S. L. W. 2004, Nature 428, 724
\bibitem[\protect\citeauthoryear{Portegies Zwart}{2006}]{SPZ2006} Portegies Zwart, S., 2006, ApJ, 641, 319
\bibitem[\protect\citeauthoryear{Rauch \&{}Tremaine}{1996}]{RT1996} Rauch, K.P., Tremaine, S. 1996, NewA, 1, 149
\bibitem[\protect\citeauthoryear{Rea et al.}{2013}]{Rea2013} Rea, N., Esposito, P., Pons, J. A., Turolla, R., Torres, D. F., Israel, G. L., Possenti, A., Burgay, M., Vigan\`o, D., Papitto, A., et al. 2013, ApJ, 775, L34
\bibitem[\protect\citeauthoryear{Reid \&{} Brunthaler}{2004}]{ReidBru2004} Reid, M.~J., Brunthaler, A. 2004, ApJ, 616, 872
\bibitem[\protect\citeauthoryear{Reid et al.}{2007}]{Reid2007} Reid, M. J., Menten, K. M., Trippe, S., Ott, T., Genzel, R. 2007, ApJ, 659, 378
\bibitem[\protect\citeauthoryear{Rieke \&{} Rieke}{1988}]{Rieke88} Rieke, G. H., Rieke, M. J. 1988, ApJ, 330, L33
\bibitem[\protect\citeauthoryear{Rieke \&{} Rieke}{1989}]{Rieke89} Rieke, G. H., Rieke, M. J. 1989, ApJ, 344, L
\bibitem[\protect\citeauthoryear{Rubilar \&{} Eckart}{2001}]{RubilarEckart2001} Rubilar,G.F., Eckart,A. 2001, A\&{}A, 374, 95 
\bibitem[\protect\citeauthoryear{Salpeter}{1955}]{Salpeter55} Salpeter, E. E. 1995, ApJ, 121, 161
\bibitem[\protect\citeauthoryear{Sanchez-Bermudez et al.}{2014}]{SanchezBermudez2014}Sanchez-Bermudez, J., Sch\"odel, R., Alberdi, A., Muzi\'c, K., Hummel, C. A., Pott, J.-U. 2014, A\&{}A, 567, 16
\bibitem[\protect\citeauthoryear{Sanders}{1998}]{Sanders98} Sanders, R. H. 1998, MNRAS 294, 35
\bibitem[\protect\citeauthoryear{Schartmann et al.}{2012}]{Schartmann12} Schartmann, M., Burkert, A., Alig, C., Gillessen, S., Genzel, R., Eisenhauer, F., Fritz, T. K. 2012, ApJ, 755, 155
\bibitem[\protect\citeauthoryear{Sch\"odel et al.}{2002}]{Schodel02} Sch\"odel, R., Ott, T., Genzel, R., Hofmann, R., Lehnert, M., Eckart, A., Mouawad, N., Alexander, T., Reid, M. J., Lenzen, R., et al. 2002, Nature, 419, 694
\bibitem[\protect\citeauthoryear{Sch\"odel et al.}{2003}]{Schodel03}Sch\"odel, R., Ott, T., Genzel, R., Eckart, A., Mouawad, N., Alexander, T. 2003, ApJ, 596, 1015
\bibitem[\protect\citeauthoryear{Sch\"odel  et al.}{2007}]{Schoedel07}  Sch\"odel, R., Eckart, A., Alexander, T., Merritt, D., Genzel, R., Sternberg, A., Meyer, L., Kul, F., Moultaka, J., Ott, T., Straubmeier, C. 2007, A\&{}A, 469, 125
\bibitem[\protect\citeauthoryear{Sch\"odel et al.}{2009}]{Schoedel09} Sch\"odel, R., Merritt, D., Eckart, A. 2009, A\&{}A, 502, 91
\bibitem[\protect\citeauthoryear{Sch\"odel et al.}{2010}]{Schoedel10} Sch\"odel, R., Najarro, F., Muzic, K., Eckart, A. 2010, A\&{}A, 511, 18 
\bibitem[\protect\citeauthoryear{Scoville et al.}{2003}]{Scoville03} Scoville, N. Z., Stolovy, S. R., Rieke, M., Christopher, M., Yusef-Zadeh, F. 2003, ApJ, 594, 294
\bibitem[\protect\citeauthoryear{Scoville \&{} Burkert}{2013}]{Scoville13} Scoville, N., Burkert, A. 2013, ApJ, 768, 108
\bibitem[\protect\citeauthoryear{Sellgren et al.}{1990}]{Sellgren90} Sellgren, K., McGinn, M. T., Becklin, E. E., Hall, D. N. 1990, ApJ, 359, 112
\bibitem[\protect\citeauthoryear{Serabyn et al.}{1986}]{Serabyn86} Serabyn, E., Guesten, R., Walmsley, J. E., Wink, J. E., Zylka, R. 1986, A\&{}A, 169, 85
\bibitem[\protect\citeauthoryear{Serabyn et al.}{1992}]{Serabyn92} Serabyn, E., Lacy, J. H., Achtermann, J. M. 1992, ApJ, 395, 166
\bibitem[\protect\citeauthoryear{Shcherbakov}{2014}]{Shcherbakov14} Shcherbakov, R. V. 2014, ApJ, 783, 31
\bibitem[\protect\citeauthoryear{Shlosman \&{} Begelman}{1989}]{Shlosman89} Shlosman, I., Begelman, M. C. 1989, ApJ, 341, 685
\bibitem[\protect\citeauthoryear{Solomon et al.}{1972}]{Solomon72} Solomon, P. M., Scoville, N. Z., Penzias, A. A., Wilson, R. W., Jefferts, K. B. 1972, ApJ, 178, 125
%\bibitem[\protect\citeauthoryear{Springel et al.}{2001}]{Springel01} Springel V., Yoshida N., White S. D. M., 2001,New Astronomy, 6, 79
%\bibitem[\protect\citeauthoryear{Springel}{2005}]{Springel05} Springel V., 2005, MNRAS, 364, 1105
\bibitem[\protect\citeauthoryear{\v{S}ubr  et al.}{2009}]{Subr09} \v{S}ubr, L., Schovancov\'a, J., Kroupa, P. 2009, A\&{}A, 496, 695
\bibitem[\protect\citeauthoryear{\v{S}ubr \&{} Haas}{2012}]{Subrhaas12} \v{S}ubr, L., Haas, J. 2012, Journal of Physics: Conference Series, 372
\bibitem[\protect\citeauthoryear{\v{S}ubr  \&{} Haas}{2014}]{Subr14} \v{S}ubr, L., Haas, J. 2014, ApJ, 786, 121
\bibitem[\protect\citeauthoryear{Sunyaev et al.}{1993}]{Sunyaev93} Sunyaev, R. A., Markevitch, M., Pavlinsky, M. 1993, ApJ, 407, 606
\bibitem[\protect\citeauthoryear{Sutton et al.}{1990}]{Sutton90} Sutton, E. C., Danchi, W. C., Jaminet, P. A., Masson, C. R. 1990, ApJ, 348, 503
\bibitem[\protect\citeauthoryear{Tamblyn et al.}{1996}]{Tamblyn96} Tamblyn, P., Rieke, G. H., Hanson, M. M., Close, L. M., McCarthy, D. W., Jr., Rieke, M. J. 1996, ApJ, 456, 206
\bibitem[\protect\citeauthoryear{Telesco et al.}{1996}]{Telesco96} Telesco, C. M., Davidson, J. A., Werner, M. W. 1996, ApJ, 456, 541
\bibitem[\protect\citeauthoryear{Thompson et al.}{2005}]{Thompson05} Thompson, T. A., Quataert, E., Murray, N. 2005, ApJ, 630, 167
\bibitem[\protect\citeauthoryear{Tillich et~al.}{2009}]{Tillich2009} Tillich, A., Przybilla, N., Scholz, R.-D., Heber, U. 2009, A\&{}A, 507, L37
\bibitem[\protect\citeauthoryear{Toomre}{1964}]{Toomre64} Toomre, A. 1964, ApJ, 139, 1217
\bibitem[\protect\citeauthoryear{Trippe et al.}{2008}]{Trippe08} Trippe, S., Gillessen, S., Gerhard, O. E., Bartko, H., Fritz, T. K., Maness, H. L., Eisenhauer, F., Martins, F., Ott, T., Dodds-Eden, K., Genzel, R. 2008, A\&{}A, 492, 419
\bibitem[\protect\citeauthoryear{Tsiklauri \&{} Viollier}{1998}]{Tsiklauri98} Tsiklauri, D., Viollier, R. D. 1998, ApJ, 500, 591
\bibitem[\protect\citeauthoryear{Tsuboi et al.}{2009}]{Tsuboi09} Tsuboi, M., Miyazaki, A., Okumura, S. K. 2009, PASJ, 61, 29
\bibitem[\protect\citeauthoryear{Tsuboi et al.}{2011}]{Tsuboi11} Tsuboi, M.,Tadaki, K.-I., Miyazaki, A., Handa, T. 2011, PASJ, 63, 763
\bibitem[\protect\citeauthoryear{Tsuboi \&{} Miyazaki}{2012}]{Tsuboi12} Tsuboi, M., Miyazaki, A. 2012, PASJ, 64, 111
\bibitem[\protect\citeauthoryear{Vollmer \&{} Duschl}{2001}]{Vollmer01} Vollmer, B., Duschl, W. J. 2001, A\&{}A, 367, 72
\bibitem[\protect\citeauthoryear{Wadsley et al.}{2004}]{Wadsley04} Wadsley, J. W., Stadel, J., Quinn, T. 2004, New Astronomy, 9, 137
\bibitem[\protect\citeauthoryear{Wang \&{} Silk}{1994}]{Wang94} Wang, B., Silk, J. 1994, ApJ, 427, 759
\bibitem[\protect\citeauthoryear{Wardle \&{} Yusef-Zadeh}{2008}]{Wardle08} Wardle, M., Yusef-Zadeh, F. 2008, ApJ, 683, L37
\bibitem[\protect\citeauthoryear{Whiteoak et al.}{1974}]{Whiteoak74} Whiteoak, J. B., Rogstad, D. H., Lockhart, I. A. 1974, A\&{}A, 36, 245
\bibitem[\protect\citeauthoryear{Witzel et al.}{2014}]{Witzel14}Witzel, G., Ghez, A. M., Morris, M. R., Sitarski, B. N., Boehle, A., Naoz, S., Campbell, R., Becklin, E. E., Canalizo, G., Chappell, S., et al. 2014, ApJ, 796, L8
\bibitem[\protect\citeauthoryear{Wright et al.}{2001}]{Wright01} Wright, M. C. H., Coil, A. L., McGary, R. S., Ho, P. T. P., Harris, A. I. 2001, ApJ, 551, 254
\bibitem[\protect\citeauthoryear{Yelda et al.}{2012}]{Yelda12} Yelda, S., Ghez, A. M., Lu, J. R., Do, T., Meyer, L., Morris, M. R. 2012, Adaptive Optics Systems III. Proceedings of the SPIE, 8447, 7 pp.
\bibitem[\protect\citeauthoryear{Yelda et al.}{2013}]{Yelda2013} Yelda, S., Meyer, L., Ghez, A. M., Do, T. 2013, Proceedings of the Third AO4ELT Conference. Firenze, Italy, May 26-31, 2013, Eds.: Simone Esposito and Luca Fini Online at {\tt http://ao4elt3.sciencesconf.org/}, id. \#83
\bibitem[\protect\citeauthoryear{Yelda et al.}{2014}]{Yelda2014} Yelda, S., Ghez, A. M., Lu, J. R., Do, T., Meyer, L., Morris, M. R., Matthews, K. 2014, ApJ, 783, 131
\bibitem[\protect\citeauthoryear{Yu et al.}{2011}]{Yu11} Yu, Y.-W., Cheng, K. S., Chernyshov, D. O., Dogiel, V. A. 2011, MNRAS, 411, 2002
\bibitem[\protect\citeauthoryear{Yu \&{} Tremaine}{2003}]{YuTr2003}{Yu}, Q., \& {Tremaine}, S. 2003, ApJ, 599, 1129
\bibitem[\protect\citeauthoryear{Yungelson et al.}{2008}]{Yungelson2008} Yungelson, L. et al. 2008, A\&A, 477, 223
\bibitem[\protect\citeauthoryear{Yusef-Zadeh \&{} Morris}{1987}]{Yusef-zadeh87a} Yusef-Zadeh, F., Morris, M. 1987, ApJ, 320, 545
\bibitem[\protect\citeauthoryear{Yusef-Zadeh et al.}{2004}]{Yusef-zadeh04} Yusef-Zadeh, F., Hewitt, J. W., Cotton, W. 2004, ApJS, 155, 421
\bibitem[\protect\citeauthoryear{Yusef-Zadeh et al.}{2008}]{Yusef-zadeh08} Yusef-Zadeh, F., Braatz, J., Wardle, M., Roberts, D. 2008, ApJ, 683, L147
\bibitem[\protect\citeauthoryear{Yusef-Zadeh et al.}{2013}]{Yusef-zadeh13} Yusef-Zadeh, F., Royster, M., Wardle, M., Arendt, R., Bushouse, H., Lis, D. C., Pound, M. W., Roberts, D. A., Whitney, B., Wootten, A. 2013, ApJ, 767, L32
\bibitem[\protect\citeauthoryear{Zhao et al.}{2009}]{Zhao09} Zhao, J.-H., Morris, M. R., Goss, W. M., An, T. 2009, ApJ, 699, 186
\bibitem[\protect\citeauthoryear{Zhao et al.}{2010}]{Zhao10} Zhao, J.-H., Blundell, R., Moran, J. M., Downes, D., Schuster, K. F., Marrone, D. P. 2010, ApJ, 723, 1097
\bibitem[\protect\citeauthoryear{Zylka \&{} Mezger}{1988}]{Zylka88} Zylka, R., Mezger, P. G. 1988, A\&{}A, 190, L25
\end{thebibliography}
%

\printindex

\end{document}